\documentclass[10pt,preprint]{aastex}
\usepackage{bm}
\usepackage{threeparttable}
\usepackage{amsmath,amssymb}
\usepackage{textcomp}
\usepackage{booktabs}
\usepackage{lscape}
\usepackage{url}
\makeatletter
\newcommand{\tblcaption}[1]{\def\@captype{table}\caption{#1}}
\def\VEV#1{\left\langle #1\right\rangle}
\makeatother
\usepackage{subfigure}
\usepackage[colorinlistoftodos]{todonotes}
\usepackage{ulem}

\newcommand\ltsima{$\; \buildrel <\over\sim \;$}
\newcommand\simlt{\lower.5ex\hbox{\ltsima}}
\newcommand\gtsima{$\; \buildrel >\over\sim \;$}
\newcommand\simgt{\lower.5ex\hbox{\gtsima}}
\newcommand\msun {M_\odot}

\newcommand\rep {\tilde{r}_E}
\newcommand{\mathbold}[1]{\mbox{\boldmath $\bf#1$}}
\newcommand\piEbold{{\mathbold \pi_{\rm E}}}

\begin{document}


\title{Bayesian Approach for Determining Microlens System Properties with High-Angular-Resolution Follow-up Imaging }


\author{Naoki Koshimoto\altaffilmark{1,2,3}, David P. Bennett\altaffilmark{2,3} and Daisuke Suzuki\altaffilmark{4} 
}


\altaffiltext{1}{Department of Astronomy, Graduate School of Science, The University of Tokyo, 7-3-1 Hongo, Bunkyo-ku, Tokyo 113-0033, Japan}
\altaffiltext{2}{Laboratory for Exoplanets and Stellar Astrophysics, NASA/Goddard Space Flight Center, Greenbelt, MD 20771, USA}
\altaffiltext{3}{Department of Astronomy, University of Maryland, College Park, MD 20742, USA}
\altaffiltext{4}{Institute of Space and Astronautical Science, Japan Aerospace Exploration Agency, 3-1-1 Yoshinodai, Chuo, Sagamihara, Kanagawa, 252-5210, Japan}
 

\begin{abstract}
We present the details of the Bayesian analysis on the planetary microlensing event MOA-2016-BLG-227, whose excess flux is likely due to a source/lens companion or an unrelated ambient star, as well as of the assumed prior distributions.
Furthermore, we apply this method to four reported planetary events, MOA-2008-BLG-310, 
MOA-2011-BLG-293, OGLE-2012-BLG-0527, and OGLE-2012-BLG-0950, where adaptive optics observations
have detected excess flux at the source star positions. For events with small angular Einstein
radii, our lens mass estimates are more uncertain than those of previous analyses who assumed that the excess was due to the lens.
Our predictions for MOA-2008-BLG-310 and OGLE-2012-BLG-0950 are consistent with recent results on these events
obtained via Keck and Hubble Space Telescope observations when the source star is resolvable from the lens star.
For events with small angular Einstein radii, we find that it is generally difficult to conclude whether the excess flux comes from the host star.
Therefore, it is necessary to identify the lens star by measuring its
proper motion relative to the source star to determine whether the excess flux comes from the lens star. 
Even without such measurements, our method can be used to statistically
test the dependence of the planet-hosting probability on the stellar mass.
\end{abstract}


\section{Introduction}
Gravitational microlensing, which has gained a unique niche in the study of extrasolar planetary systems,
enables us to statistically investigate planetary systems down to sub-Earth masses
\citep{ben96} beyond the snow line \citep{suz16} as a function of the galactocentric distance.
It is also sensitive to unbound planets that have been ejected from the systems of their
formation \citep{ben97,sum11,mro17}. A major challenge for the microlensing method is the 
determination of the lens and planetary host star mass, $M_L$. One microlensing light 
curve parameter that is directly related to the host star mass is the Einstein radius crossing 
time $t_{\rm E} = \theta_{\rm E} / \mu_{\rm rel}$, where $\theta_{\rm E}$ is the angular Einstein radius and 
$\mu_{\rm rel}$ is the relative lens-source proper motion. The angular Einstein radius is
given by $\theta_{\rm E} = \sqrt{(4GM_L/c^2)(D_S-D_L)/(D_S D_L)}$, where $D_L$ and $D_S$ are
the distances to the lens and source, respectively. The quantities $t_{\rm E}$ and $\mu_{\rm rel}$ are 
commonly measured in an inertial reference frame that moves with the Earth near the
time of peak magnification of the event. Because $t_{\rm E}$ depends on the lens mass and distance, 
as well as the lens-source relative proper motion, $\mu_{\rm rel}$, measurement of $t_{\rm E}$ does
not yield the lens mass measurement. However, the planet-to-star mass ratio is usually well determined
from the microlensing light curve \citep{gau12}; hence, the planet masses are generally known when the 
host star mass can be measured.

There are three methods for relating the lens mass $M_L$ and distance $D_L$. 
When the microlensing light curve has sharp features, as is the case for most planetary 
events and many stellar binary events, the source radius crossing time, $t_*$, can be measured.
Because the angular source star radius, $\theta_*$, can generally be determined from the de-reddened 
magnitude and color of the source \citep{kervella_dwarf,boy14,ada17}, the
measurement of $t_*$ generally allows the determination of the angular Einstein radius,
$\theta_{\rm E} = \theta_* t_{\rm E}/t_*$, and the lens-source relative proper motion, $\mu_{\rm rel} = \theta_*/t_*$.
Alternatively, the lens-source relative proper motion can also be measured directly from 
high-angular-resolution follow-up observations \citep{ben06,ben15,bat15}. These follow-up
observations can also be used to determine $\theta_{\rm E} = \mu_{\rm rel} t_{\rm E}$, although it is important
to ensure that $\mu_{\rm rel}$ and $t_{\rm E}$ are measured in the same
coordinate system. Direct measurements of the relative proper motion, $\mu_{\rm rel}$, 
are generally performed in a nearly heliocentric coordinate system, while $t_{\rm E}$ is usually
measured in an inertial ``geocentric'' coordinate system that moves with the Earth
near the time of peak magnification. In any case, once $\theta_{\rm E}$ is measured, we have the
following mass-distance relation:
\begin{equation}
M_L = {c^2\over 4G} \theta_{\rm E}^2 {D_S D_L\over D_S - D_L} \ .
\label{eq-m_thetaE}
\end{equation}
Another light curve parameter that can provide the mass-distance relation is the microlensing
parallax \citep{gou92,alc95}, which can be parameterized by the Einstein radius projected from the 
source to the position of the observer, $\rep$. However, it is usually parameterized by the microlensing
parallax parameter, $\pi_{\rm E} = {\rm AU}/\rep$. Actually, $\pi_{\rm E}$ is a two-dimensional vector, $\piEbold$, 
in the same direction as the lens-source relative motion; however, only the length of this vector 
appears in the mass-distance relation: 
\begin{equation}
M_L = {c^2\over 4G} \left({{\rm AU}\over \pi_{\rm E}}\right)^2 { D_S - D_L \over D_S D_L} \ .
\label{eq-m_piE}
\end{equation}
When $\theta_{\rm E}$ and $\pi_{\rm E}$ are both measured, we can directly determine the lens
mass \citep{an-eros2000blg5,gould-lmc5,mur11} by multiplying Eq. (\ref{eq-m_thetaE}) by
Eq. (\ref{eq-m_piE}) and taking the square root to obtain
\begin{equation}
M_L = {\theta_{\rm E} c^2 {\rm AU}\over 4G\pi_{\rm E}}
    = {\theta_{\rm E} \msun\over (8.1439\,{\rm mas}) \pi_{\rm E}} \ .
\label{eq-mass}
\end{equation}

The third method for relating the mass and distance is to detect
and measure the lens star flux, $F_L$. This requires the use of a mass-luminosity relation,
${\cal M}(M_L)$, where ${\cal M}$ is the absolute magnitude in the passband in which the lens star flux is measured \citep{del00}. The measured
lens flux corrected for extinction is $\propto 10^{-0.4 {\cal M}(M_L)}/D_L^2$.
Owing to the extreme crowding in the galactic bulge fields where microlensing events
are observed, the detection of the lens flux requires high-angular-resolution imaging that can be
realized with adaptive optics (AO) systems or the Hubble Space Telescope ({\it HST}). 
Measurement
of the lens flux provides two additional methods for determining the lens mass and distance. The first method is by measurement of the lens flux and $\theta_{\rm E}$
\citep{ben06,ben15,bat15}. The second method is
by measurement of the microlensing parallax and lens flux \citep{kubas12,kos17b,bea18}.
The lens flux plus $\theta_{\rm E}$ method is expected to be the primary exoplanet
system mass measurement method for the WFIRST mission \citep{ben02,ben07,spe15}.

There are two approaches for measuring the lens flux. Both require high-angular-resolution follow-up observations by large 
ground-based telescopes with AO systems or the {\it HST}.
First, even before the lens and source have a
sufficiently large separation to be measured separately, it is possible to obtain the excess 
flux at the position of the event 
because the source flux is readily determined by light curve modeling.
If we can confirm that the excess flux comes from the lens, then we can use the
excess flux as the lens flux, which yields a mass--distance relation.
However, stars other than
the lens star, such as an unrelated star or a companion to the source or lens, may contribute to or even dominate 
this excess flux. This possibility has been considered in some previous analyses 
\citep{jan10,bat14,fuk15,kos17b} of planetary microlensing events; however, these studies have not
always included a consistent treatment of prior and posterior constraints.
In this paper, we present a new systematic Bayesian approach for determining lens star masses and distances
from measurements of the excess flux at the location of the source stars. This new method was used and
briefly explained in the analysis of the MOA-2016-BLG-227 microlensing event \citep{kos17}. Here,
we present the details of this method and apply it to some previously reported events in which excess flux was detected at the position of the source:
MOA-2008-BLG-310 \citep[M08310,][]{jan10}, MOA-2011-BLG-293 \citep[M11293,][]{yee12, bat14}, 
OGLE-2012-BLG-0563 \citep[O120563,][]{fuk15}, 
OGLE-2012-BLG-0950 \citep[O120950,][]{kos17b}, and MOA-2016-BLG-227 \citep[M16227,][]{kos17}. Although the calculation results for M16227 have already been presented previously by \citet{kos17}, we provide
further details in this paper.

Second, if sufficient time has elapsed since the microlensing event such that the lens and source have a
sufficiently large separation to be resolved, then it is possible to directly measure the lens flux unless the lens is too faint.
In this case, the observable quantity is not only the flux, but also the separation between the source and a lens candidate; hence, we can
confirm our prediction on the possible origin of the excess by comparing the 
measured separation and the lens-source relative proper motion obtained through light curve fitting.
If the two values are consistent with each other, then the candidate is probably the lens or a lens companion; otherwise, the candidate is a source companion or an unrelated ambient star.
By this approach, the lens star is shown to be too faint to produce excess flux at the source position \citep{bha17}, or the
lens-source separation can be measured by resolving the lens and source \citep{bat15,van19}, measuring
the elongation of the blended lens-source image \citep{ben07,ben15,ben19,bha18}, or measuring the color-dependent centroid shift of the blended image \citep{ben06}.
In fact, our predictions on the possible origins of the excess fluxes for M08310 and 
O120950 provided in this paper are consistent with that revealed by recent follow-up observations after sufficient time had elapsed for those events \citep{bha17, bha18}.

The main aim of this paper is to provide a Bayesian approach for determining microlensing system properties that consistently treat the prior and posterior probabilities, 
which were not properly treated in previous studies, when excess flux at the position of a microlensing event is measured (the first approach above).
Although we explain the details of our prior choices, they are shown as an example, and some are optimized for the events to which we apply the method; hence, 
one can apply their own choices depending on the character of the event, purpose, preferences, or knowledge.

The remainder of this paper is organized as follows. Section \ref{sec-pro} discusses some problems in the previous analyses of 
the potential ``contamination'' of the flux attributed to the lens star by other stars.
Section \ref{sec-meth} presents the concept of our new Bayesian approach and outline of calculations, while
Section \ref{sec-pri} presents our detailed assumptions and models used to calculate the
prior probability density functions. Section \ref{sec-app} describes the application of this method to 
previously reported events and present the results. Section \ref{sec-interp} discusses the interpretation of the results.
Section \ref{sec-pre} shows how the detectability of the lens flux can be predicted when planning high-angular-resolution follow-up observations.
Section \ref{sec-phost} describes the dependency of our results on the unknown planet hosting probability, while 
Section \ref{sec-vali} describes the dependency on other priors.
Section \ref{sec-valiSLC} tests the binary distribution used in this study by comparing the number of detectable companions predicted by the model 
and the actual number of detected companions.
Section \ref{sec-dis} discusses the overall findings of this study. Finally, Section \ref{sec-summ} gives a summary.

\section{Previous Prior Probability Calculations} \label{sec-pro}
Several previous studies have considered the probability of the observations of excess flux 
being ``contaminated'' by excess flux due to a star or stars other than the lens star
\citep{jan10,bat14,fuk15,kos17b}. These studies considered four possible origins of the 
excess flux: the lens star, unrelated ambient stars, and companions to the source and lens stars.
They calculated the ``prior'' probability that each of the three alternatives other than the lens has 
a brightness in a certain range including the observed excess flux. This range was taken to be the range of the measurement uncertainty in some cases, while it was larger in other cases. Then, 
the sum of these three excess flux ``contamination'' probabilities was subtracted from 1,
and the resulting value was treated as the probability that all of the observed excess flux originates 
from the lens. In most of these cases, the resulting value was relatively large, and it was claimed that the excess flux likely originated from the lens star.

The justification often given for this type of calculation is that we know that the lens star exists, while
stellar companions to the source and lens or ambient stars unresolved from the source may not exist. However, there is no reason to assume that the lens star is likely to be sufficiently bright to be detected, and for some events, such as those with small $\theta_{\rm E}$ values, it is reasonable to expect that the lens star is too faint to be detected.
This is because $\theta_{\rm E} \propto \sqrt{M_L \pi_{\rm rel}}$, where $\pi_{\rm rel} \equiv {\rm AU}\,(1/D_L - 1/D_S)$, and small $\theta_{\rm E}$ indicate 
a small mass and/or distant star, such as an M-dwarf in the Galactic bulge, which is the most common in our Galaxy and too faint to be detected on the position of a much brighter source star.
Moreover, the choice of a particular flux
range that is selected to include the measured excess flux value is not a prior choice, as
it depends on the measured excess flux. The contamination probability determined in this manner also
depends on the somewhat arbitrary flux range that is considered. If this range is taken to be the 
measurement uncertainty, then the difficulty becomes clearer. If the measurement
uncertainty is small, then the contamination probability tends to zero. However, this is just a reflection of 
the fact that the probability of any particular value included in a small uncertainty of a 
precise measurement is small. With
consistent comparison of {\it a priori} and {\it a posteriori} probabilities, the improbability of
a particular precise measurement would affect the probabilities of the different excess flux sources
in a similar manner. A correct Bayesian analysis needs to include the {\it a priori} probabilities for
the detectable flux from the lens star, an unrelated blended star, and companions to the lens
and source stars, and then apply the excess flux measurement constraint to the combined
probability distribution. The previous analyses were flawed because they applied a version
of the measurement constraints to only the ``contamination'' priors while ignoring the possibly
small prior probability that the lens star is detectable.

\section{Method} \label{sec-meth}
We must consider all possible contributions to the excess flux in our analysis. Accordingly, we consider four different
contributions, $F_i$, to the excess flux: flux from the lens star, $F_L$, flux from blended ambient
star(s), $F_{\rm amb}$, flux from a companion to the source, $F_{SC}$, and flux from a companion to the lens, $F_{LC}$. 
Thus, the index $i$ takes four values: $L$, ${\rm amb}$, $SC$, and $LC$. With this notation, the joint posterior
probability density function (PDF) for the excess flux is given by
\begin{align}
f_{post} (&F_L, F_{\rm amb}, F_{SC}, F_{LC} | F_{\rm excess} = F_{\rm ex, obs}) \notag\\  
&\propto {\cal L} (F_{\rm excess} = F_{\rm ex, obs} | F_L, F_{\rm amb}, F_{SC}, F_{LC} ) f_{pri} (F_L, F_{\rm amb}, F_{SC}, F_{LC}), \label{eq-postPDF}
\end{align}
where $f_{pri} (F_L, F_{\rm amb}, F_{SC}, F_{LC})$ is the joint prior PDF of 
$F_L, F_{\rm amb}, F_{SC}$, and $F_{LC}$ and 
\begin{align}
F_{\rm excess} \equiv F_L + F_{\rm amb} + F_{SC} + F_{LC} \ . 
\end{align}
${\cal L} (F_{\rm excess} = F_{\rm ex, obs} | F_L, F_{\rm amb}, F_{SC}, F_{LC} )$ is the likelihood of observed excess flux $F_{\rm ex, obs}$. We use the Gaussian distribution with the measured $F_{\rm ex, obs}$ value and its error in flux unit for it.
The observed excess flux is obtained through subtraction,
\begin{align}
F_{\rm ex, obs} = F_{\rm tar, obs} - F_{S, {\rm obs}},  \label{eq-ftar}
\end{align}
where $F_{\rm tar, obs}$ is the target flux measured in AO or {\it HST} imaging and $F_{S, {\rm obs}}$ is the source flux in the same band-pass as 
the one that conducted the imaging. $F_{S, {\rm obs}}$ is measured from light curve fitting if there is light curve data
in the corresponding band-pass; otherwise, it is converted from the source flux in different band-pass using a color-color relation.
In this study, we apply our method to five events that $H$-band AO imaging observations conducted in previous studies; hence, all the fluxes above are defined as brightness in the $H$-band.
We use $H_{\rm amb}$, $H_L$, $H_{SC}$, $H_{LC}$,  $H_{\rm excess}$, and $H_{\rm ex, obs}$ to denote
the $H$-band magnitudes corresponding to fluxes $F_{\rm amb}$, $F_L$, $F_{SC}$, $F_{LC}$, $F_{\rm excess}$, and $F_{\rm ex, obs}$ respectively.

We note that excess flux can be also measured in any optical band with which the survey observations are conducted, usually in $I$-band or $V$-band, and 
we call this blending flux to distinguish from the excess flux obtained through high-angular-resolution imaging.
We do not use blending flux in our calculation because of the following two reasons.
First, blending flux usually gives information that is irrelevant to the excess in the AO image because 
the angular resolution of the survey observations is $5$--$10$ times worse than that of AO imaging.
This makes the expected number of ambient stars contained in blending flux $25$--$100$ times larger than that in excess flux. 
Given the extreme crowding of the included bulge field, blending flux is very likely to be contaminated by ambient stars that are not included in the excess flux.
A more serious problem with blending flux is that it could be underestimated.
Sky brightness in the bulge field is usually overestimated in a seeing-limited image because a lot of unresolved stars contribute to it, which leads to underestimation of the blending flux.
In fact, \citet{van19} found that the lens flux measured by AO imaging for MOA-2013-BLG-220 was brighter than the upper limit based on the blending flux constrained by \citet{yee14}.

\subsection{Outline of calculation}
We use a Monte Carlo method to calculate the joint prior and joint posterior PDFs.
Fig. \ref{fig-flow} shows a flowchart of our calculation procedure of the two joint PDFs. 
Table \ref{tab-outputs} shows the PDFs and parameters that are modeled in the Monte Carlo simulation, while 
Tables \ref{tab-inputs} and \ref{tab-info} summarize all the models and inputs  that are needed to calculate those parameters, respectively.
In this section, we outline our calculation process to give a perspective, 
where all input parameters in Table \ref{tab-info} are introduced.
Section \ref{sec-pri} describes the details of the assumptions and parameters in the calculation explained here.

\subsubsection{Calculation of the prior probability density function}\label{sec-expri}
In our analysis, we define the prior probabilities as those that do not depend on the measurement of the target flux in the AO image
to be the observed value $F_{\rm tar, obs}$.
Thus, we use any other available information about the target we are analyzing to 
calculate the prior probability, such as the microlensing light curve parameters and 
the full width at half maximum (FWHM) value of the AO image for measuring the target flux.

In every trial of the Monte Carlo simulation, we simulate each of the four objects using the models given in Table \ref{tab-inputs} 
under the constraints from the input parameters given in Table \ref{tab-info}.
The left box in Fig. \ref{fig-flow} shows this procedure. 
We calculate the joint prior PDF $f_{pri} (F_L, F_{\rm amb}, F_{SC}, F_{LC})$ by repeating the trials many times.
Below we briefly summarize calculations of each of the four brightnesses in each trial.

\begin{description}

\item {\it Ambient star flux}

We simulate the ambient star flux by combining the luminosity function (LF) for a field star 
and the distribution of the number of field stars within the resolution element of the AO image where each excess flux was measured.
For the LF of a field star $L_1 (F)$, we use the $H$-band LF from the {\it HST} observations of the Galactic bulge by \citet{zoc03}.

The number of field stars within the size of the resolution element of the AO image follows the Poisson distribution with the mean of $\lambda_{\rm amb}$.
We need the number density of ambient stars in the target field $n_{\rm amb}$ and we characterize the resolution element by a radius of a circle 
where a star within it cannot be resolved from the target $\phi_{\rm wide}$ to calculate the mean $\lambda_{\rm amb} = n_{\rm amb} \pi \phi_{\rm wide}^2$.
We derive $n_{\rm amb}$ for each field by counting the field stars in the AO images or counting the red clump stars in the OGLE-III catalog \citep{szy11} in Section \ref{sec-namb}, and 
we use the radius $\phi_{\rm wide}$ used by the previous studies.
Section \ref{sec-priamb} describes the details of the ambient star flux prior.

\item {\it Lens flux}

The prior lens flux distribution depends on three microlensing parameters observed: the Einstein radius crossing time, 
$t_{\rm E, obs}$, the angular Einstein radius, $\theta_{\rm E, obs}$, and the microlens parallax, $\pi_{\rm E, obs}$.
Combining the Galactic model with those constraints, we derive the joint prior PDF of the lens mass $M_L$, lens distance $D_L$, 
source distance $D_S$, and transverse velocity $v_t$, $f_{pri} (M_L, D_L, D_S, v_{\rm t} | t_{\rm E} = t_{\rm E, obs}, \theta_{\rm E} = \theta_{\rm E, obs}, \pi_{\rm E} = \pi_{\rm E, obs})$, with which 
we can simulate $M_L$, $D_L$ and $D_S$ in every trial of the Monte Carlo simulation.
Section \ref{sec-priML} describes the details.

Given the lens mass and distance from $f_{pri} (M_L, D_L, D_S, v_{\rm t} | t_{\rm E} = t_{\rm E, obs}, \theta_{\rm E} = \theta_{\rm E, obs}, \pi_{\rm E} = \pi_{\rm E, obs})$, 
we can calculate the lens magnitude $H_L$ using both the mass-luminosity 
relation that is described in Section \ref{sec-masslumi} and the extinction for the lens system $A_{H, L}$ that is described in Section \ref{sec-pricomp}.
The calculation of $A_{H, L}$ requires two input parameters for each target field: the mean extinction value for the 
red clump in the vicinity of the target, $A_{H, {\rm rc}}$, and the mean distance modulus to these bulge red clump stars in this field, ${\rm DM}_{\rm rc}$.
$A_{H, {\rm rc}}$ is taken from the previous published paper for each event while ${\rm DM}_{\rm rc}$ is from the value at the nearest grid point to each event from Table 3 of \citet{nat13}.

\item {\it Source companion flux}

In this paper, we analyze events where no stellar companion is detected through the light curve or AO imaging; thus 
we simulate the source companion using the undetected binary distribution, which combines the full binary distribution and detection efficiency for a companion.

Because the source star can be either a single star or the primary or secondary star in 
a binary system, we use the binary distribution for such an arbitrary star, $f_{\rm arb} (q, a | M)$, introduced in Section \ref{sec-bindis}, as the full binary distribution.
In each trial of the Monte Carlo simulation, this function gives mass ratio of either $q_{SC} = 0$, $0 < q_{SC} \leq 1$, or $q_{SC} > 1$, meaning that 
the arbitrary star (i.e., the source star here) is a single, primary, or secondary star, respectively, in addition to the semi-major axis $a_{SC}$.
Because $f_{\rm arb} (q, a | M)$ depends on the arbitrary star mass $M$, we need to input the source mass $M_S$ obtained by applying the 
mass-luminosity relation to the source absolute magnitude ${\cal M}_{H,S} = H_{S, {\rm obs}} - A_{H,S} - {\rm DM}_S$.
We have the source distance $D_S$ from $f_{pri} (M_L, D_L, D_S, v_{\rm t} | t_{\rm E} = t_{\rm E, obs}, \theta_{\rm E} = \theta_{\rm E, obs}, \pi_{\rm E} = \pi_{\rm E, obs})$ 
and use it to calculate the extinction $A_{H, S}$ and distance modulus ${\rm DM}_S$.
The source magnitude $H_{S, {\rm obs}}$ is another input parameter needed and we take the value from the previous paper for each event.

For the detection efficiency, we use $\epsilon_{SC} = \Theta[(\phi-\phi_{\rm wide})(\phi-\phi_{{\rm close},SC})]$ where $\Theta$ is the Heaviside step function.
That is, we assume that a source companion whose projected angular separation $\phi$ is smaller than $\phi_{{\rm close}, SC} = \theta_{\rm E}/4$ or 
larger than the resolution element size of the AO image $\phi_{\rm wide}$ is detectable through the light curve or AO imaging, respectively.
By accepting a combination of $q_{SC}$ and $a_{SC}$ that gives $\epsilon_{SC} = 0$ in the simulation, we have the undetected binary distribution and 
can calculate the magnitude $H_{SC}$ for the accepted companion using the mass-luminosity relation.

\item {\it Lens companion flux}

The lens companion flux is simulated by a similar process to that of the source companion. 
Given the lens mass $M_L$ from $f_{pri} (M_L, D_L, D_S, v_{\rm t} | t_{\rm E} = t_{\rm E, obs}, \theta_{\rm E} = \theta_{\rm E, obs}, \pi_{\rm E} = \pi_{\rm E, obs})$ as input of $f_{\rm arb} (q, a | M)$, 
we have $q_{LC}$ and $a_{LC}$ in each trial of the Monte Carlo simulation.
If the projected angular separation $\phi$ satisfies $\phi_{{\rm close},LC} < \phi < \phi_{\rm wide}$, we accept the trial and calculate magnitude $H_{LC}$.
This means that we use the detection efficiency $\epsilon_{LC} =  \Theta[(\phi-\phi_{\rm wide})(\phi-\phi_{{\rm close},LC})]$ for the lens companion, and 
the closer limit $\phi_{{\rm close}, LC} = \theta_{\rm E} (\sqrt{q_{LC}/u_{\rm 0} + 1} + \sqrt{q_{LC}/u_{\rm 0}})$, where 
$u_{\rm 0}$ is the microlensing impact parameter and we use $u_{\rm 0, obs}$ taken from the previous paper for each event in Table \ref{tab-info}.
This limit comes from the comparison between central caustic size created by the lens companion and impact parameter, which is described in Section \ref{sec-undetbin}.
 
\end{description}
 
\subsubsection{Calculation of the posterior probability density function}
Once we have the joint prior PDF $f_{pri} (F_L, F_{\rm amb}, F_{SC}, F_{LC})$, it is straightforward
to derive the joint posterior PDF of Eq. (\ref{eq-postPDF}) numerically by randomly
selecting combinations of the four flux values following the prior PDF and accepting only the 
combinations that satisfy the condition $F_{\rm excess}= F_{\rm ex, obs}$ within the measurement uncertainty (see the right box in Fig. \ref{fig-flow}).
Here we use the Gaussian distribution in flux unit to judge whether $F_{\rm excess}$ is consistent with $F_{\rm ex, obs}$. 
This requires the last input parameter, the excess flux measured by AO imaging (in magnitude), $H_{\rm ex, obs}$, and the value taken from
the previous paper for each event listed in the last line in Table \ref{tab-info}.

This procedure automatically determines the posterior PDFs of all the parameters that are required to 
calculate these four fluxes listed in Table \ref{tab-outputs} including the lens mass $M_L$ and distance $D_L$.

\section{Prior Probability Density Function} \label{sec-pri}

In this section, we describe the details of the calculation of the joint prior PDF $f_{pri} (F_L, F_{\rm amb}, F_{SC}, F_{LC})$, that is summarized in Section \ref{sec-expri}.
Because $F_{\rm amb}$ is independent of the other variables, we can split the prior joint PDF into
two different functions: 
$f_{pri} (F_L, F_{\rm amb}, F_{SC}, F_{LC}) = f_{pri} (F_L, F_{SC}, F_{LC}) f_{pri} (F_{\rm amb})$.
We discuss these two prior distributions in two different subsections. In Section~\ref{sec-priamb},
we discuss the prior PDF for the ambient star flux, $F_{\rm amb}$. The joint prior PDF for the three
parameters describing the flux of the lens and companion stars, namely $F_L$, $F_{SC}$, and $F_{LC}$, are 
discussed in Section~\ref{sec-pricomp}. 

As described in Section \ref{sec-expri}, the calculation of $f_{pri} (F_L, F_{SC}, F_{LC})$ requires 
the joint prior PDF of the lens mass $M_L$, distance $D_L$, source distance $D_S$, 
a mass-luminosity relation, and the undetected binary distribution.
We describe the joint prior PDF of $M_L$, $D_L$, and $D_S$ and the Galactic model which is needed to calculate the PDF in Section \ref{sec-priML}.
Then we describe the mass-luminosity relation used in this paper in Section \ref{sec-masslumi}. 
Section \ref{sec-bindis} describes the undetected binary distribution which is needed to simulate $F_{SC}$ and $F_{LC}$.

Again, Tables \ref{tab-inputs} and \ref{tab-info} summarize all models and input parameters that are needed to conduct the Bayesian analysis.
These tables also show that calculation of which parameters or models require those models or input parameters in the row denoted as "To model."
The (a) components of Figs. \ref{fig-227}--\ref{fig-0950} show the results of the prior distributions calculated by the Monte Carlo simulation for each of the five events.
 
\subsection{Ambient star flux prior} \label{sec-priamb}
\subsubsection{Number density of ambient stars}\label{sec-namb}
We can derive the number density of ambient stars $n_{\rm amb}$ by counting the stars in a region 
in the vicinity of the target in a high-angular-resolution image; however, such a count can be contaminated
by incompleteness. We can correct for incompleteness with artificial star tests \citep{fuk15} or 
comparison with a luminosity function (LF) with high completeness \citep{kos17}. We use the latter method in this paper, 
i.e., we adopt the LF of \citet{zoc03}. Thus, our number density, $n_{\rm amb}$, includes 
only stars in the magnitude range covered by \citet{zoc03}, $H_0 =$ 7.7--23.5 mag, where $H_0$ is the
extinction-free magnitude. Fainter stars will make no significant contribution to the high-angular-resolution follow-up observations that we consider. A star with $H_0 > 23.5$ would make a contribution
of only 2.5\% for $H_{\rm ex, obs} = 19.7$ mag for M16227 and a smaller fraction for the
other targets. Even if three $H_0 > 23.5$ stars were to exist, this would be still up to a 7.5\% contribution, and
 much smaller than the magnitude excess ($H_{\rm ex, obs}$) uncertainty of 0.4 mag for M16227. 
 Our method includes a correction of the input LF to match the extinction, 
$A_{H, {\rm rc}}$, and mean distance modulus, ${\rm DM}_{\rm rc}$, for the field of each event. 
That is, we add both the extinction $A_{H, {\rm rc}}$ and the difference of ${\rm DM}_{\rm rc}$ from 
the value for Zoccali's field, 14.51 \citep{nat13}, to the extinction-free magnitude of the \citet{zoc03} LF, where we use $A_{H, {\rm rc}}$ and 
${\rm DM}_{\rm rc}$ in Table \ref{tab-info} for each event and ignore their uncertainty.

\begin{figure}
\centering
\epsscale{0.7}
\plotone{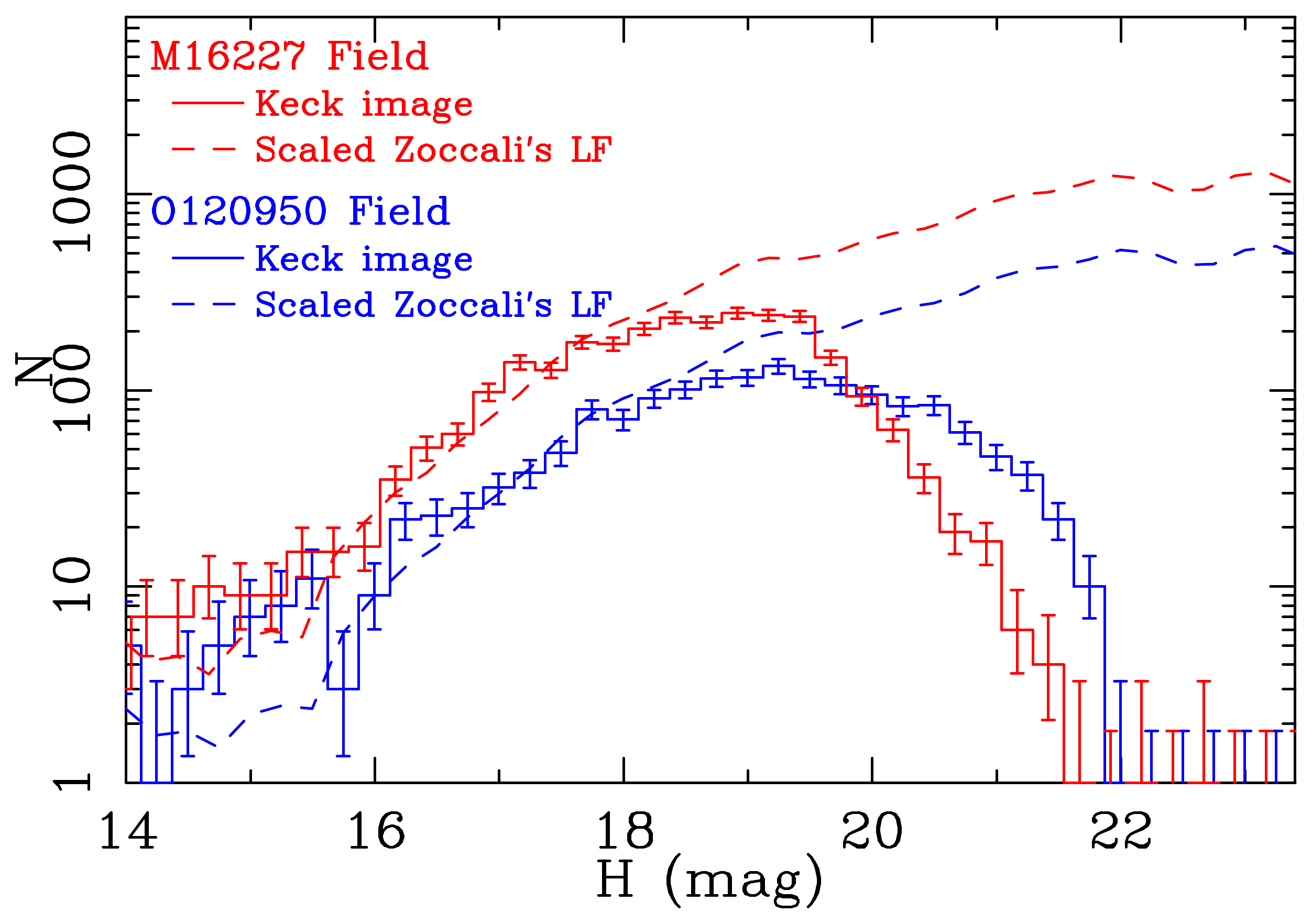}
\caption{Luminosity function in Keck AO images (histogram curves) and that of \citet{zoc03} scaled to the field's area and number density (dashed curves) 
for the fields of M16227 (red) and O120950 (blue). 
The dashed curves are scaled so that the number of stars equals the histogram curves in the magnitude range of $H < 17.91$ for
 M16227 and $H < 17.99$ for O120950.
See Section \ref{sec-priamb} for the details. }
\label{fig-ZocKeck}
\end{figure}

We derive the $n_{\rm amb}$ values for M16227 and O120950 following 
\citet{kos17}, using the number density of stars in high-angular-resolution Keck AO images around each target. 
This can be done because we have an access to the data of Keck images for these two events unlike the other three events.
First, we plot the observed $H$-band LFs for the Keck AO images for these two targets 
(see the solid histogram curves in Fig.~\ref{fig-ZocKeck}), and we see that the LFs 
decrease at $H \simgt 19.5$ mag in both cases, indicating severe  incompleteness.
We count the stars with $H < 17.91$ in the AO image for M16227 and those with $H < 17.99$ 
in the AO image for O120950, and we assume that the detection completeness is 100\% for these stars.
Then, we use these numbers for the normalization of the LF of \citet{zoc03} for each image (see the dashed curves in Fig.~\ref{fig-ZocKeck}). 
For the Subaru AO image of O120563, 
\citet{fuk15} measured the detection completeness of stars with $18.55 < H < 19.11$ as 72\%.
The depth of the Keck AO images is similar to that of the Subaru images. The field of 
O120563 has a star density similar to that of M16227, and it is
denser than that of O120950. Hence, we believe that our completeness assumption is reasonable.
Fig.~\ref{fig-ZocKeck} compares the observed $H$-band LFs for the Keck AO images for these
two targets (solid histogram curves) with the scaled LFs of \citet{zoc03} (dashed curves) for each
field. By integrating the scaled LF and dividing it by the area of each field, we 
find that $n_{\rm amb} = 7.3 \pm 1.0$ as$^{-2}$ for M16227 and 
 $n_{\rm amb} = 4.0 \pm 0.6$ as$^{-2}$ for O120950.
 
We also compared the number of red clump stars in the M08310 field with that in the field 
observed by \citet{zoc03} using the OGLE-III catalog \citep{szy11} to obtain an ambient star density 
of $n_{\rm amb} = 5.5 \pm 1.7$ as$^{-2}$.
We use the same $n_{\rm amb}$ value as that of M16227 for M11293 and 
O120563 because 
M11293 is not located in the OGLE-III survey field and the O120563 coordinate is 
very close to the M16227 coordinate, with a separation of 53 arcsecs. 
Note that a large uncertainty in the $n_{\rm amb}$ value for M11293 does not affect the 
result because of the very small $\phi_{\rm wide}$ value of 
the AO observation for this event, which leads to a negligible contribution from ambient stars to the excess flux.\footnote
{This statement could be false if the assumed $n_{\rm amb}$ value for M11293 is significantly underestimated. However, it is likely overestimated.
\citet{bat14} reported number density of $0.106$ as$^{-2}$ for stars of $18.66 < H < 19.66$ in the AO image for M11293, which 
is more than three times smaller than the number density of $0.358$ as$^{-2}$ for stars in the same brightness range 
in AO image for M16227. 
Although neither of these values are corrected for detection completeness, the completeness of the M11293 image is likely to be higher than that for the M16227 image considering 
its $\sim 2.5$ times smaller $\phi_{\rm wide}$ value than that for M16227.}

\subsubsection{Ambient star flux distribution} \label{sec-ambdis}
If an ambient star is sufficiently close to the source star in our high-resolution images, we will not
be able to resolve it from the source. We denote the separation angle that is the boundary between
resolved and unresolved stars by $\phi_{\rm wide}$. 
All the papers that measured the excess fluxes analyzed in this paper set this limit as constant. We use the value from the previous paper as $\phi_{\rm wide}$ for each event, as listed in Table \ref{tab-info}.
For example, \citet{kos17} used $\phi_{\rm wide} = 0.8\,$FWHM, where the FWHM of $186$ mas is the full width at half 
maximum of objects of the Keck AO image for M16227.
These limits were conservatively set by each study in which the authors were able to resolve an object with excess brightness.
However, objects much fainter than the excess brightness can be missed even at $> \phi_{\rm wide}$ because most of those previous studies have considered only possibility of contamination by a star with the excess brightness.

Nevertheless, we assume that this
limit does not depend on the brightness of the source and the hypothetical blended star.
This is for simplicity, to fairly compare our results with previous studies who used this limit as constant, 
and above all because our results are little affected by the ambient stars as shown in Section \ref{sec-valiamb}.
This is equivalent to setting the detection efficiency of ambient stars located at an angular 
separation $\phi$ from the source star centroid as
\begin{align}
\epsilon_{\rm amb} = \Theta(\phi - \phi_{\rm wide}) \ , \label{eq-epsamb}
\end{align}
where $\Theta$ is the Heaviside step function.

Under the assumption that stars are uniformly randomly distributed in the image with a constant number density $n_{\rm amb}$, 
the number of unrelated ambient stars $N_{\rm amb}$ within a circle of radius $\phi_{\rm wide}$ 
will follow the Poisson distribution 
\begin{align}
Po (N_{\rm amb}; \lambda_{\rm amb}) = \frac{\lambda_{\rm amb}^{N_{\rm amb}} \, e^{-\lambda_{\rm amb}}}{N_{\rm amb}!}, \label{eq-poi}
\end{align}
where $\lambda_{\rm amb} $ is the mean value of $N_{\rm amb}$ 
and 
\begin{align}
\lambda_{\rm amb} =  n_{\rm amb} \, \pi \, \phi_{\rm wide}^2.  \label{eq-lamamb}
\end{align}

Next,  we consider the distribution of the total flux from $N$ stars that are blended together, 
$L_{N} (F_{\rm total})$,
where $F_{\rm total} = \sum \limits_{i=1}^{N} F_i$ and $N$ is fixed.
Because we derived $n_{\rm amb}$ using the LF of \citet{zoc03}, the flux distribution of 
a single ambient star, $L_{1} (F)$, follows this LF. 
Because the flux from each component of the stellar blend, $F_i$, also follows the LF 
$L_{1} (F_i)$, 
the LF for the total flux $F_{\rm total} = \sum \limits_{i=1}^{N} F_i$ can be calculated 
using the recurrence formula
\begin{align}
L_{N} (F_{\rm total}) = \int_{0}^{F_{\rm total}} L_{N - 1} (F) \, L_{1} (F_{\rm total} - F) \, dF \ . \label{eq-LFN}
\end{align}

By combining Eq. (\ref{eq-poi}) and Eq. (\ref{eq-LFN}), we can 
derive the prior PDF of the ambient star flux $F_{\rm amb}$:
\begin{align}
f_{pri} (F_{\rm amb}) = Po (0; \lambda_{\rm amb}) \, \delta (F_{\rm amb}) + \sum\limits_{N_{\rm amb} 
= 1}^{\infty} Po (N_{\rm amb}; \lambda_{\rm amb}) \, L_{N_{\rm amb}} (F_{\rm amb}), \label{eq-priFamb} 
\end{align}
where $\delta (F_{\rm amb})$ is  the Dirac delta function and we define $L_{N_{\rm amb}} (F = 0) = 0$.
Although this distribution seems to depend only on $\lambda_{\rm amb}$, we note that it also 
depends on the average distance modulus ${\rm DM}_{\rm rc}$ and the extinction of the stars in the selected field.

The cyan solid line in the bottom right panel of each
(a) component of each of Figs. \ref{fig-227}-\ref{fig-0950} represents the prior 
probability distribution for the $H$-band ambient star flux, $f_{pri} (F_{\rm amb})$. This requires the use of
the $n_{\rm amb}$, $\phi_{\rm wide}$, $A_{H, {\rm rc}}$, and ${\rm DM}_{\rm rc}$ values for each event, where
we do not include the uncertainties of those parameters for $F_{\rm amb}$.
We do not include a bin for $F_{\rm amb} = 0$ because this corresponds to $H_{\rm amb} = \infty$. 
Instead, we denote the probability of $N_{\rm amb} > 0$ as $P_{\rm exist}$  
in the cyan label in each figure, where 
$P_{\rm exist} = 1 - Po (0, \lambda_{\rm amb})$ for ambient stars.

\subsection{Priors for the lens mass and distance and the source distance} \label{sec-priML}
Many previous studies have estimated event properties via Bayesian analysis 
based on a standard Galactic model and the observed Einstein radius crossing time, $t_{\rm E, obs}$, angular Einstein radius,
$\theta_{\rm E, obs}$, and microlensing parallax, $\pi_{\rm E, obs}$ \citep{alc95, bea06, kos14,ben14}.
These studies produced PDFs for the lens system properties that are referred to as ``posterior'' distributions;
however, in this study, we consider these distributions to be  ``prior'' distributions, because we are considering the
effect of high-angular-resolution follow-up observations on the inferred properties of the lens systems.

We employ the input Galactic model, described below in Section \ref{sec-gal}, to provide our prior PDF for the lens mass $M_L$, the distances to 
the lens and source stars, $D_L$ and $D_S$, respectively, and the relative transverse velocity 
between the lens and the line-of-sight to the source, $v_{\rm t}$, i.e.,
$f_{pri} (M_L, D_L, D_S, v_{\rm t} | t_{\rm E} = t_{\rm E, obs}, \theta_{\rm E} = \theta_{\rm E, obs}, \pi_{\rm E} = \pi_{\rm E, obs})$,
using the same method as that used in the above-mentioned studies, where a 3D normal distribution with no correlation among the 
three is assumed for the likelihood of $t_{\rm E, obs}, \theta_{\rm E, obs},$ and $\pi_{\rm E, obs}$.
When light curve fitting is conducted, there is often a correlation especially among $t_{\rm E}$, $\pi_{\rm E}$, the source flux $F_{\rm S}$, and the blending flux.
Although we do not use the blending flux in our calculation as explained in Section \ref{sec-meth}, we do use $t_{\rm E}$, $\pi_{\rm E}$, 
and also $F_{\rm S}$, which is used to calculate $F_{\rm ex, obs}$.
Therefore, one should ideally apply the joint probability distribution of the fitting parameters which is an output of the 
Markov Chain Monte Carlo fitting to the event light curve data instead of the normal distribution.
Including the correlation might lead a different result when the uncertainty of each parameter is very large. 
However, it is not the case for the five events analyzed in this paper and we do not expect much difference in our results due to the correlation.
Hereafter, we use the notation $f_{pri}' (M_L, D_L, D_S, v_{\rm t})$ instead of
$f_{pri} (M_L, D_L, D_S, v_{\rm t} | t_{\rm E} = t_{\rm E, obs}, \theta_{\rm E} = \theta_{\rm E, obs}, \pi_{\rm E} = \pi_{\rm E, obs})$ because this expression is less cumbersome.

Note that we do not consider remnants, assuming that their probability of hosting planets detectable by microlensing is low. 
We discuss this assumption in Section \ref{sec-rem}.

\subsubsection{Galactic model} \label{sec-gal}
We use the S11 model from \citet{kos19} as our fiducial Galactic model  while we apply different models in Section \ref{sec-galtest}.
The S11 model is a slightly modified version of the \citet{sum11} model, who constructed the model based on \citet{han95}.
The density model consists of the boxy-shaped 
bulge model \citep{dwe95}
\begin{align}
\rho_B =  \rho_{0, B} \exp (-0.5 r_s^2) \ , \label{eq-rhob}
\end{align}
where $r_s =  \{ [(x'/x_0)^2 + (y'/y_0)^2]^2 + (z'/z_0)^4\}^{1/4}$ and the origin of the $(x', y', z')$ 
coordinate is the galactic center. 
The $x'$ axis is along the long axis of the bar, which is inclined at 20$^{\circ}$ to the sun’s direction, 
the $y'$ axis is perpendicular to the $x'$ axis on the 
galactic plane, and the $z'$ axis is toward the galactic north pole.
Moreover, $R = (x^2 + y ^2)^{1/2}$ and the $(x, y, z)$ coordinate rotates the $(x', y', z')$ coordinate 
such that the $x$ axis is toward the sun’s direction.
The S11 model uses galactic bar parameters of $\rho_{0,B} = 2.07~M_{\odot}~$pc$^{-3}$, 
$x_0 = 1580~$pc, $y_0 = 620~$pc, and $z_0 = 430~$pc for the parameters in $\rho_B$ \citep{han95,alc97}.

For the galactic disk, we use the model of \citet{bah86}, i.e., 
\begin{align}
\rho_D =  \rho_{0, D} \exp \left[- \left( \frac{R-R_{GC}}{R_0} + \frac{z}{z_{0,D}} \right) \right], \label{eq-rhod}
\end{align}
with $\rho_{0,D} = 0.06~M_{\odot}~$pc$^{-3}$, $R_{GC} = 8000~$pc, 
$R_0 = 3500~$pc, and $z_{0,D} = 325~$pc for the parameters in $\rho_D$.

For velocity distribution of disk stars, we use the disk rotation speed of 220 km/s and velocity dispersions of 30 km/s and 30 km/s along the azimuthal axis and z-axis, respectively.
A streaming velocity of 50 km/s along $x'$ axis is included and velocity dispersions of (113.6, 77.4, 66.3) km/sec are used along $x'$, $y'$ and $z'$ axes for bar stars.

We consider the present-day mass function $\Phi_{\rm PD} (M)$ as follows.
First we take the initial mass function (IMF) to be
\begin{equation}
\Phi_{\rm IMF} (M) \, dM \propto
\begin{cases}
M^{-2.0} \, dM & \text{ when $M > 0.7 \, M_{\odot}$} \\
M^{-1.3} \, dM & \text{ when $0.08 \, M_{\odot} < M \leq 0.7 \, M_{\odot}$} \\
M^{-0.5} \, dM & \text{ when $0.01 \, M_{\odot} < M \leq 0.08 \, M_{\odot}$} \ ,  \label{eq-IMF}
\end{cases}
\end{equation}
where slopes and breaks are taken from model 4 presented in the Supplementary Information of \citet{sum11}, 
but the high-mass end at $M=1 \, M_{\odot}$ is taken away to make it an IMF. 
We set minimum mass limit to $0.01~M_{\odot}$ to avoid the controversial extension
of this mass function into the planetary mass regime \citep{mro17}.

To construct a present-day mass function from the IMF, we need an age distribution.
The original high-mass cutoff at $M=1 \, M_{\odot}$ in \citet{sum11} corresponds to an assumption that all stars are at $\sim 10$ Gyr because a star with 
initial mass of $M=1 \, M_{\odot}$ evolves into a white dwarf at $\sim 10$ Gyr.
In this paper, we assume a normal distribution for stellar age $T$ instead of the original mono-age assumption.
Let $\overline{T}$ and $\sigma_T$ to be respectively the mean age and the standard deviation., We use $\overline{T} = 5~{\rm Gyr}$ and 
$\sigma_T = 2~{\rm Gyr}$ for the disk component, and $\overline{T} = 9~{\rm Gyr}$ and $\sigma_T = 1~{\rm Gyr}$ for the bulge component. We also limit the disk
stars to lie within the age range $1\,{\rm Gyr}\, < T < 10$ Gyr  and the bulge stars to lie within the range
$4\,{\rm Gyr}\, < T < 11$ Gyr.
We assume solar metallicity for the stellar metallicity, which somewhat affects the lifetime of a star.

By combining the age and metallicity assumptions with the IMF, we construct the present-day mass function $\Phi_{\rm PD} (M)$, where 
we use the PARSEC isochrones \citep{bre12, che14, tan14, che15} model to determine whether the star with the picked initial mass and age has evolved into a remnant or not.
If it is not a remnant yet, then we accept the star with the picked mass; otherwise, we reject it because we do not consider compact objects in this paper.
This Monte Carlo procedure automatically provides a reasonable edge of the high-mass side of the mass function 
corresponding to the age distribution used.

Another assumption that we need to consider is the planet hosting probability, which is previously unknown, because we select lens stars that host planets.
First we perform the Bayesian analysis assuming all stars host planets with the same probability in Section \ref{sec-app}.
In Section \ref{sec-phost}, we apply different priors to the planet hosting probability and show the extent to which our results depend on the prior.

\subsection{Mass-luminosity relation} \label{sec-masslumi}
The next component needed to calculate the prior PDF $f_{pri} (F_L, F_{SC}, F_{LC})$ is the mass-luminosity relation.
In this study, we use different mass-luminosity relations depending on the mass of a star.
For high-mass stars ($M \geq 0.80~M_{\odot}$) where the stellar evolution has a large effect on the luminosity, 
we use PARSEC isochrones \citep{bre12, che14, tan14, che15}, where 
we assume solar metallicity and select age $T$ from the age distribution described in Section \ref{sec-gal}.
For the mass range $0.12 M_{\odot} < M \leq 0.78 M_{\odot}$, we use the empirical relation used 
by \citet{ben15}, which combines the relations of \citet{hen93} and \citet{del00}; 
we use the \citet{hen93} relation for $M > 0.66 \, M_{\odot}$ and we use the \citet{del00} relation for $0.12 M_{\odot} < M < 0.54 M_{\odot}$.
For low-mass stars ($M < 0.1~M_{\odot}$) near brown dwarf transition, we use isochrone models of \citet{bar03} for sub-stellar objects at an age of 5 Gyr.
We linearly interpolate between the two relations used in the boundaries between these mass ranges.

Our choice of using the empirical relation for the intermediate mass range rather than isochrones is motivated by the suggestion of \citet{ben18a} who 
compared PARSEC isochrones with the same empirical mass-luminosity relation as ours and found disagreement between them.

\begin{figure}
\centering
\epsscale{0.5}
\plotone{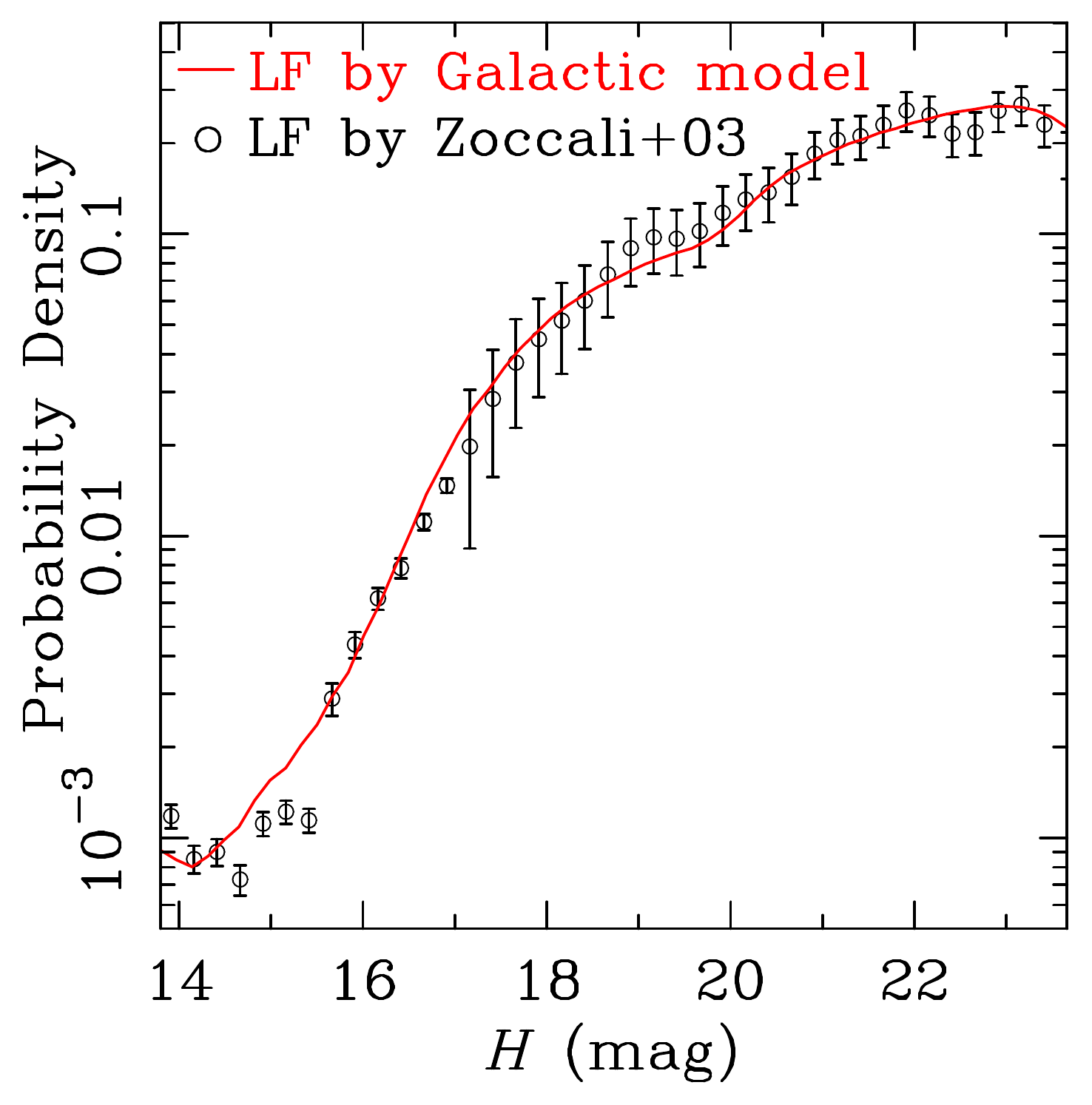}
\caption{Comparison of luminosity function calculated by the mass-luminosity relation and the Galactic model (the S11 model) with
 the luminosity function of \citet{zoc03} in the $H$-band, 
where the extinction, mean distance modulus, and galactic coordinate toward M16227 are used.
The error bars of the LF of \citet{zoc03} are from the Poissonian errors reported by them. }
\label{fig-compLF}
\end{figure}

To test the validity of our choices for the mass-luminosity relation and also for the Galactic model  briefly, we calculate the bulge LF in the $H$-band using those distributions.
Fig. \ref{fig-compLF} compares the model LF with the observed LF of \citet{zoc03}, which is used as $L_1 (F)$ in the ambient star flux prior\footnote{
The LF calculated by the Galactic model  is another choice for $L_1 (F)$ instead of the \citet{zoc03} LF.
However, we believe that using an observed LF as $L_1 (F)$ is more direct and less model-dependent way because some parameters in the Galactic model, such as slopes of 
the mass function, were originally determined by the observation of the LF.}.
The error bars of the LF of \citet{zoc03} are from the Poissonian errors reported by them.
Because they combined near-IR data from observations using different instruments with different fields of view (FOVs) to derive the LF, 
the relative error does not increase monotonically with the value of the vertical axis.
This plot shows that these two observational and model LFs are consistent with each other in the entire range related to lens stars in this paper, $H \simgt 17.5$,
because $H_{\rm ex, obs} = 17.52 \pm 0.10$ is the brightest observed excess flux value in this paper.

We denote the $H$-band absolute magnitude as a function of the mass 
by ${\cal M}_{H} (M)$ from 
the mass-luminosity relations described above, and we denote the mass derived from the 
mass-luminosity relation and absolute magnitude by $M ({\cal M}_{H})$.

\subsection{Binary distribution} \label{sec-bindis}
In this section, we discuss the stellar binary distribution used in our calculations, as this distribution 
is another crucial component of the calculation of the prior PDF for $f_{pri} (F_L, F_{SC}, F_{LC})$. 
We use the term binary distribution
to refer to the probability that a star has a bound companion as a function of its mass ratio and 
semi-major axis.

As described in Section \ref{sec-expri}, we use the undetected binary distribution, which is a combination of full binary distribution and detection 
efficiency, to calculate the source companion flux $F_{SC}$ and lens companion flux $F_{LC}$ in the Monte Carlo simulation to determine the prior PDF $f_{pri} (F_L, F_{SC}, F_{LC})$.
We describe the binary distribution for an ambient star, $f_{\rm arb} (q, a \,|\, M)$, in Section \ref{sec-binarb} because we use it as our full binary distribution.
The calculation of $f_{\rm arb} (q, a \,|\, M)$ requires the binary distribution for a non-secondary star, which is described in Section \ref{sec-binpri}.
We show examples of the full binary distributions by applying the binary distribution for an ambient star to the source and lens systems in M16227 and M11293 in Section \ref{sec-binplot}.
Finally, Section \ref{sec-undetbin} describes the detection efficiencies for a source companion $\epsilon_{SC}$ and a lens companion $\epsilon_{LC}$, then combines them with the full binary distribution to derive the undetected binary distribution.

We make the following simplifying assumptions:
\begin{enumerate}
\item[(i)] We consider only binary systems and ignore the possibility of third-order and higher-order systems.
\item[(ii)] We do not consider the case where lens systems consisting of close binary stars 
have the gravitational lensing effect that closely resembles that of a single star \citep{ben16}. 
We treat such systems as a resolvable binary system and they are ``rejected'' in our Monte Carlo simulation shown in Fig. \ref{fig-flow}.
\item[(iii)] We assume that the existence of the detected planet or planets does not affect the binary distribution.
\item[(iv)] We assume that the location of the star in our galaxy (i.e., differences of stellar number density surrounded, metallicity, age, etc.) does not affect the binary distribution.
\end{enumerate}

\subsubsection{Binary distribution for an arbitrary star}  \label{sec-binarb}
For each of the events that we consider, there is a source star, a lens star, and a planet orbiting the lens star.
The properties of the source star are clearly independent of the planet orbiting the lens star,
and we assume that the properties of the lens star do not depend on the properties or existence of 
the detected planet or planets. 
This allows us to use the same distributions to describe the lens and source systems.

In microlensing, the source and lens stars are selected randomly owing to the alignment with 
the other star (the lens or source star, respectively).  Hence, the source or lens star could be
a secondary star with a more massive companion. We consider the following cases
for the target star, which can be either the source star or the lens star:
\begin{enumerate}
\item The target star is a single star with no stellar companion.
\item The target star is the primary star in a stellar binary system.
\item The target star is the secondary star in a stellar binary system.
\end{enumerate}
The prior information about the target star is different from that in existing observational studies of
binary star systems with nearby stars \citep{duq91, all07, rag10, war15}. 
The systems in these studies are selected on the basis of their brightness; hence, it is common to classify these 
systems on the basis of the properties of the brightest star in the system, which would be either the primary star
or a single star. Hence, these are non-secondary stars.
For microlensing events, we must include the possibility that the source and lens stars are secondary
stars; hence, we cannot simply apply the observational results for non-secondary stars.
In this paper, we refer to a star that could be in any of the three above-mentioned categories, such 
as the source star or the lens star, as an ``arbitrary star'' to distinguish it from a non-secondary star.

We represent the number density of systems that consist of a star of mass $M$ and a second star of mass $qM$, 
separated by a semi-major axis $a$, by $\nu_{\rm arb} (M, q, a) dM dq da$, where
$0 < M < \infty$, $0 \leq q < \infty$ and $0 \leq a < \infty$. We use $(q, a) = (0, 0)$ to
indicate the frequency of single stars with mass $M$. Of course, 
binary systems with $0 < M < \infty$ and $0 < q \leq 1$ can also be represented
by $0 < M < \infty$ and $1 < q < \infty$.
Thus, $\nu_{\rm arb} (M, q, a)$ counts each binary system twice when it is integrated over $0 < M < \infty$, $0 \leq q < \infty$ and $0 \leq a < \infty$. 
However, this double-counting does not exist in what we pursue below, $f_{\rm arb} (q, a | M)$, the binary distribution for a given arbitrary star mass $M$.
With this number density, the binary distribution for an arbitrary star (that is known to exist) with mass 
$M$ is given by a conditional probability:
\begin{equation}
 f_{\rm arb} (q, a \,|\, M) = \frac{\nu_{\rm arb} (M, q, a)}{\nu_{\rm arb} (M)},   \label{eq-confarb1}
\end{equation}
where
\[ \nu_{\rm arb} (M) \equiv \int_{0}^{\infty} \int_{0}^{\infty} \nu_{\rm arb} (M, q, a)\, dq da \ .  \]
We consider target stars that are either source or lens stars such that $M = M_S$ or $M = M_L$.
We can calculate the probability that the source or the lens has a companion as well 
as the probability distribution of the mass ratio $q$ 
and the semi-major axis $a$ of such companions with $f_{\rm arb} (q, a \,|\, M)$.
We consider the number density of arbitrary systems, $\nu_{\rm arb} (M, q, a)$, in the following.

The function $\nu_{\rm arb} (M, q, a)$ represents the number density in each of the three categories 
depending on the mass ratio $q$: 
\begin{equation}
\nu_{\rm arb} (M, q, a) = 
\begin{cases}
\nu_{\rm single} (M)\delta(q)\delta(a) \, , & \text{$q = 0$, $a = 0$} \\
\nu_{\rm prim} (M, q, a) \, , & \text{$0 < q \leq 1$} \\
\nu_{\rm second} (M, q, a) \, , & \text{$1 < q < \infty$} \ . \label{eq-farb1}
\end{cases}
\end{equation}
The number density of binary systems consisting of a primary star whose mass is in the range of $M$--$M+dM$ and a secondary star whose mass is in the range of $qM$--$(q + dq)M$, separated by a
 semi-major axis in the range of $a$--$a + da$, is given by
$\nu_{\rm prim} (M, q, a) \, dM \, dq \, da$.
With changes in the variables $q' = 1/q$ and $M' = q M$, this function also indicates the frequency of 
binary systems with a secondary star whose mass is in the range of $M'$--$M' + dM'$ and a primary star 
whose mass is in the range of $q'M'$--$(q' + dq')M'$, separated by a semi-major axis in the range of $a$--$a + da$.
This is the same as $\nu_{\rm second} (M', q', a) dM' dq' da$. 
This implies a relationship between $\nu_{\rm second} (M, q, a)$ and $\nu_{\rm prim} (M, q, a)$ given by
\begin{align}
\nu_{\rm second} (M, q, a) &= \nu_{\rm prim} (q M, q^{-1}, a) \left| \frac{\partial (qM, q^{-1})}{\partial (M, q)} \right| \notag\\
&=  \nu_{\rm prim} (q M, q^{-1}, a)\, q^{-1}.  \label{eq-fsec1}
\end{align}
This allows us to combine the three expressions in Eq. (\ref{eq-farb1}) into a single expression,
\begin{align}
\nu_{\rm arb} (M, q, a) &= \nu_{\rm single} (M) \, \delta (q) \, \delta (a) + \nu_{\rm prim} (M, q, a) + \nu_{\rm second} (M, q, a) \notag\\
&= \nu_{\rm single} (M) \, \delta (q) \, \delta (a) + \nu_{\rm prim} (M, q, a) + \nu_{\rm prim} (q M, q^{-1}, a)\, q^{-1},  \label{eq-farb2}
\end{align}
where we define the $\nu_{\rm prim}$, $\nu_{\rm second}$, and $\nu_{\rm single}$  functions to be zero outside the 
ranges specified in Eq. ~(\ref{eq-farb1}).\footnote{We might consider 
an additional term $\nu_{\rm single} (q M) \, \delta (1/q) \, \delta (a) \, q^{-1}$ to include the frequency of 
a single star system with mass $q M$ in $\nu_{\rm arb} (M, q, a)$. This would allow
$\nu_{\rm arb} (M, q, a)$ to be invariant after the changes in the variables used in 
Eq. (\ref{eq-fsec1}), i.e., $q' = 1/q$ and $M' = q M$.
However, this term has a non-zero value only when $M = 0$ and $q \rightarrow \infty$, and we will never
consider these values. 
Hence, we do not include this term in $\nu_{\rm arb} (M, q, a) $ in this paper.}

As mentioned above, existing studies on the binary distribution of nearby stars 
\citep{duq91, all07, rag10, war15} presented the binary distribution for non-secondary stars as a
function of their mass $M$. In particular, we refer to two functions given in these studies, namely the 
multiplicity fraction ${\cal F}_{\rm mult} (M)$ and the joint PDF for a secondary star with mass 
ratio $q$ and semi-major axis $a$ orbiting a primary star of mass
$M$, $f_{\rm prim} (q, a \, | \, M)$. This is the binary distribution for a non-secondary star.
The multiplicity fraction ${\cal F}_{\rm mult} (M)$ is the probability that a non-secondary star 
with mass $M$ is a primary star, i.e.,
\begin{align}
{\cal F}_{\rm mult} (M) = \frac{\nu_{\rm prim} (M)}{\nu_{\rm single} (M) + \nu_{\rm prim} (M)}\ ,  \label{eq-MF1}
\end{align}
where
\[ \nu_{\rm prim} (M) \equiv \int_{0}^{1} \left( \int_{0}^{\infty} \nu_{\rm prim} (M, q, a)\, da \right) dq \, .  \]
The function $f_{\rm prim} (q, a \, | \, M)$ is the conditional probability for a secondary star with mass ratio
$q< 1$ and semi-major axis $a$, given a primary star of mass $M$, and it is given by
\begin{equation}
 f_{\rm prim} (q, a \,|\, M) = \frac{\nu_{\rm prim} (M, q, a)}{\nu_{\rm prim} (M)}\ .  \label{eq-confpri1}
\end{equation}
We will present the form of the functions ${\cal F}_{\rm mult} (M)$ and $ f_{\rm prim} (q, a \,|\, M)$ in
Section \ref{sec-binpri}.

To express $\nu_{\rm arb} (M, q, a)$ in terms of ${\cal F}_{\rm mult} (M)$ and $ f_{\rm prim} (q, a \,|\, M)$,
we insert two relations from Eqs. (\ref{eq-MF1})--(\ref{eq-confpri1}), i.e.,
\[ \nu_{\rm prim} (M, q, a) = [\nu_{\rm single} (M) + \nu_{\rm prim} (M)]\, {\cal F}_{\rm mult} (M) \, f_{\rm prim} (q, a \, | \, M) \]
and 
\[ \nu_{\rm single} (M) = [\nu_{\rm single} (M) + \nu_{\rm prim} (M)]\, [1 - {\cal F}_{\rm mult} (M)] \ , \]
into Eq. (\ref{eq-farb2}), and we find that
\begin{align}
\nu_{\rm arb} (M, q, a) =  [\, \nu_{\rm single} (M) &+ \nu_{\rm prim} (M)]\, [\, 1 - {\cal F}_{\rm mult} (M)]  \, \delta (q) \, \delta (a) \notag \\
 +\,  [\, \nu_{\rm single} (M)  &+ \nu_{\rm prim} (M)]\, {\cal F}_{\rm mult} (M) \, f_{\rm prim} (q, a \, | \, M)    \notag\\
 +\,  [\, \nu_{\rm single} (qM) &+ \nu_{\rm prim} (qM)]\, {\cal F}_{\rm mult} (q M) \, f_{\rm prim} (q^{-1}, a \, | \, q M) \, q^{-1} \ . \label{eq-farb3}
\end{align}

Now, we need an expression for $\left[\, \nu_{\rm single} (M) + \nu_{\rm prim} (M) \right ]$. 
We use the stellar present-day mass function $\Phi_{\rm PD} (M)$ defined in Section \ref{sec-gal} for it.
In this paper, we assume that
\begin{align}
\left [\, \nu_{\rm single} (M) + \nu_{\rm prim} (M) \right ] = \nu_0 \, \Phi_{\rm PD} (M), \label{eq-assumphi}
\end{align}
where $\nu_0$ is the number density of stellar systems at the location in question, which is canceled between the denominator and the numerator of 
$f_{\rm arb}(q, a \,|\, M)$.
With this assumption, we have
\begin{align}
\nu_{\rm arb} (M, q, a)/\nu_0 =  \,  \Phi_{\rm PD} (M) \,& [\, 1 - {\cal F}_{\rm mult} (M)]  \, \delta (q) \, \delta (a) \notag \\
 +\,  \Phi_{\rm PD} (M) \, & {\cal F}_{\rm mult} (M) \, f_{\rm prim} (q, a \, | \, M)  \notag \\
 +\,   \Phi_{\rm PD} (q M) \,  & {\cal F}_{\rm mult} (q M) \, f_{\rm prim} (q^{-1}, a \, | \, q M) \, q^{-1} \ ,  \label{eq-farb4}
\end{align}
and thus
\begin{align}
\nu_{\rm arb} (M)/\nu_0 = \,  \Phi_{\rm PD} (M)  +  \int_{1}^{\infty} \left( \int_{0}^{\infty} \Phi_{\rm PD} (q M) \,  {\cal F}_{\rm mult} (q M) \, f_{\rm prim} (q^{-1}, a \, | \, q M) \, q^{-1} da \right) dq  \label{eq-farbM}
\end{align}
by integrating $\nu_{\rm arb} (M, q, a)$ over $q$ and $a$.

Inserting these into Eq. (\ref{eq-confarb1}), we find the binary distribution for an arbitrary star with 
mass $M$:
\begin{align}
f_{\rm arb}(q, a \,|\, M) = & P_{\rm single} (M) \, \delta (q) \, \delta (a) 
+ P_{\rm prim} (M) \, f_{\rm prim} (q, a \, | \, M) \notag\\
&+ P_{\rm second} (M) \, \frac{\Phi_{\rm PD} (q M) \, {\cal F}_{\rm mult} (q M) \, f_{\rm prim} (q^{-1} , a \, | \, q M) \,
   q^{-1}}{ \int_{1}^{\infty} \left[ \int_{0}^{\infty} \Phi_{\rm PD} (q' M) \,  {\cal F}_{\rm mult} (q' M) \, f_{\rm prim} (q'^{-1}, a' \, | \, q' M) \, q'^{-1} da' \right] dq'}\ , \label{eq-fqac}
\end{align}
where $P_{\rm single} (M)$, $P_{\rm prim} (M)$, and $P_{\rm second} (M)$ are the probabilities that an arbitrary star with mass $M$
is a single, primary, and secondary star, respectively. These are given by
\begin{align}
P_{\rm single} (M)  = \frac{\Phi_{\rm PD} (M) \, \left[ 1 - {\cal F}_{\rm mult} (M) \right] }{\nu_{\rm arb} (M)/\nu_0}\ , \label{eq-psingle}
\end{align}
\begin{align}
P_{\rm prim} (M)  = \frac{\Phi_{\rm PD} (M) \,  {\cal F}_{\rm mult} (M)}{\nu_{\rm arb} (M)/\nu_0}\ , \label{eq-phost}
\end{align}
\begin{align}
P_{\rm second} (M)  = \frac{\int_{1}^{\infty} \left[ \int_{0}^{\infty} \Phi_{\rm PD} (q M) \,  {\cal F}_{\rm mult} (q M) \, f_{\rm prim} (q^{-1}, a \, | \, q M) \, q^{-1} da \right] dq}{\nu_{\rm arb} (M)/\nu_0} \ , \label{eq-psecond}
\end{align}
where $P_{\rm single} (M) + P_{\rm prim} (M) + P_{\rm second} (M) = 1$ and the denominators are given by Eq. (\ref{eq-farbM}).
Eq. (\ref{eq-fqac}) gives the complete probability distribution of an arbitrary star of mass $M$ having
a binary companion of any mass ratio and separation, including the case of $q=0$ and $a=0$, which
represents single stars.

\subsubsection{Binary distribution for a non-secondary star} \label{sec-binpri}

To calculate the binary distribution for an arbitrary star $f_{\rm arb}(q, a \,|\, M)$ given by Eq. (\ref{eq-fqac}), 
we need to determine the forms of the binary distribution for a non-secondary star with mass $M$. This is 
the multiplicity fraction ${\cal F}_{\rm mult} (M)$, that is the fraction of primary stars with respect to non-secondary stars defined by Eq. (\ref{eq-MF1}), times the joint PDF at mass ratio $q$ and 
semi-major axis $a$ for a primary star of mass $M$,  $f_{\rm prim} (q, a \, | \, M)$, that is defined by Eq. (\ref{eq-confpri1}).

Following \citet{duc13}, we assume that the mass ratio obeys a power-law distribution and the 
semi-major axis obeys a log-normal distribution. Hence, we use
\begin{align}
f_{\rm prim} (q, a \, | \, M) \propto q^{\gamma} \, \Lambda (a; \eta_{\log a}, \sigma_{\log a}^2)\ ,  \label{eq-fqa}
\end{align}
where 
$\Lambda (a; \eta_{\log a}, \sigma_{\log a}^2)$
is the log-normal distribution and $\eta_{\log a}$ and $\sigma_{\log a}$ are the mean and 
standard deviation of the associated normal distribution, respectively.
Thus, there are four parameters that characterize the binary distribution for a non-secondary star of mass $M$,  
namely the multiplicity fraction ${\cal F}_{\rm mult} (M)$, the slope of the mass-ratio distribution, 
$\gamma = \gamma (a, M)$, and the mean $\eta_{\log a} =  \eta_{\log a} (M)$ and 
standard deviation $\sigma_{\log a} = \sigma_{\log a} (M)$ of the logarithm of the semi-major axis.
All these parameters are considered to be functions of $M$. 
The slope of the mass-ratio distribution $\gamma (a, M)$ depends on the primary 
mass $M$ and its value depends on 
whether the logarithm of the semi-major axis is larger or smaller than its mean value: 
\begin{equation}
\gamma (a, M) = 
\begin{cases}
\gamma_c (M)  & \text{ when $\log [a/{\rm AU}] < \eta_{\log a} (M)$} \\
\gamma_w (M) & \text{ when $\log [a/{\rm AU}] \geq \eta_{\log a} (M)$}\ .  \label{eq-gamma}
\end{cases}
\end{equation}
We set $f_{\rm prim} (q, a \, | \, M) = 0$ with $q < 0.1$ because binaries with $q < 0.1$ are thought to be
rare \citep{duc13} and because such systems are rarely important for our calculations. \citet{duc13}
derived the slope of the mass-ratio distribution, $\gamma$, using binary systems with $0.1 < q \leq 1$.

The mass dependence of these four parameters that characterize our $f_{\rm prim} (q, a \, | \, M)$ 
function is not well understood thus far. For our analysis, we fit each of these parameters to the data
summarized by \citet{duc13} with two models that are linear in $M$ and $\log M$, and we use the
one that gives better fit to the data for each parameter.
In addition to the data summarized by \citet{duc13}, we add the data at $M = 0.4 \pm 0.2 M_{\odot}$ given by \citet{war15}  to 
determine the $M$ dependence of ${\cal F}_{\rm mult} (M)$, $\eta_{\log a} (M)$, and $\sigma_{\log a} (M)$.

\begin{figure}
\centering
\begin{minipage}{0.35\hsize}
\begin{center}
\includegraphics[width=58mm]{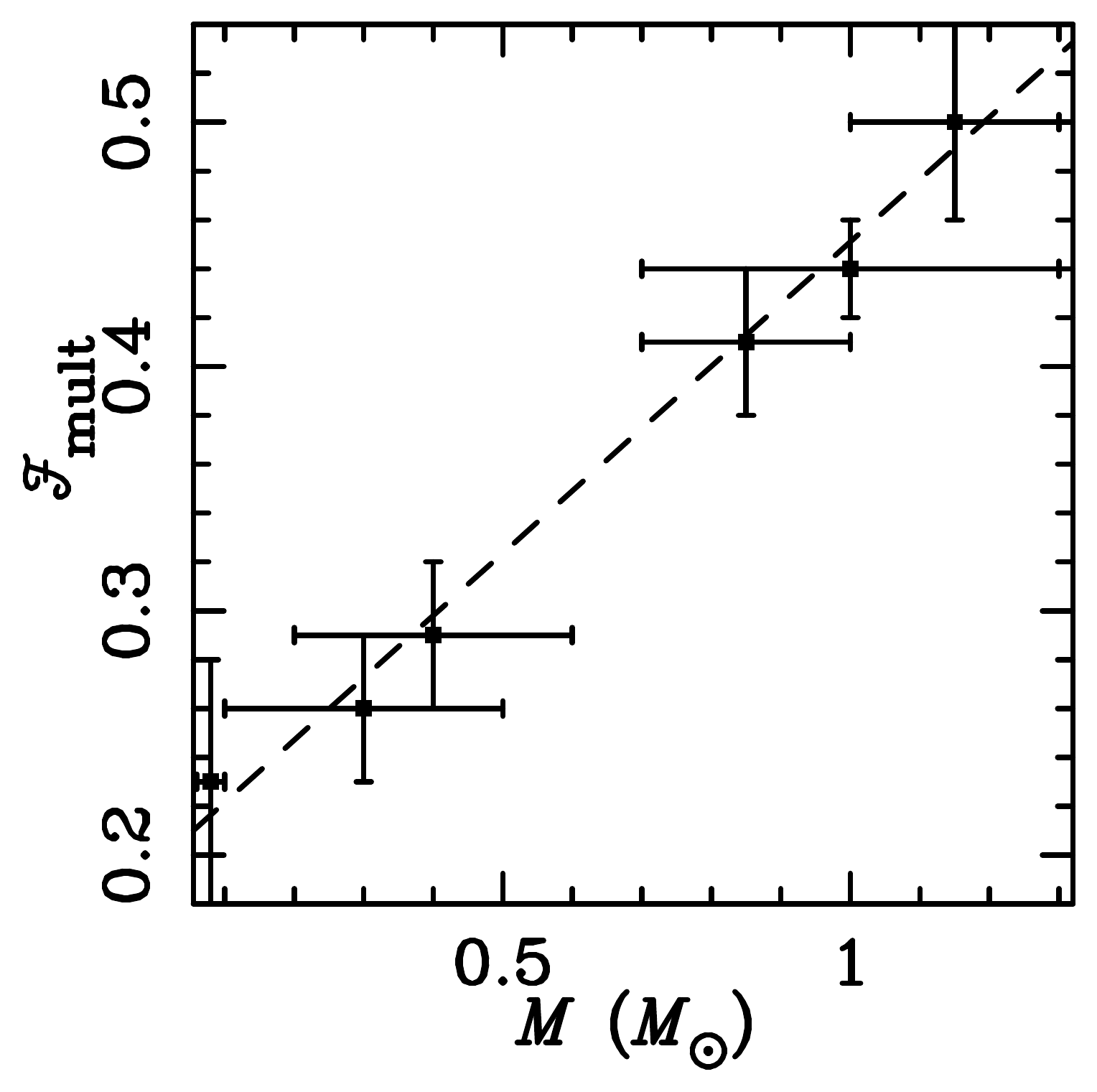}
\end{center}
\end{minipage}
\begin{minipage}{0.35\hsize}
\begin{center}
\includegraphics[width=58mm]{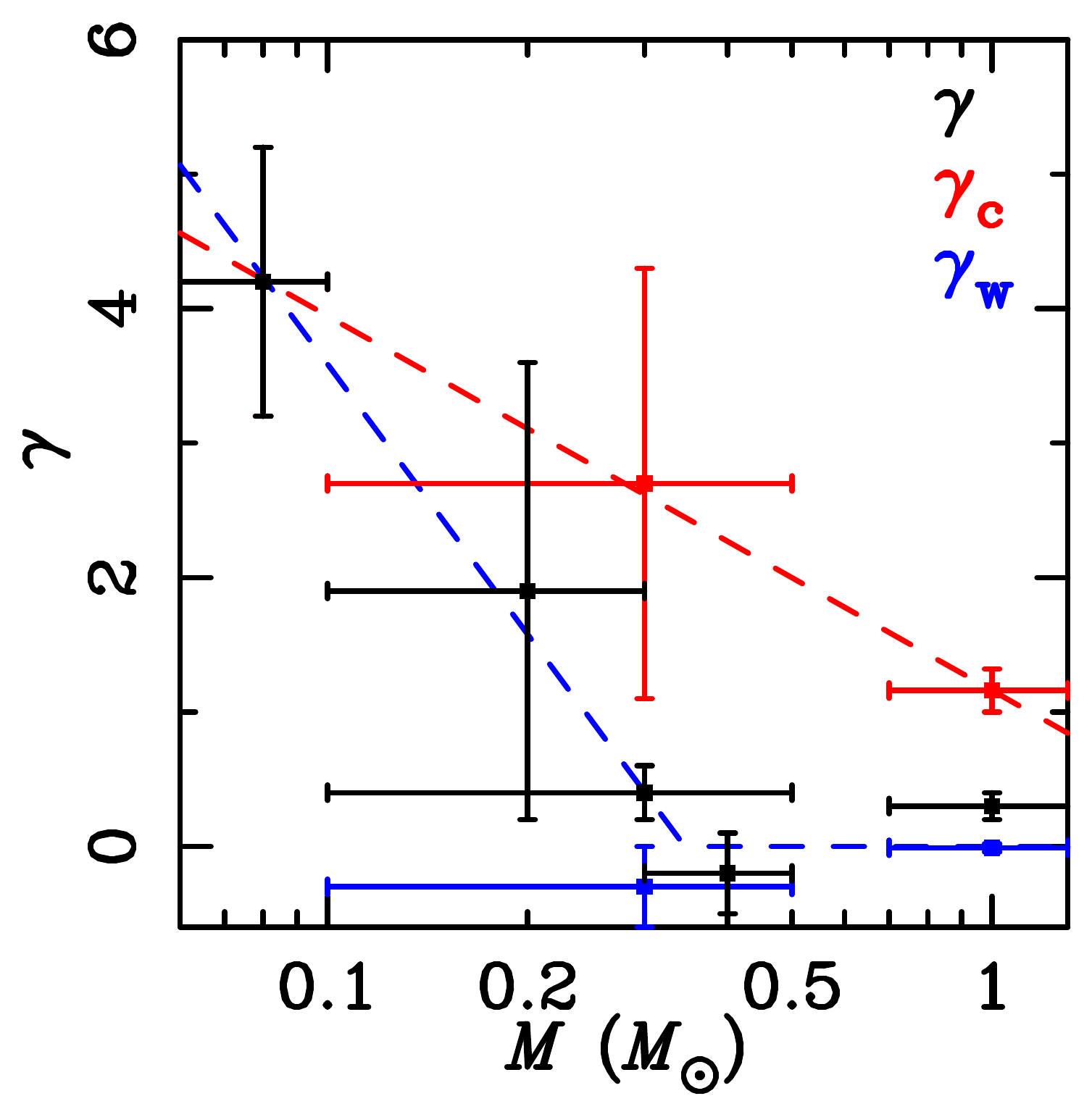}
\end{center}
\end{minipage}

\begin{minipage}{0.35\hsize}
\begin{center}
\includegraphics[width=58mm]{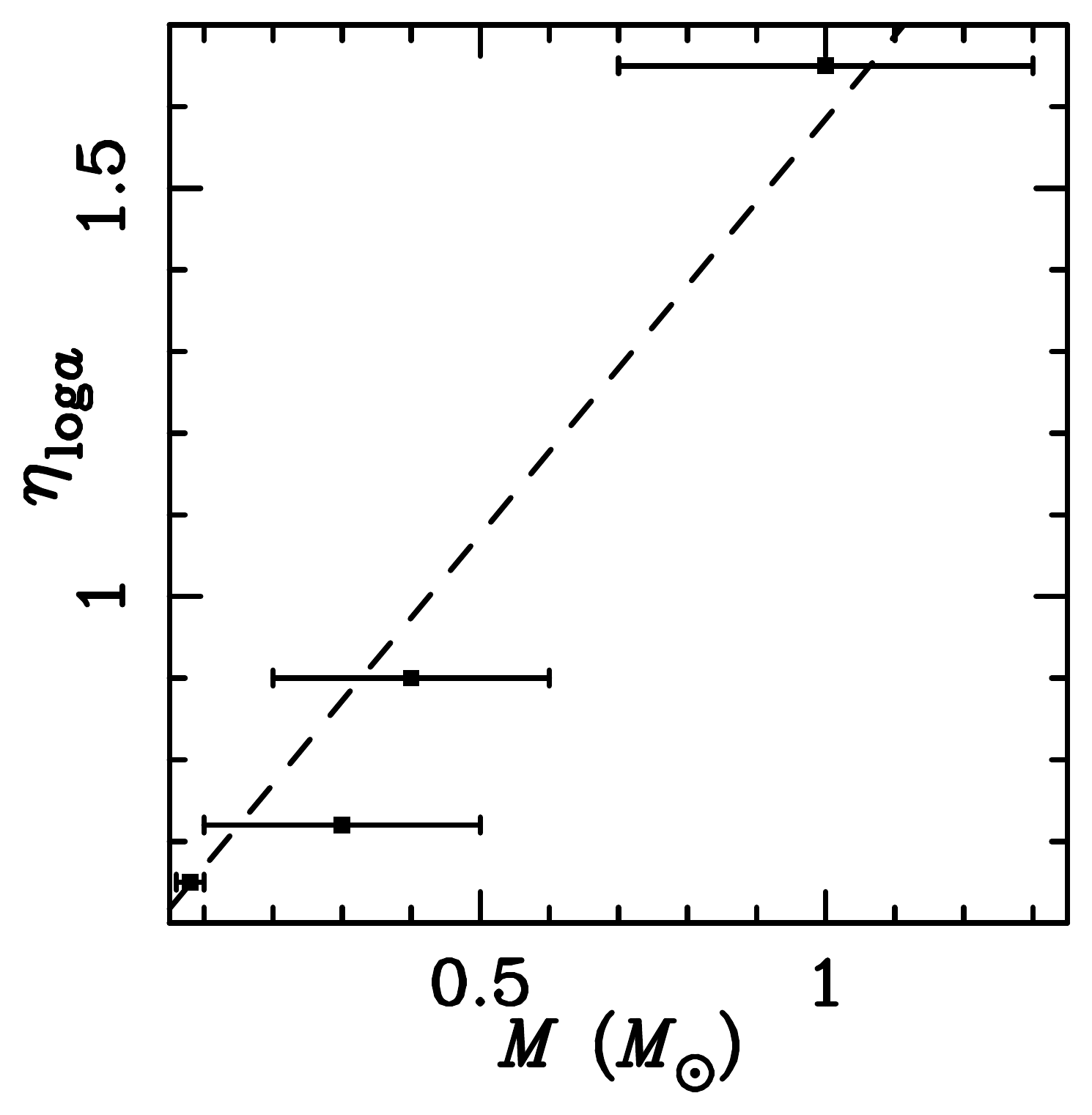}
\end{center}
\end{minipage}
\begin{minipage}{0.35\hsize}
\begin{center}
\includegraphics[width=58mm]{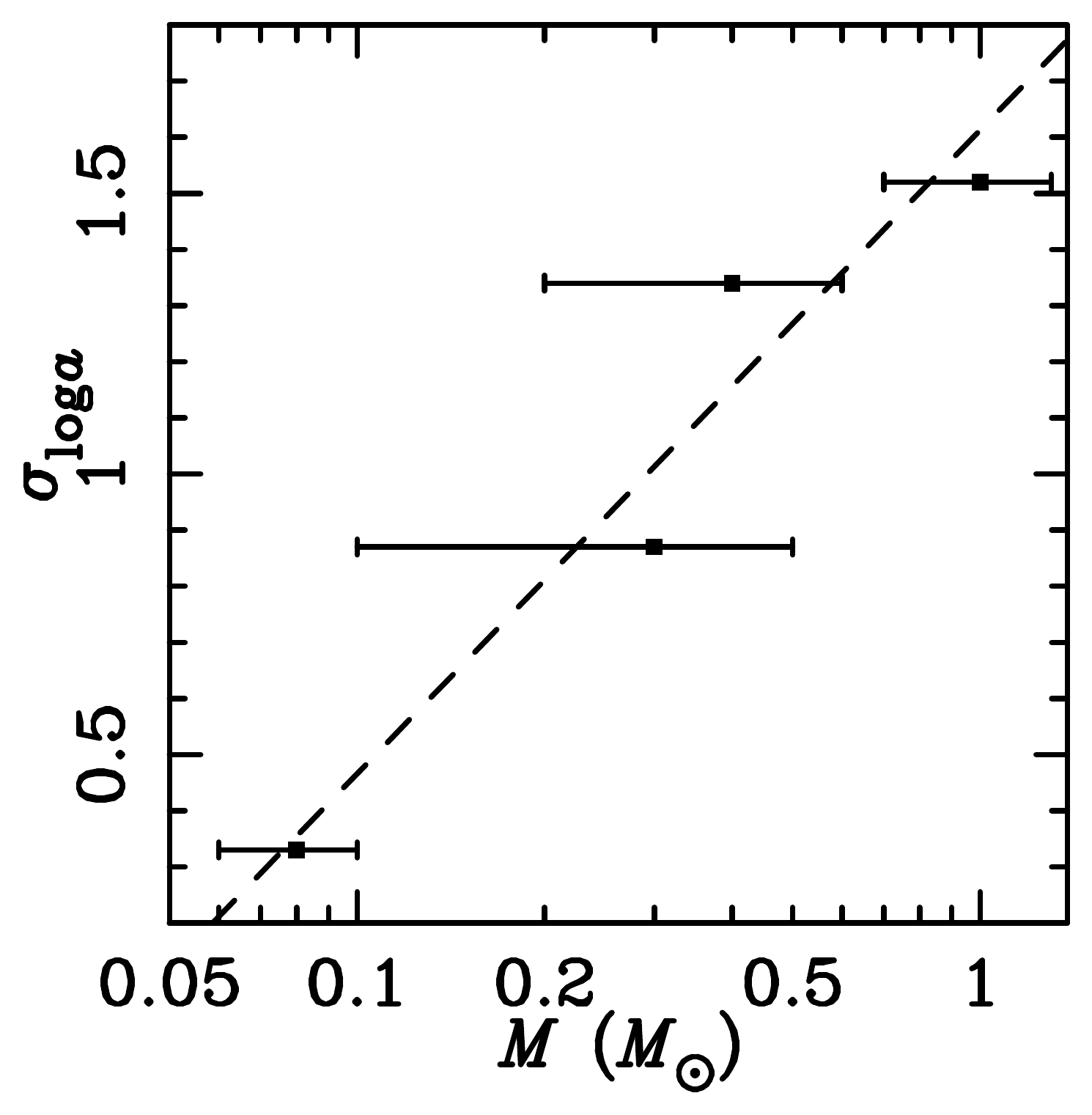}
\end{center}
\end{minipage}
\caption{Parameters for the description of the binary distribution for a non-secondary star as functions of the star mass $M$.
The multiplicity fraction ${\cal F}_{\rm mult}$, the slope of the mass-ratio distribution, $\gamma$, 
the mean and the standard deviation of the log semi-major axis distribution, $\eta_{\log a}$ and $\sigma_{\log a}$, respectively,
are plotted.
The data are from \citet{duc13}, but the values 
of \citet{war15} are added at $M = 0.4 \pm 0.2 M_{\odot}$ for the three parameters ${\cal F}_{\rm mult}$, $\eta_{\log a}$, and $\sigma_{\log a}$.
In the top right panel, the red ($\gamma_c$), blue ($\gamma_w$), and black ($\gamma$) points 
with error bars indicate the slopes of the mass-ratio distribution
for close ($\log a < \eta_{\log a}$), wide ($\log a \geq \eta_{\log a}$), and all binary systems, respectively. 
The dashed lines indicate our best linear fit models.}
\label{fig-parafit}
\end{figure}

We plot the values of these parameters as a function of $M$ and show each of the best-fit models
for each parameter in Fig. \ref{fig-parafit}. Further, we summarize the best-fit models in Table \ref{tab-inputs}.
We conduct fitting of the slopes of the mass-ratio distribution, $\gamma_c (M)$ and $\gamma_w (M)$, as follows.
\citet{duc13} derived the slope of the mass-ratio distribution for companions of primary stars by 
fitting the mass-ratio distributions to a power law with a region of $0.1 < q < 1$.
They derived the slope $\gamma $ using their full sample within $0.1 < q < 1$, and they also determined
the power-law exponents for close ($\gamma_c$) and wide ($\gamma_w$) binaries with
semi-major axes logarithms that are smaller and larger than the mean value ${\eta_{\log a}}$, 
respectively. 
We show the values of $\gamma$, $\gamma_c$, and $\gamma_w$, based on the work of \citet{duc13}, as a function
of $M$ as the black, red, and blue dots in the top right panel in Fig.  \ref{fig-parafit}, respectively.
We do not plot $\gamma_c$ and $\gamma_w$ values for $M \leq 0.2 M_{\odot}$, because \citet{duc13}
reported only $\gamma$ and not $\gamma_c$ or $\gamma_w$ in this mass range.
Consequently, we use $\gamma$ values represented by the black dots 
at low masses when fitting $\gamma_c (M)$ (the red dashed line) and $\gamma_w (M)$ (the blue dashed line).
We use $\gamma = 0.42$ at $0.08 M_{\odot}$ in our fit to determine $\gamma_c (M)$. 
To determine $\gamma_w (M)$, we assume that $\gamma_{w} = \gamma$ for $M < 0.3M_{\odot}$ because 
this condition is true when $M > 0.3M_{\odot}$. Then, we conduct linear fitting of $\gamma(M)$ in the region
$M < 0.3~M_{\odot}$ and we use the result for $\gamma_w (M)$ at $M < 0.34~M_{\odot}$ 
(the sloping part of the blue dashed line). For $M \geq 0.34~M_{\odot}$, $\gamma_w$ seems to be 
approximately constant; hence, we use $\gamma_w = 0$ for this mass range (indicated by the 
flat part of the blue dashed line). 

\begin{figure}
\begin{minipage}{0.5\hsize}
\begin{center}
\includegraphics[width=78mm]{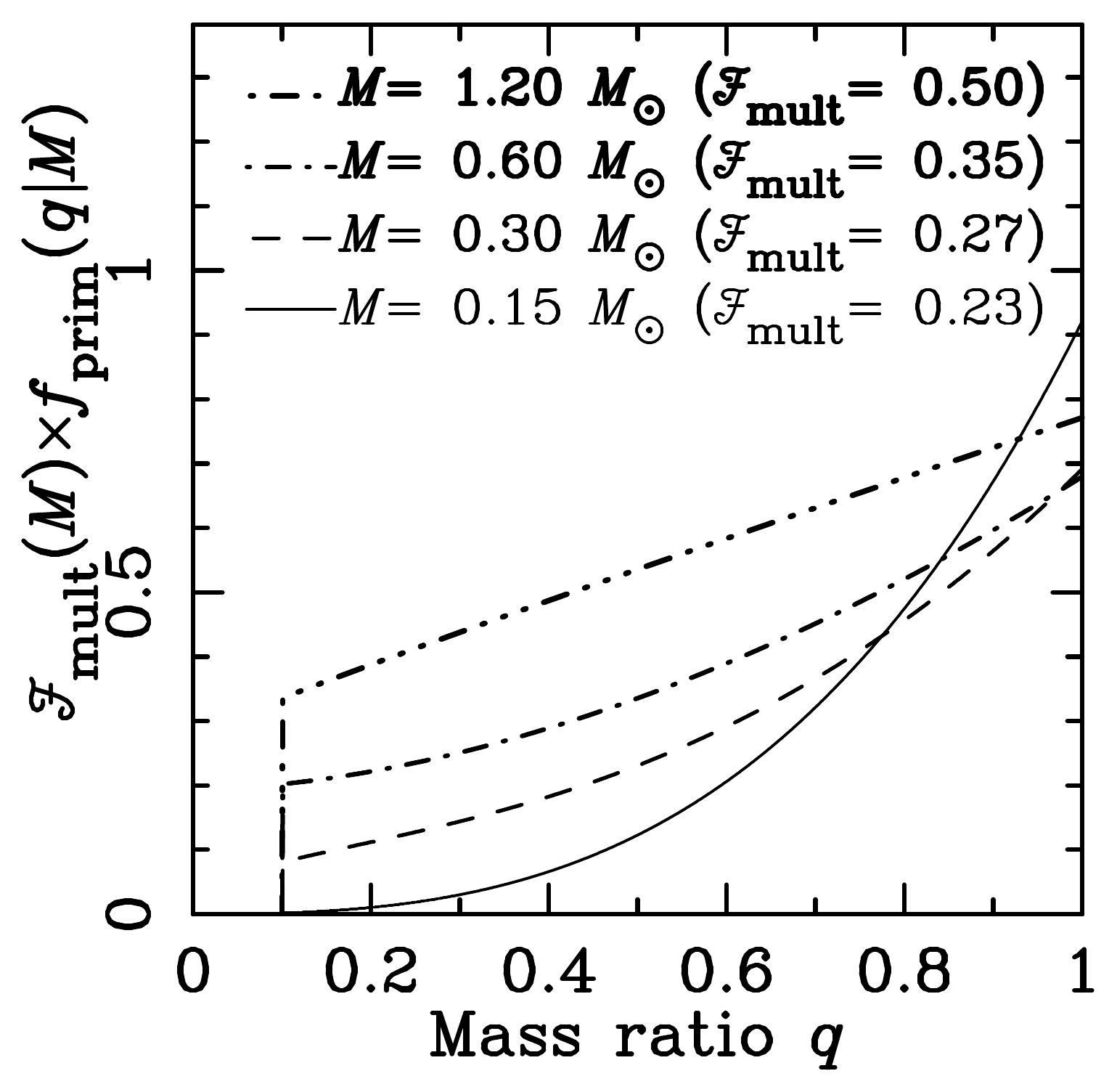}
\end{center}
\end{minipage}
\begin{minipage}{0.5\hsize}
\begin{center}
\includegraphics[width=82mm]{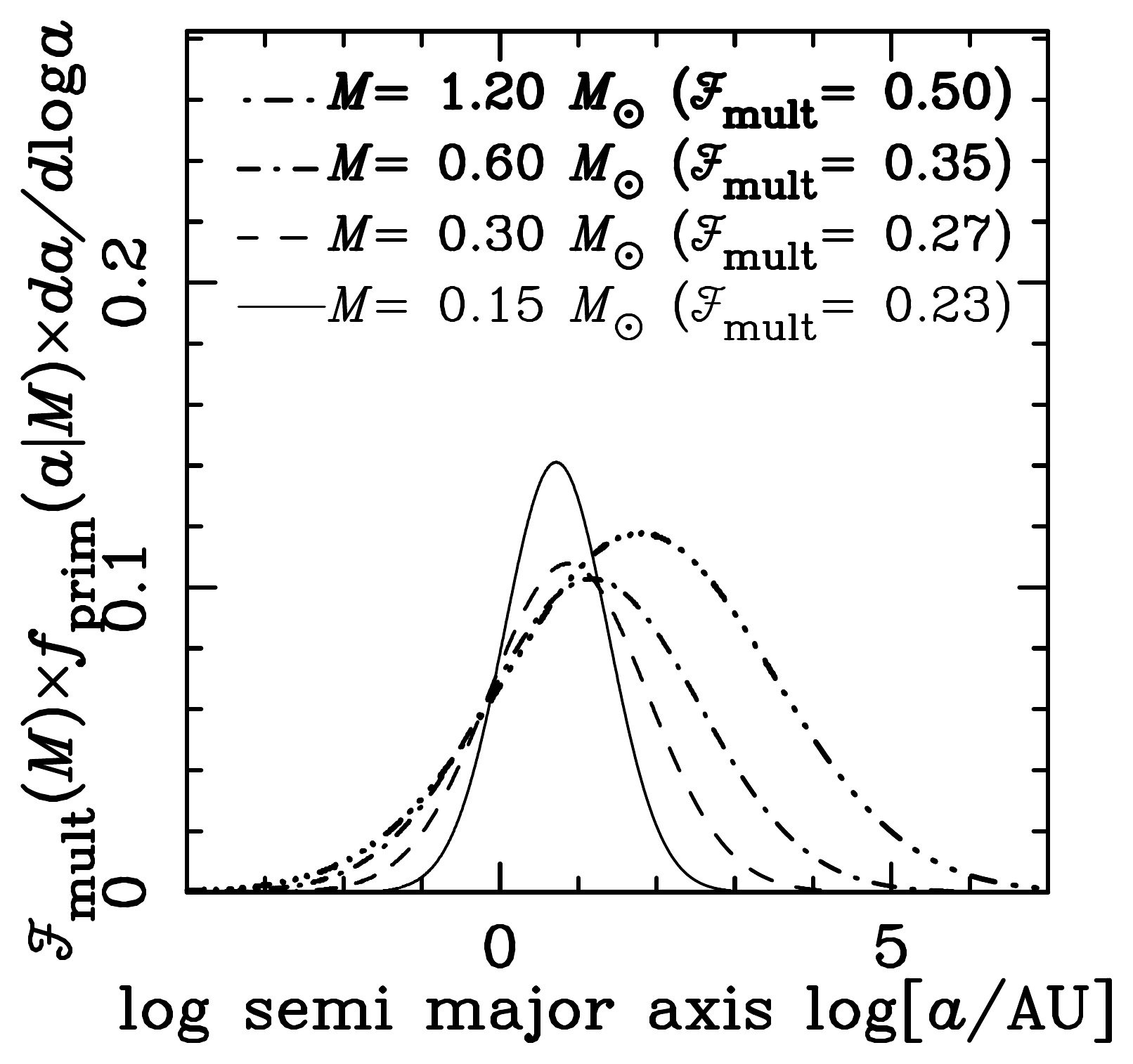}
\end{center}
\end{minipage}
\caption{Binary distributions for non-secondary stars of mass $M = 0.15\,M_{\odot}$, $0.30\,M_{\odot}$, $0.60\,M_{\odot}$, and $1.20\,M_{\odot}$.
Left: Mass-ratio distributions $f_{\rm prim} (q \, | \, M) = \int_{a} f_{\rm prim}(q, a \, | \, M) \, d a$. 
Right: Logarithm of semi-major axis distributions $f_{\rm prim} (a \, | \, M) \times (da/d\log a) = \int_{q} f_{\rm prim}(q, a \, | \, M) \, d q \times (da/d\log a)$, where $(da/d\log a)$ is required for the change of variable from $a$ to $\log a$.
The total area of each distribution in both panels is normalized by the multiplicity fraction ${\cal F}_{\rm mult} (M)$.}
\label{fig-nonsec}
\end{figure}

We note that these models simply attempt to provide a
convenient description of the empirical data; they do not have any theoretical basis.
We extrapolate these relations to mass $M$ out of the region plotted in Fig. \ref{fig-parafit}, but such very high or 
low mass stars are too bright or too faint to contribute to the excess flux analyzed in this paper, respectively.
Fig. \ref{fig-nonsec} shows examples of the binary distribution for a non-secondary star using these models.

\subsubsection{Full binary distribution}  \label{sec-binplot}

\begin{figure}
\centering
\epsscale{0.4}
\plotone{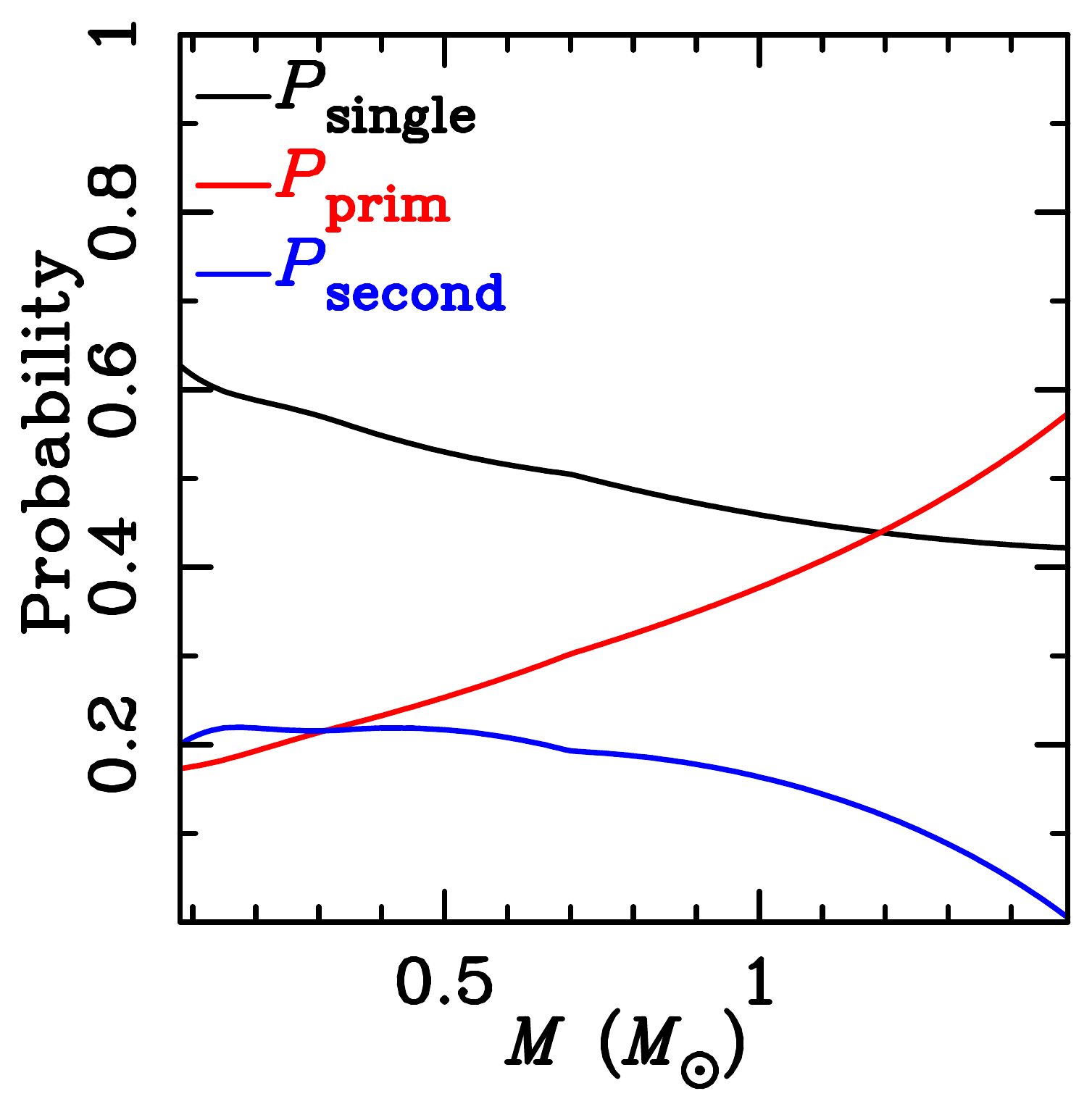}
\caption{Probabilities  of an arbitrary star with mass $M$ being a single star, $P_{\rm single} (M)$, a primary 
star, $P_{\rm prim} (M)$, and a secondary star, $P_{\rm second} (M)$, where 
$P_{\rm single} (M) + P_{\rm prim} (M) + P_{\rm second} (M) = 1$.
The black, red, and blue lines represent $P_{\rm single} (M)$, $P_{\rm prim} (M)$, and $P_{\rm second} (M)$, respectively.
The mass function $\Phi_{\rm PD} (M)$ with a cutoff at $1.5 M_{\odot}$ is used for the calculation.}
\label{fig-pmult}
\end{figure}

Fig.~\ref{fig-pmult} shows the probabilities that an arbitrary star is a single ($P_{\rm single} (M)$), 
primary ($P_{\rm prim} (M)$), and secondary ($P_{\rm second} (M)$) star, given by Eqs. 
(\ref{eq-psingle})--(\ref{eq-psecond}), 
where the binary distributions for a non-secondary star described in Section~\ref{sec-binpri} are applied.
For $\Phi_{\rm PD} (M)$ for the plot, we just apply a cutoff at $1.5 M_{\odot}$ to the IMF given by Eq. (\ref{eq-IMF}).
This simplified cutoff is used only for this plot and we apply cutoff with the age distribution described in Section \ref{sec-gal} in our Bayesian analysis.
Because ${\cal F}_{\rm mult} (M)$ increases with $M$, 
$P_{\rm prim} (M) \propto {\cal F}_{\rm mult} (M)$ also increases.
Meanwhile, $P_{\rm second} (M)$ decreases as the mass $M$ approaches the upper cutoff mass 
(1.5 $M_{\odot}$ in this plot) in the stellar mass function. 

\begin{figure}
  \begin{center}
  \subfigure[]{
   \includegraphics[width=110mm]{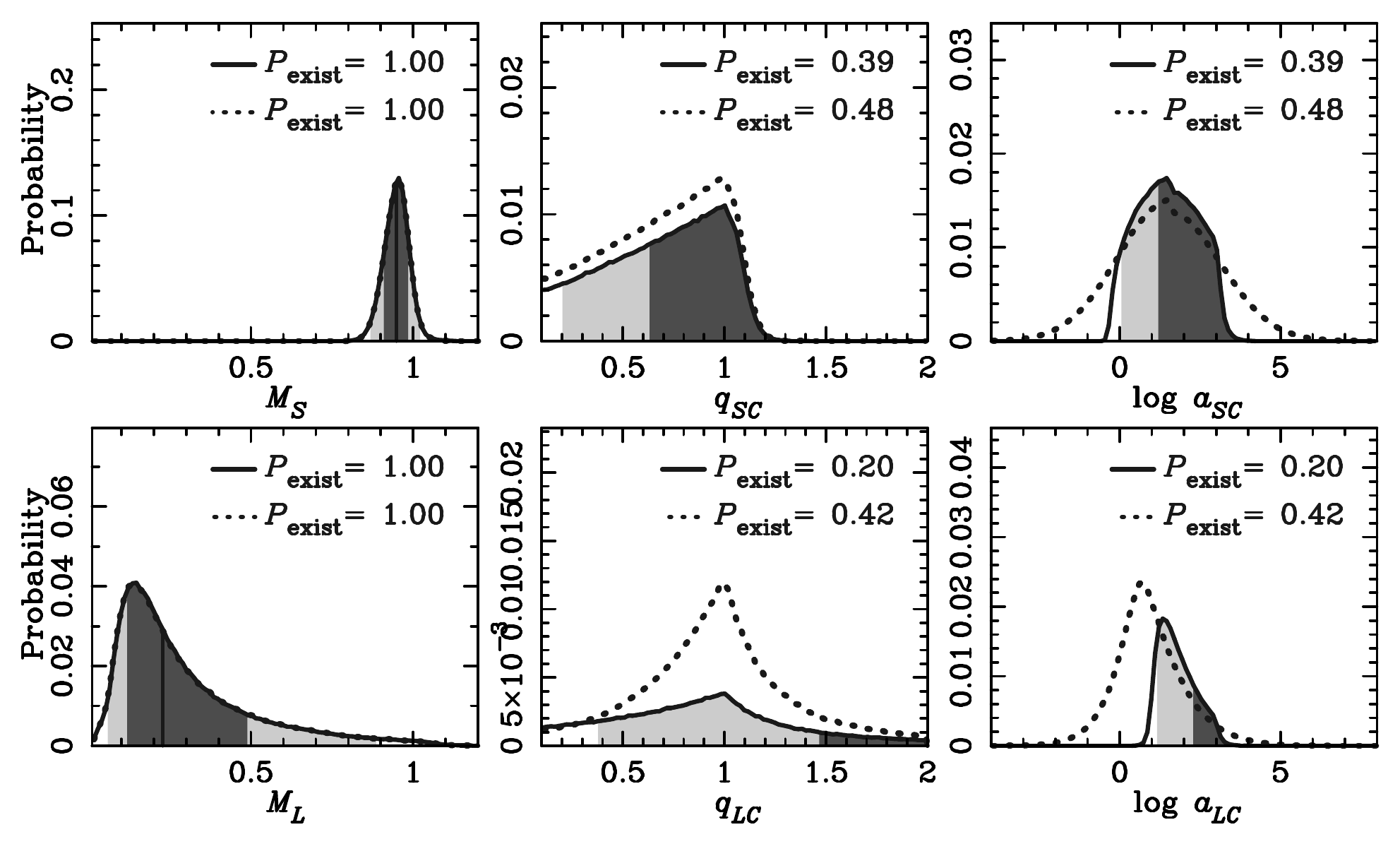}
 \label{fig-bindis-227}

  }
  \hfill
  \subfigure[]{

  \includegraphics[width=110mm]{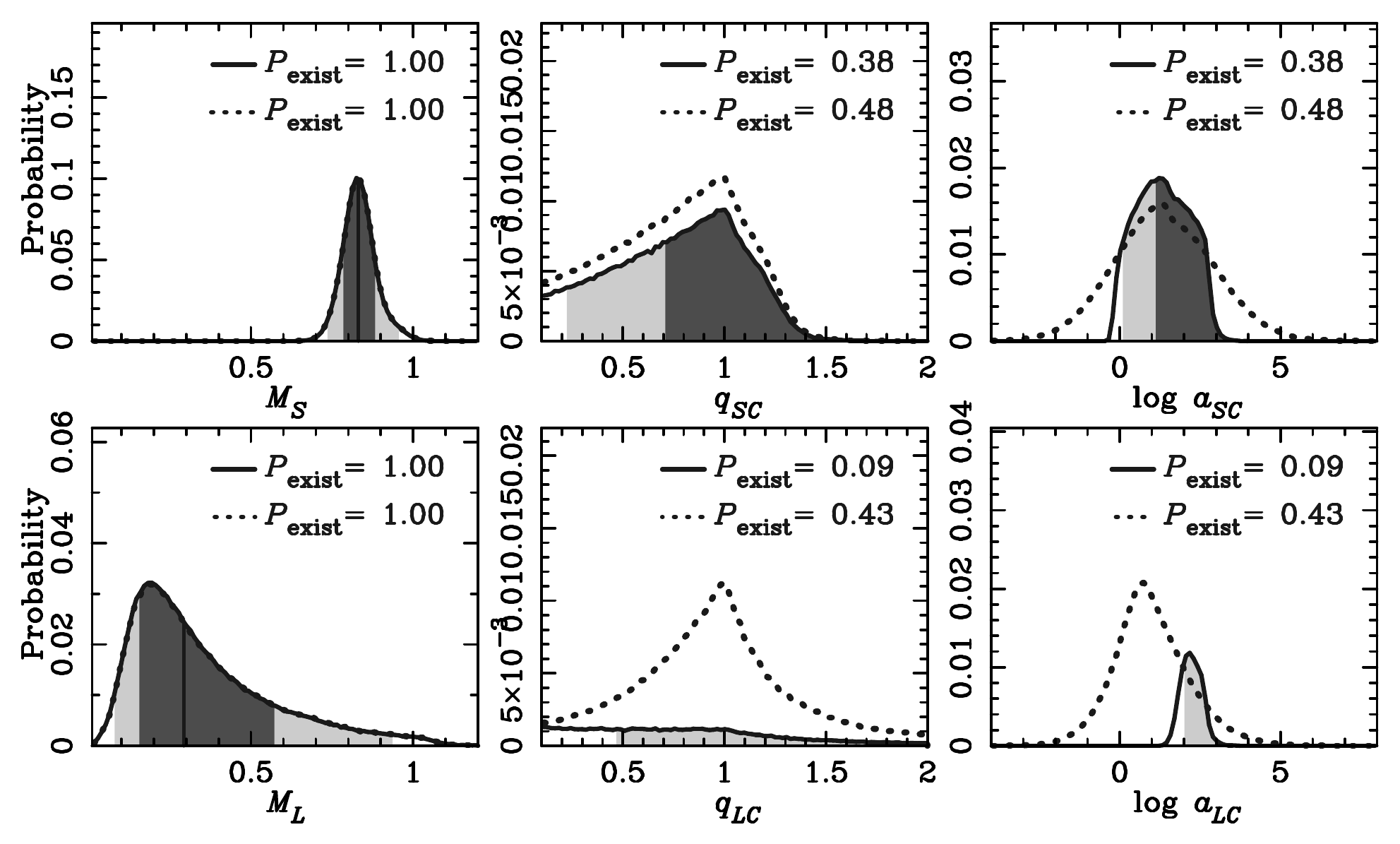}
    \label{fig-bindis-293}
  }
  \end{center}
  \vspace{-0.6cm}
  \caption{
The undetected (solid) and full (dotted) binary distributions calculated using parameters of (a) M16227 ($\phi_{\rm wide} = 148$ mas, $\theta_{\rm E, obs} = 0.23$ mas, 
  $u_{\rm 0, obs} = $ 0.08) and (b) M11293 ($\phi_{\rm wide} = 60$ mas, $\theta_{\rm E, obs} = 0.26$ mas, $u_{\rm 0, obs} = $ 0.0035).
The undetected binary distribution is part of the full binary distribution $f_{\rm arb}(q_i, a_i \, | \, M_*)$ ($i = SC$ for $M_* = M_S$ and $i = LC$ for $M_* = M_L$)
 that gives the detection efficiency $\epsilon_{i } = 0$. See Section \ref{sec-undetbin} for the details.
The $P_{\rm exist}$ values denote the probability that the corresponding object exists.
For the undetected distributions, the borders between different colors indicate the 2.3, 16, 84, and 97.7 percentiles from left to right. 
The thick vertical lines between the 16 and 84 percentiles indicate the median. 
Some percentiles are out of the plot in the case of $P_{\rm exist} < 1$. 
}
  \label{fig-bindis}
\end{figure}

We show the binary distribution for arbitrary stars, $f_{\rm arb}(q, a \, | \, M)$, in Eq. (\ref{eq-fqac}), 
projected to the mass ratio and semi-major axis axes in Fig. \ref{fig-bindis} as dotted lines.
Recall that $f_{\rm arb}(q, a \, | \, M)$ depends on the arbitrary star mass $M$ that is usually given by a probability distribution.
The two panels to the far left in Fig. \ref{fig-bindis} (a) show the prior probability distributions of the source 
mass $M_S$ and the lens mass $M_L$ for event M16227, where $M_S$ is calculated by 
applying the mass-luminosity relation to the observed source mag $H_{S, {\rm obs}}$ (see details in Section \ref{sec-pricomp}), and $M_L$ is given by 
the prior PDF $f_{pri}' (M_L, D_L, D_S, v_{\rm t})$ described in Section \ref{sec-priML}.
We combine the mass and binary distributions through a Monte Carlo simulation that selects the $M_*$ ($M_S$ or $M_L$) 
value randomly from its probability distribution and then picks a combination of $(q_i, a_i)$ ($i= SC$ for $M_* = M_S$ or  $i= LC$ for $M_* = M_L$)  randomly 
from the appropriate $f_{\rm arb}(q_i, a_i \, | \, M_*)$ distribution to obtain the dotted lines in the panels labeled $q_{i}$ and $\log a_i$.
Fig. \ref{fig-bindis} (b) shows these distributions for event M11293 in a similar manner.
We refer to these distributions in the dotted lines as the full binary distribution to distinguish from the undetected binary 
distribution in the solid lines in the same panels described in Section \ref{sec-undetbin}.

Eq. (\ref{eq-fqac}) shows $f_{\rm arb} (q, a \, | \, M) = P_{\rm prim} (M) f_{\rm prim} (q, a \, | \, M)$ with 
$q$ in the range $0 < q < 1$; hence, the shapes of the $q_i$ and $a_i$ distributions, shown in Fig. \ref{fig-bindis}, are  
similar to that of $f_{\rm prim} (q, a \, | \, M)$ given by Eq. (\ref{eq-fqa}) and plotted in Fig. \ref{fig-nonsec}.
These would be power-law and log-normal distributions for $q_i$ and $a_i$, respectively. 
However, these shapes are somewhat distorted by the contribution from $q > 1$ and by the 
$M$ distributions that take various $M$ values instead of a fixed $M$ value.
For $q > 1$, the $q_{SC}$ and $q_{LC}$ distributions follow the third term of Eq. (\ref{eq-fqac}), 
where the $\Phi_{\rm PD} (q M)$ and $q^{-1}$ factors decrease the probability rapidly as $q$ increases
because $\Phi_{\rm IMF} (q M) \propto q^{-2}$ for $q M > 0.7 M_{\odot}$ as seen in Eq. (\ref{eq-IMF}).

While we know that the lens and source star exist for each event, the existence of a source or lens companion
is not certain. Therefore, we use $P_{\rm exist}$ to denote the probability that a lens or source companion exists, 
and for consistency in notation, we use $P_{\rm exist} =1$ for the lens and source stars.
For the full binary distributions in the dotted curves in Fig. \ref{fig-bindis}, the probability of $1 - P_{\rm exist}$ for the source 
and lens companions corresponds to $P_{\rm single} (M_*)$ with $M_* = M_S$ or $M_* = M_L$, respectively, 
as given in the first term in Eq. (\ref{eq-fqac}).

\subsubsection{Undetected binary distribution}  \label{sec-undetbin}
As described in Section \ref{sec-expri}, we use the undetected binary distribution which is the combination of the full binary distribution and the detection efficiency to 
simulate the source and lens companions in the Monte Carlo simulation to derive the prior PDF $f_{pri} (F_L, F_{SC}, F_{LC})$.
Some companions are so close to the source or lens star that they would affect the light curve 
in ways that are inconsistent with observations. Hence, unlike the case of ambient stars, we must 
include a minimum allowable separation in addition to the maximum separation to consider the detection efficiencies for the source and lens companions.
We denote the angular separations corresponding to the minimum (or close) separation limit by 
$\phi_{{\rm close}, SC}$ for the source companion and $\phi_{{\rm close}, LC}$ for the lens companion, 
while we use the same value of $\phi_{\rm wide}$ as that in Table \ref{tab-info} for the wide limits of the unresolvable 
region of these objects.
This is equivalent to adopting a detection efficiency of 
\begin{align}
 \hspace{2cm} \epsilon_{i} = \Theta [(\phi - \phi_{\rm wide})(\phi - \phi_{{\rm close}, i})]  \hspace{0.5cm}  (i = SC ~ {\rm or} ~LC) \label{eq-deslc}
\end{align}
for the companion to the source or lens located at angular separation $\phi$ from the centroid of the target. 

Following \citet{bat14}, we adopt $\theta_{\rm E}/4$ as $\phi_{{\rm close}, SC}$
and we derive $\phi_{{\rm close}, LC}$ using the inequality 
$w_{LC} < u_{\rm 0}$ as the condition for an unresolvable lens companion,
where $w_{LC}$ is the size of the central caustic created by the hypothetical companion to the lens.
We use the analytic formula $w_{LC} = 4q_{LC}/(s_{LC} - s_{LC}^{-1})^2$,
where $s_{LC}$ is the projected separation 
between the lens and the companion in units of the angular Einstein radius \citep{chu05}.
Although this formula is an approximate one that was derived for planetary mass ratios, $q_{LC} \ll 1$, 
we find that it works moderately well even for stellar mass-ratio companions, 
as discussed in Section \ref{sec-pdetlow}.
With this analytic formula for $w_{LC}$, the inequality $w_{LC} < u_{\rm 0}$ has two different 
unresolvable regions of $s_{LC}$ as its solutions:
\begin{align}
s_{LC} &< \sqrt{\frac{q_{LC}}{u_{\rm 0}} + 1} - \sqrt{\frac{q_{LC}}{u_{\rm 0}}}, \label{eq-solcau1}\\
s_{LC} &> \sqrt{\frac{q_{LC}}{u_{\rm 0}} + 1} + \sqrt{\frac{q_{LC}}{u_{\rm 0}}}. \label{eq-solcau2}
\end{align}
A companion in the former unresolvable region corresponds to the case of a close-in binary system whose total mass is $M_L$, 
which we ignored in point (ii) at the beginning of Section \ref{sec-bindis} for simplicity.
This means that when a companion that satisfies Eq. (\ref{eq-solcau1}) is selected in our Monte Carlo simulation shown in Fig. \ref{fig-flow}, then we treat 
it as a resolvable companion, and such a scenario is rejected in our simulation.
Considering this region carefully is important for studying possible circumbinary planetary systems 
such as OGLE-2007-BLG-349L(AB)c \citep{ben16}; 
however, this is beyond the scope of this study and it negligibly changes the derived lens properties because the 
probability of the lens companion in this region is small.
In summary, we decide to use 
\begin{align}
\phi_{{\rm close}, SC} &= \theta_{\rm E} / 4  \label{eq-phiclosesc}\\
\phi_{{\rm close}, LC} &= s_{{\rm close}, LC} ~ \theta_{\rm E}, \label{eq-phicloselc}
\end{align}
where $s_{{\rm close}, LC} \equiv \sqrt{q_{LC}/u_{\rm 0} + 1} + \sqrt{q_{LC}/u_{\rm 0}}$.
These decisions are summarized in Table \ref{tab-inputs}.

With the detection efficiencies $\epsilon_{SC}$ and $\epsilon_{LC}$, we calculate the undetected binary distribution as follows, which is also described in the left box in Fig. \ref{fig-flow}.
In each trial of the Monte Carlo simulation, we have a combination of ($q_{SC}$, $a_{SC}$) and ($q_{LC}$, $a_{LC}$), which are randomly selected from the full binary distributions of
$f_{\rm arb}(q_{SC}, a_{SC} \, | \, M_S)$ and $f_{\rm arb}(q_{LC}, a_{LC} \, | \, M_L)$, respectively.
Using a probability distribution $p (a_{\perp})  = a_{\perp}/a^2 (1 - (a_{\perp}/a)^2)^{-1/2}$ \citep{gouloe92} for the projection from a three-dimensional physical distance $a$
to a projected distance in the sky, $a_{\rm \perp}$, we randomly obtain the physical projected separations $a_{SC,\perp}$ and $a_{LC,\perp}$ 
from $a_{SC}$ and $a_{LC}$, respectively.
We simulate the undetected binary distribution by accepting a combination of parameters that satisfies both $\epsilon_{SC} = 0$ and $\epsilon_{LC} = 0$, i.e., when both of the generated $a_{SC,\perp}$ and $a_{LC,\perp}$ are 
located in the corresponding unresolvable regions, $\phi_{{\rm close}, SC} < a_{SC,\perp}/D_S < \phi_{\rm wide}$ and  $\phi_{{\rm close}, LC} < a_{LC,\perp}/D_L < \phi_{\rm wide}$, respectively.

The solid lines in Fig. \ref{fig-bindis} (a) and (b) represent the undetected binary distributions for M16227 and M11293, respectively.
As shown in the $\log a_i$ distributions, the shape of the undetected distributions of the semi-major axis is the same as that of the full distribution but with edges on both sides 
removed by considering them as detectable.
The borders between the colors of the shaded areas represent the 2.3, 16, 84, and 97.7 percentiles from left to right, and the thick vertical line represents the median.
The probability that each object exists is shown in the top right panel as $P_{\rm exist}$. 
The bins corresponding to $(q_i, a_i) = (0, 0)$, i.e., cases of a single star, are not shown and the integrated areas of the plotted regions are thus the same as the $P_{\rm exist}$ values.
Therefore, some percentiles are not shown in the panels with $P_{\rm exist} < 1$. 
For example, in the distributions of $q_{SC}$ and $\log a_{SC}$ in Fig. \ref{fig-bindis} (a) where the $P_{\rm exist}$ value is 39\%, 
which is larger than 16\% but smaller than 50\%,  the 2.3 and 16 percentiles are shown, but the median and the 84 and 97.7 percentiles are not shown.

The relation between the $P_{\rm exist}$ values of the undetected and full distributions for the companion to the source or lens is
\begin{align}
P_{{\rm exist, undet}, i} = \frac{P_{{\rm exist, full}, i} - P_{{\rm det}, i}}{1 - P_{{\rm det}, i}}, \label{eq-exist}
\end{align}
where $P_{{\rm exist, undet}, i}$ and $P_{{\rm exist, full}, i}$ are the $P_{\rm exist}$ values for the source companion ($i = SC$) or lens companion ($i = LC$) 
in the undetected (solid line) and full (dotted line) binary distributions, respectively.
Further, $P_{{\rm det}, i}$ is the fraction of detectable companions to the source or lens in the full binary distribution, where its 
projected separation $a_{i,\perp}$ does not satisfy the condition $\phi_{{\rm close}, i} < a_{i,\perp}/D_* < \phi_{\rm wide}$ ($D_* = D_S$ for $i = SC$ and $D_* = D_L$ for $i = LC$).
This fraction of $P_{{\rm det}, i}$ is subtracted not only from the numerator but also from the denominator in Eq. (\ref{eq-exist}), 
which causes the value of $P_{{\rm exist, undet}, i}$ to be higher than just the subtracted value of $P_{{\rm exist, full}, i} - P_{{\rm det}, i}$.
This is also why there is a part where the probability of the solid line is higher than the probability of the dotted line in the $\log a_{i}$ distribution.
In Section \ref{sec-valiSLC}, we compare these $P_{{\rm det}, i}$ values with binary fractions around source or lens stars actually detected in planetary microlensing events.

\subsection{Lens flux and source and lens companion flux priors} \label{sec-pricomp}
Now that we have the joint prior PDF of $M_L$, $D_L$, $D_S$, and $v_t$, $f_{pri}' (M_L, D_L, D_S, v_{\rm t})$, 
the mass-luminosity relation, ${\cal M}_{H} (M)$, and the undetected binary distribution, we are equipped 
to calculate the joint prior PDF for the fluxes of the lens and companions to the source and lens stars, $f_{pri} (F_L, F_{SC}, F_{LC})$ through the Monte Carlo method summarized in Section \ref{sec-expri} and Fig. \ref{fig-flow}.

Given the lens mass $M_L$, source companion mass $M_{SC}$, and lens companion mass $M_{LC}$, 
we can convert them into the apparent magnitudes in the $H$-band by
\begin{align}
H_L &= {\cal M}_{H,L} (M_L) + {\rm DM}_L + A_{H,L}, \label{eq-HL} \\
H_{SC} &= {\cal M}_{H,SC} (M_{SC}) + {\rm DM}_S + A_{H,S}, \label{eq-HSC} \\
H_{LC} &= {\cal M}_{H,LC} (M_{LC}) + {\rm DM}_L + A_{H,L} , \label{eq-HLC}
\end{align}
where ${\rm DM}_i$ ($i= L, S$) is the distance modulus corresponding to the distance of $D_i$
and ${\cal M}_{H,i}$ is the absolute $H$-band magnitude for star $i$.
To evaluate the amount of extinction, we use the formula 
$A_{H,i} =  (1- e^{-D_i/h_{\rm dust}})/(1- e^{-D_{\rm rc}/h_{\rm dust}})~A_{H,{\rm rc}}$, following \citet{ben15},
where $h_{\rm dust} = (0.1~ {\rm kpc})/\sin{|b|}$ is the dust scale length toward the galactic bulge at galactic latitude $b$. 
The average distance to the red clump stars at the event position, $D_{\rm rc}$, also 
corresponds to the distance modulus ${\rm DM}_{\rm rc}$.
Because we have a combination of $M_L$, $D_L$, and $D_S$, which are randomly extracted from 
$f_{pri}' (M_L, D_L, D_S, v_{\rm t})$,
we can immediately calculate $H_L$, ${\rm DM}_i$, and $A_{H,i}$ with these formulae in each trial of our Monte Carlo simulation.

The remaining uncertain values in Eqs. (\ref{eq-HL})--(\ref{eq-HLC}) are $M_{SC}$ and $M_{LC}$.
We calculate these values using the undetected binary distribution described in Section \ref{sec-undetbin}.
At this point, the lens star that we consider is characterized by its mass $M_L$ whereas the source star is characterized by
its $H$-band magnitude; hence, we must calculate the source mass $M_S = M_S ({\cal M}_{H,S})$
from ${\cal M}_{H,S} = H_{S, {\rm obs}} - A_{H,S} - {\rm DM}_S$ and the mass-luminosity relation.
Then, we randomly select the source and lens companion parameters ($q_{SC}$, $a_{SC}$ and $q_{LC}$, $a_{LC}$) 
from the $f_{\rm arb}(q_{SC}, a_{SC} \, | \, M_S)$ and $f_{\rm arb}(q_{LC}, a_{LC} \, | \, M_L)$ distributions.
Recall that the binary distribution $f_{\rm arb}(q, a \, | \, M)$ returns $(q, a) = (0, 0)$ with the probability of 
$P_{\rm single} (M)$ in Eq. (\ref{eq-psingle}), which implies that the star in question has no companion.
By accepting the binary parameters that satisfies $\epsilon_{SC} = 0$ and $\epsilon_{LC} = 0$, 
we have the source companion mass $M_{SC} = q_{SC} M_S$ and the lens companion mass $M_{LC} = q_{LC} M_L$, as 
described in Section \ref{sec-undetbin} and Fig. \ref{fig-flow}.
 
We calculate and plot the joint prior PDF $f_{pri} (F_L, F_{SC}, F_{LC})$ in magnitude for each event in 
the bottom right panel in each (a) component of Figs.  \ref{fig-227}--\ref{fig-0950}.
The solid lines in red, green, and purple correspond to the prior probability distributions of $H_L$, $H_{SC}$, and $H_{LC}$, respectively.
We plot them along a one-dimensional axis for clarity; however, we note that they are from a joint probability distribution and have correlations with each other.
The correlations between $H_{SC}$ and the other two parameters are weak, whereas the correlation between $H_L$ and $H_{LC}$ is moderately strong because
the mass of a lens companion is given by $M_{LC} = q_{LC} M_L \propto M_L$ and the $q_{LC}$ distribution depends on the lens mass $M_L$.
As in the case of Fig. \ref{fig-bindis}, we do not plot bins corresponding to $(q, a) = (0, 0)$, the case of no companion; instead, we show the probability that 
the companion exists (i.e., the total area of the shown distribution) as $P_{\rm exist}$ in the parentheses  in each color.

\section{Application and Results} \label{sec-app}
We apply our method to M16227, M08310, M11293, O120563, and O120950.
We calculate the prior PDF of the four possible origins flux, $f_{pri} (F_L, F_{\rm amb}, F_{SC}, F_{LC})$, for each event by repeating the procedure in 
the left box in Fig. \ref{fig-flow} using the models listed in Table \ref{tab-inputs} and the parameters for each event listed in Table \ref{tab-info}. 
Then, we calculate the posterior PDF $f_{post} (F_L, F_{\rm amb}, F_{SC}, F_{LC} | F_{\rm excess} = F_{\rm ex, obs})$ for each event by extracting 
combinations of the parameters for which the excess flux value is consistent with the observed value for each event from the prior PDF. 
This corresponds to the procedure shown in the right box in Fig. \ref{fig-flow}.
The (a) and (b) components of Figs. \ref{fig-227}--\ref{fig-0950} show the prior and posterior probability distributions, respectively, for the lens mass and distance (left panels), 
magnitude of each contributor (bottom right panels), and magnitude of excess flux (top right panels).
The horizontal axes are divided into 100 bins and the probability values integrated within a bin are indicated along the vertical axes in each panel.
We plot magnitude distributions up to $\sim 24$ mag in the bottom right and top right panels because the $H$-band LF of \citet{zoc03} 
covers up to $H_0 = 23.5$ mag, where $H_0$ is an extinction-free magnitude. 
Note that our results of the posterior probabilities are negligibly affected by the unconsidered fainter ambient stars, as discussed in Section \ref{sec-namb}.

\subsection{How to find possible origins of the observed excess} \label{sec-app1}
\begin{figure}
  \begin{center}
  \subfigure[]{
   \includegraphics[width=140mm]{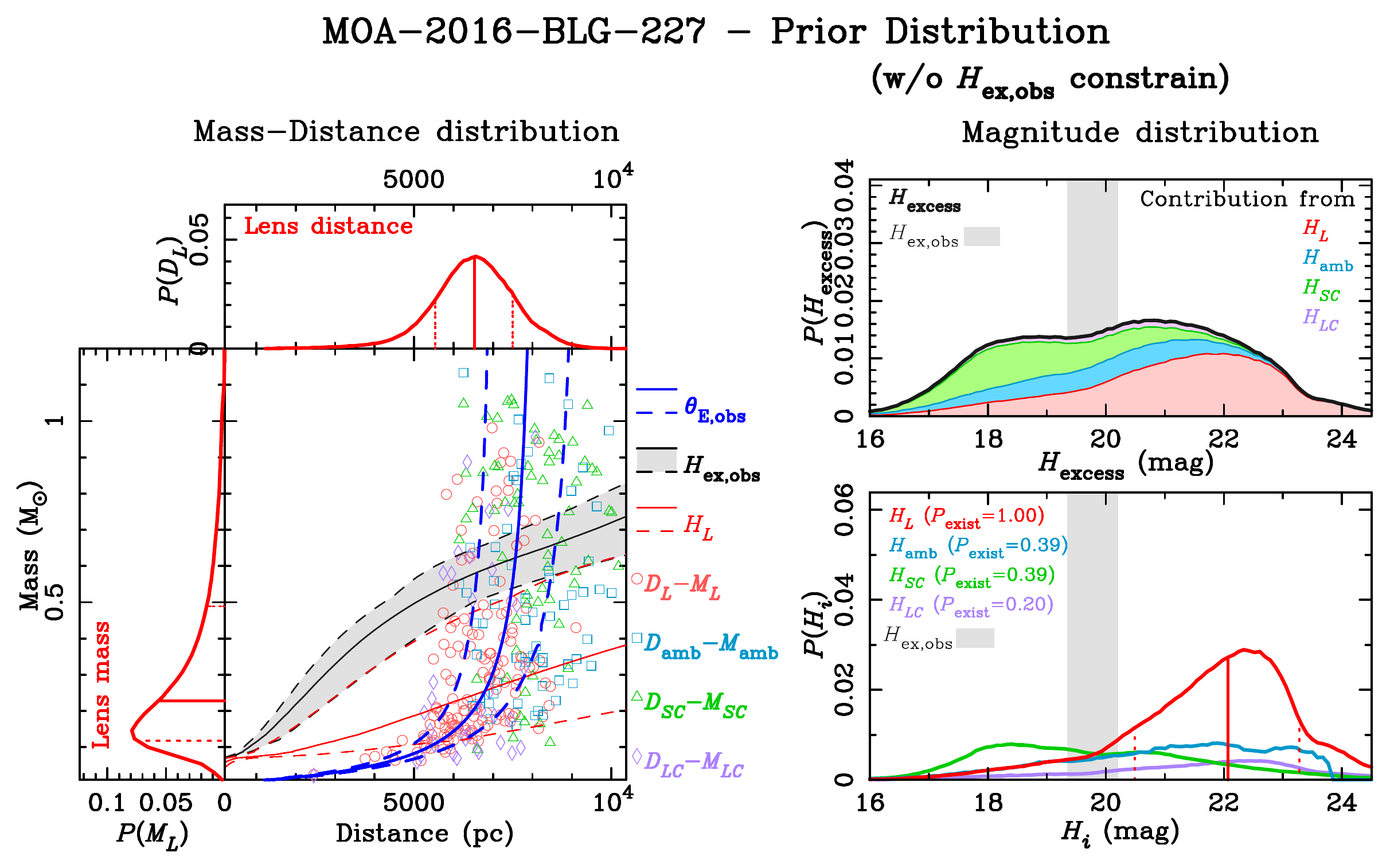}
 \label{fig-227-pri}       
  }
  \hfill
  \subfigure[]{
   \includegraphics[width=140mm]{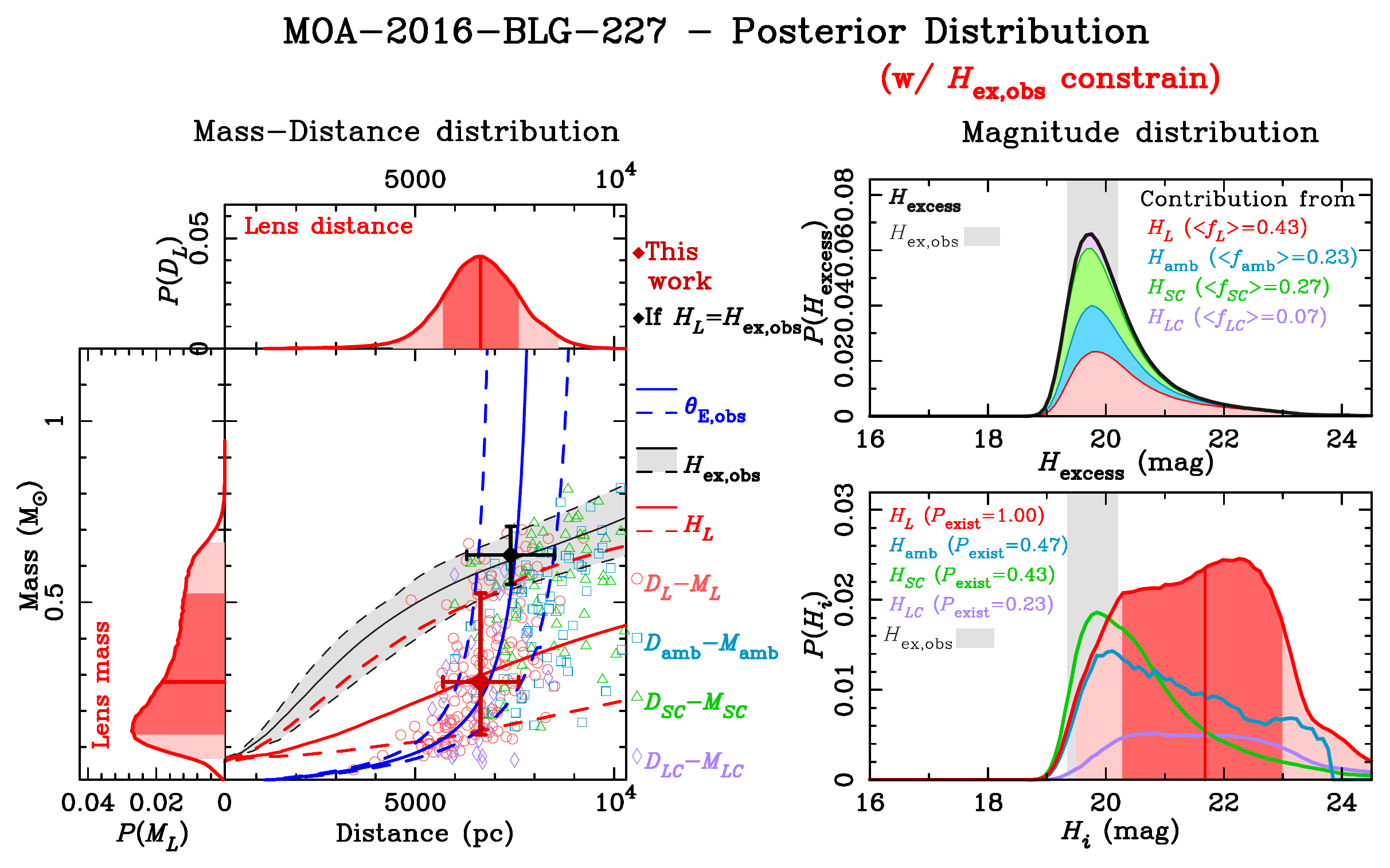}
   \label{fig-227-post}
  }
  \end{center}
  \vspace{-0.6cm}
  \caption{(a) Prior and (b) posterior probability distributions for M16227. 
  On the left plane, some accepted combinations of mass and distance of the four contributors are shown in different symbols and colors,
  in addition to the mass-distance relation curves for each parameter indicated on the right.
  The left and top histogram curves along the plane represent the $M_L$ and $D_L$ probability distributions, respectively.
The $H_{\rm excess}$ probability distribution is shown in the top right panel and the $H_L$, $H_{\rm amb}$, $H_{SC}$ and $H_{LC}$ distributions are 
shown in the bottom right panel.
The probability of existence of each object is shown as $P_{\rm exist}$ in the bottom right panel.
Each color area of the top right panel indicates a contribution from each possible origin when the excess flux is the value at the horizontal axis.
See Section \ref{sec-app1} for the details. 
The black point with the error bar in the left panel in (b) indicates the lens mass and distance when $H_L = H_{\rm ex, obs}$.}
  \label{fig-227}
\end{figure}

\begin{figure}
  \begin{center}
  \subfigure[]{
   \includegraphics[width=140mm]{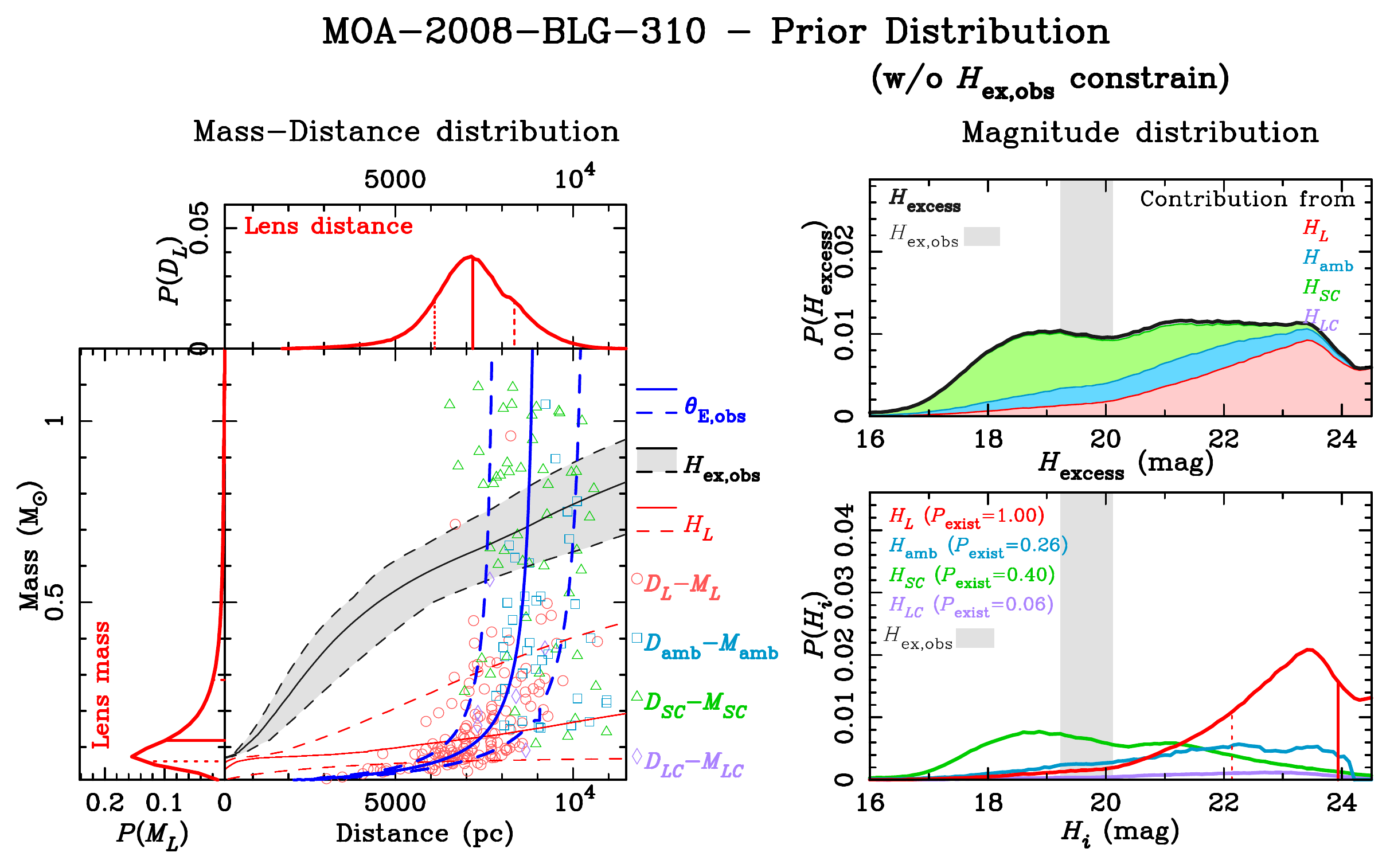}
 \label{fig-310-pri}  
  }
  \hfill
  \subfigure[]{
   \includegraphics[width=140mm]{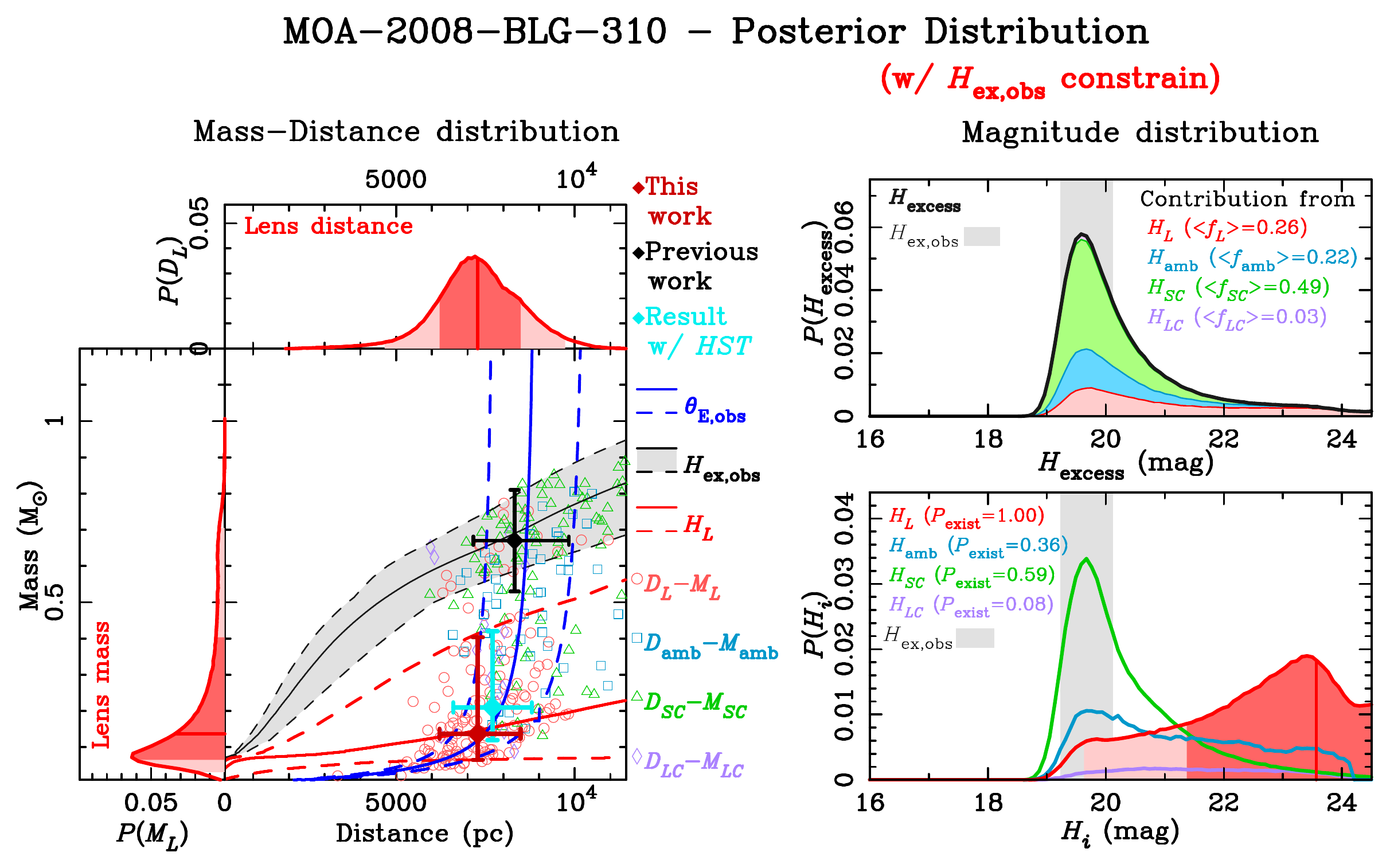}
    \label{fig-310-post}
  }
  \end{center}
  \vspace{-0.6cm}
  \caption{Same as Fig. \ref{fig-227}, but for M08310. (a) Prior and (b) posterior probability distributions. 
  The lens mass and distance estimated by \citet{jan10} and \citet{bha17} are indicated as ``Previous work'' and ``Result w/ {\it HST}'', respectively, 
  on the mass-distance plane in the posterior distribution.}
  \label{fig-310}
\end{figure}

\begin{figure}
  \begin{center}
  \subfigure[]{
   \includegraphics[width=140mm]{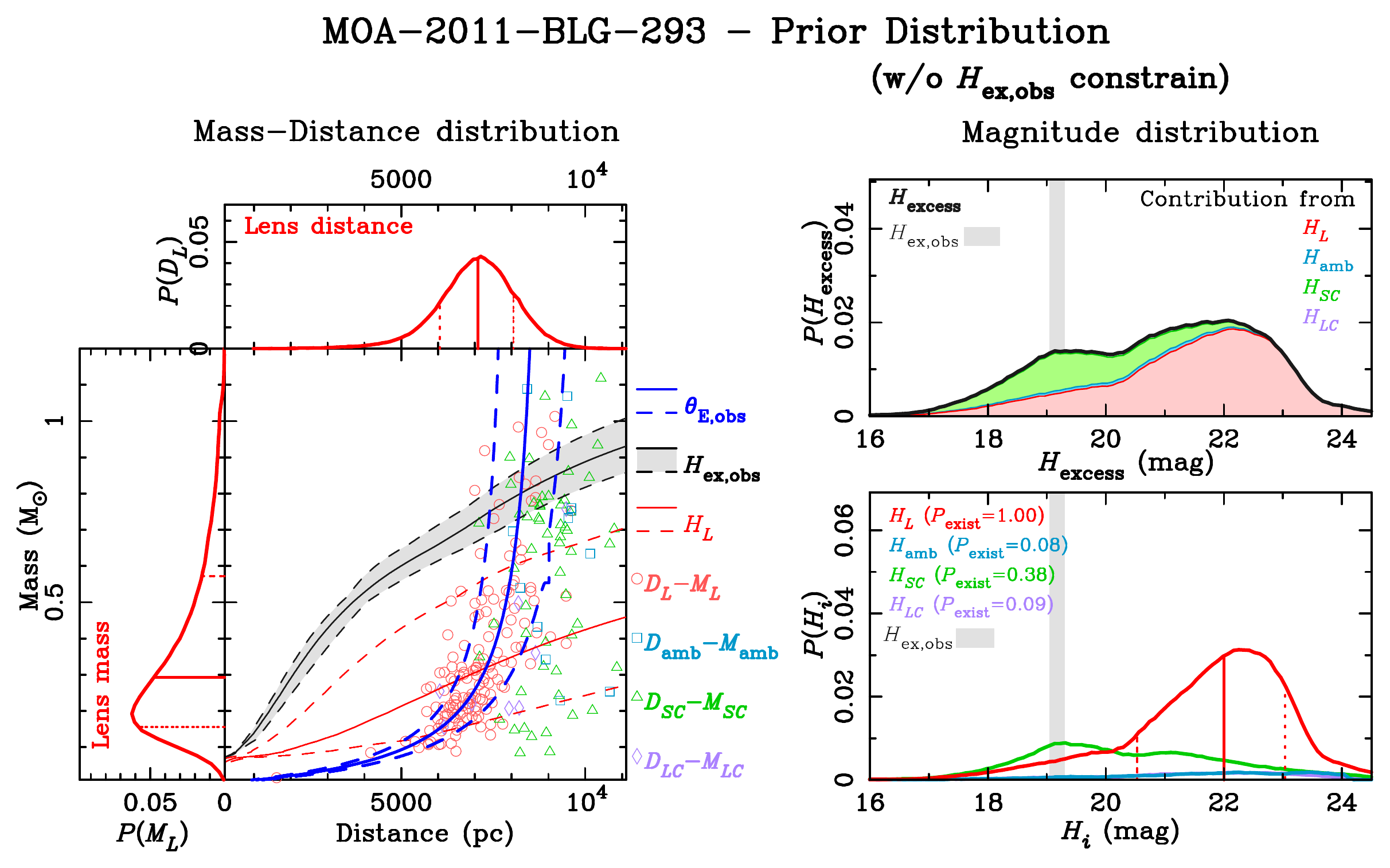}
 \label{fig-293-pri}      
    
  }
  \hfill
  \subfigure[]{
   \includegraphics[width=140mm]{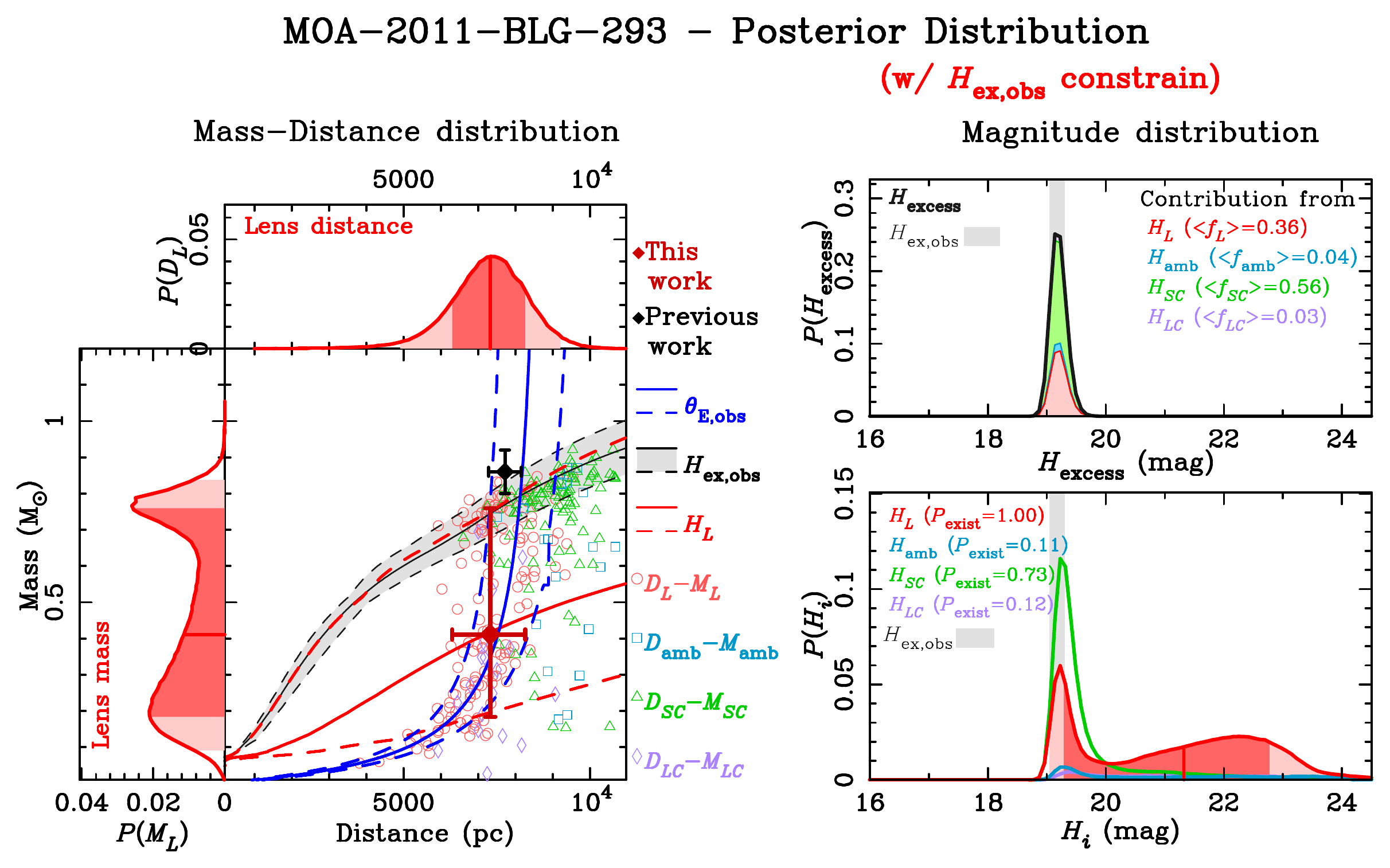}
    \label{fig-293-post}
  }
  \end{center}
  \vspace{-0.6cm}
  \caption{Same as Fig. \ref{fig-227}, but for M11293. (a) Prior and (b) posterior probability distributions. 
  The lens mass and distance estimated by \citet{bat14} is indicated as ``Previous work'' on the mass-distance plane in the posterior distribution. }
  \label{fig-293}
\end{figure}

\begin{figure}
  \begin{center}
  \subfigure[]{
   \includegraphics[width=140mm]{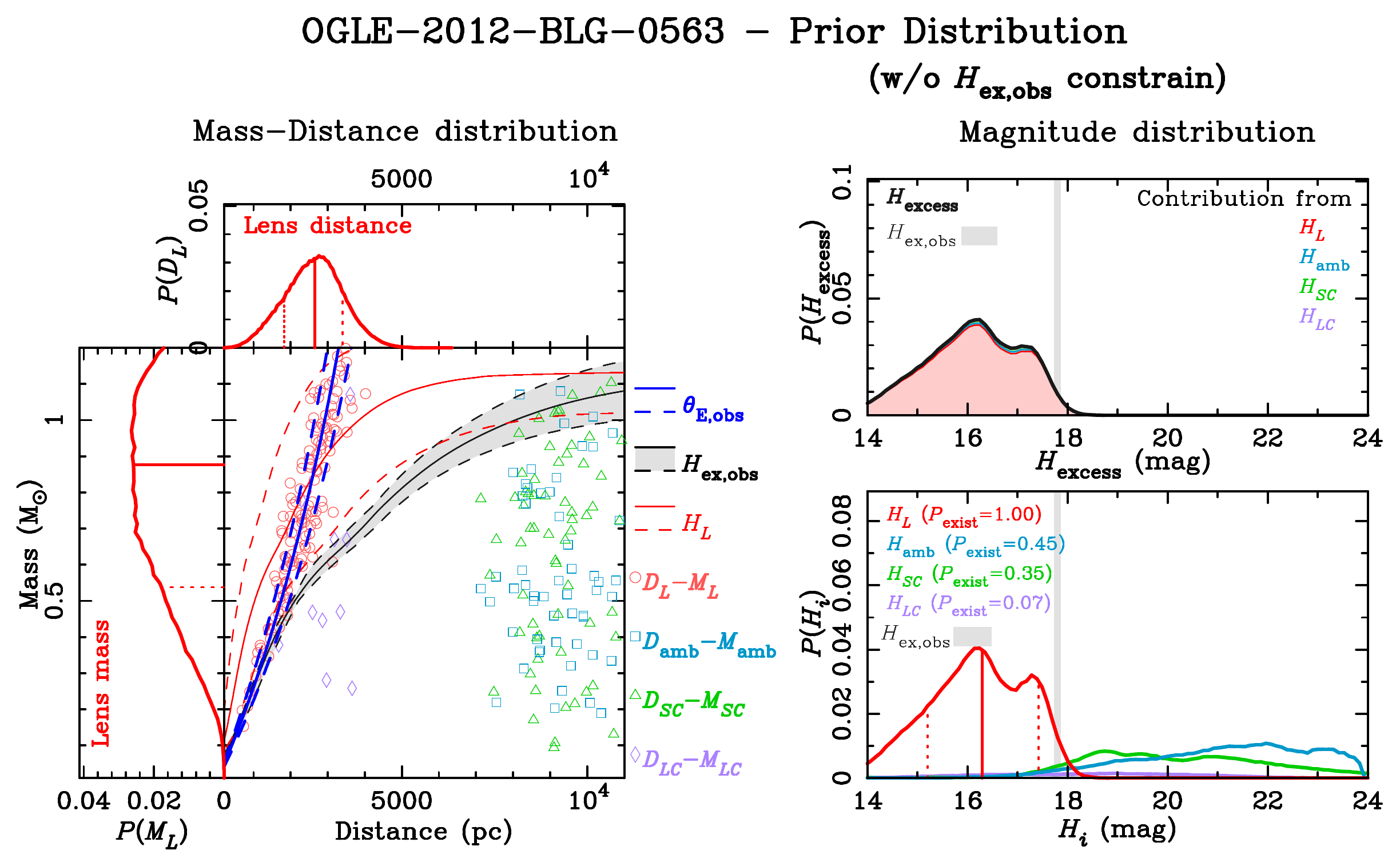}
 \label{fig-0563-pri}  
  }
  \hfill
  \subfigure[]{
   \includegraphics[width=140mm]{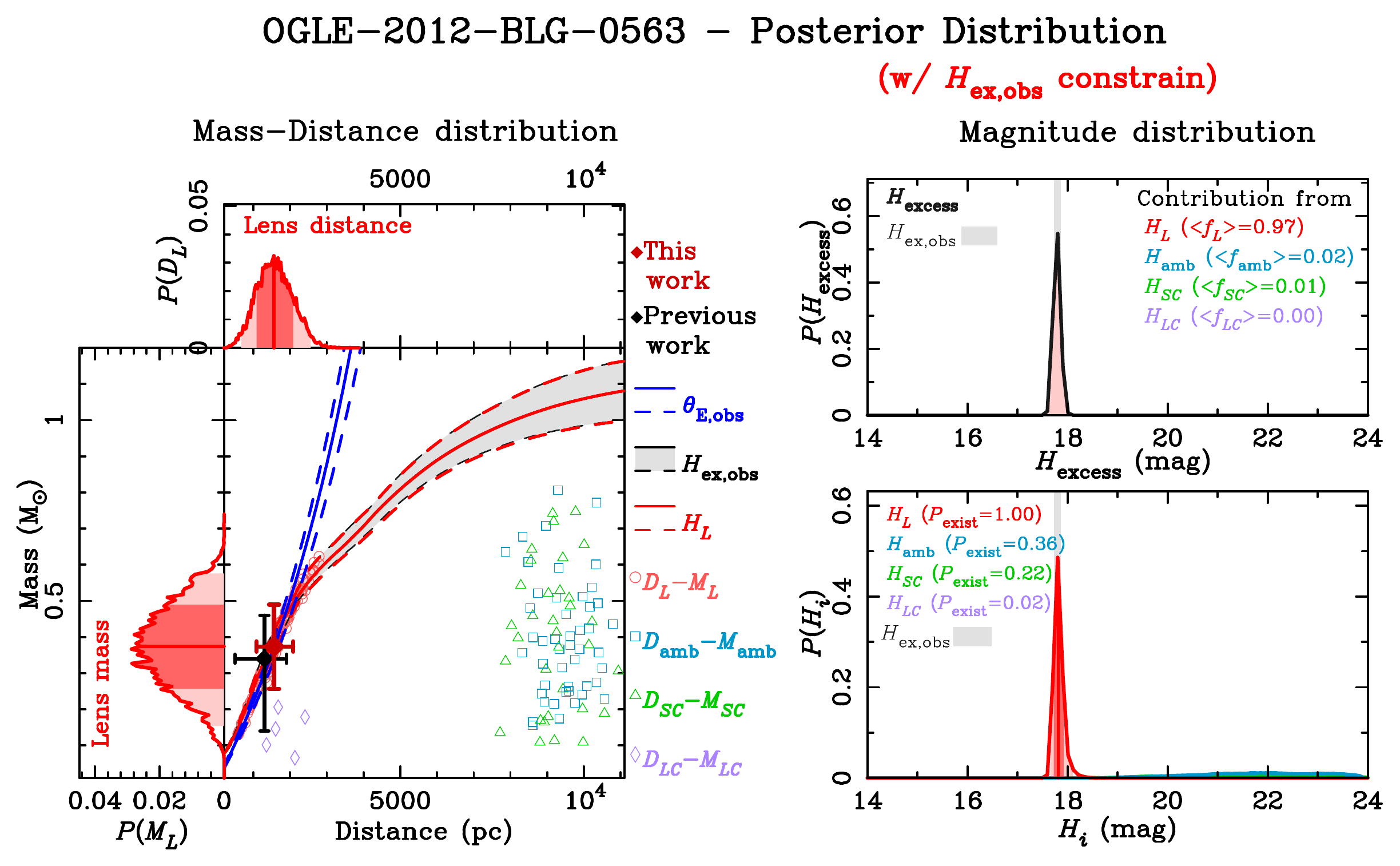}
    \label{fig-0563-post}
  }
  \end{center}
  \vspace{-0.6cm}
  \caption{Same as Fig. \ref{fig-227}, but for O120563. (a) Prior and (b) posterior probability distributions. 
  The lens mass and the distance estimated by \citet{fuk15} is indicated as ``Previous work'' on the mass-distance plane in the posterior distribution.}
  \label{fig-0563}
\end{figure}

\begin{figure}
  \begin{center}
  \subfigure[]{
   \includegraphics[width=140mm]{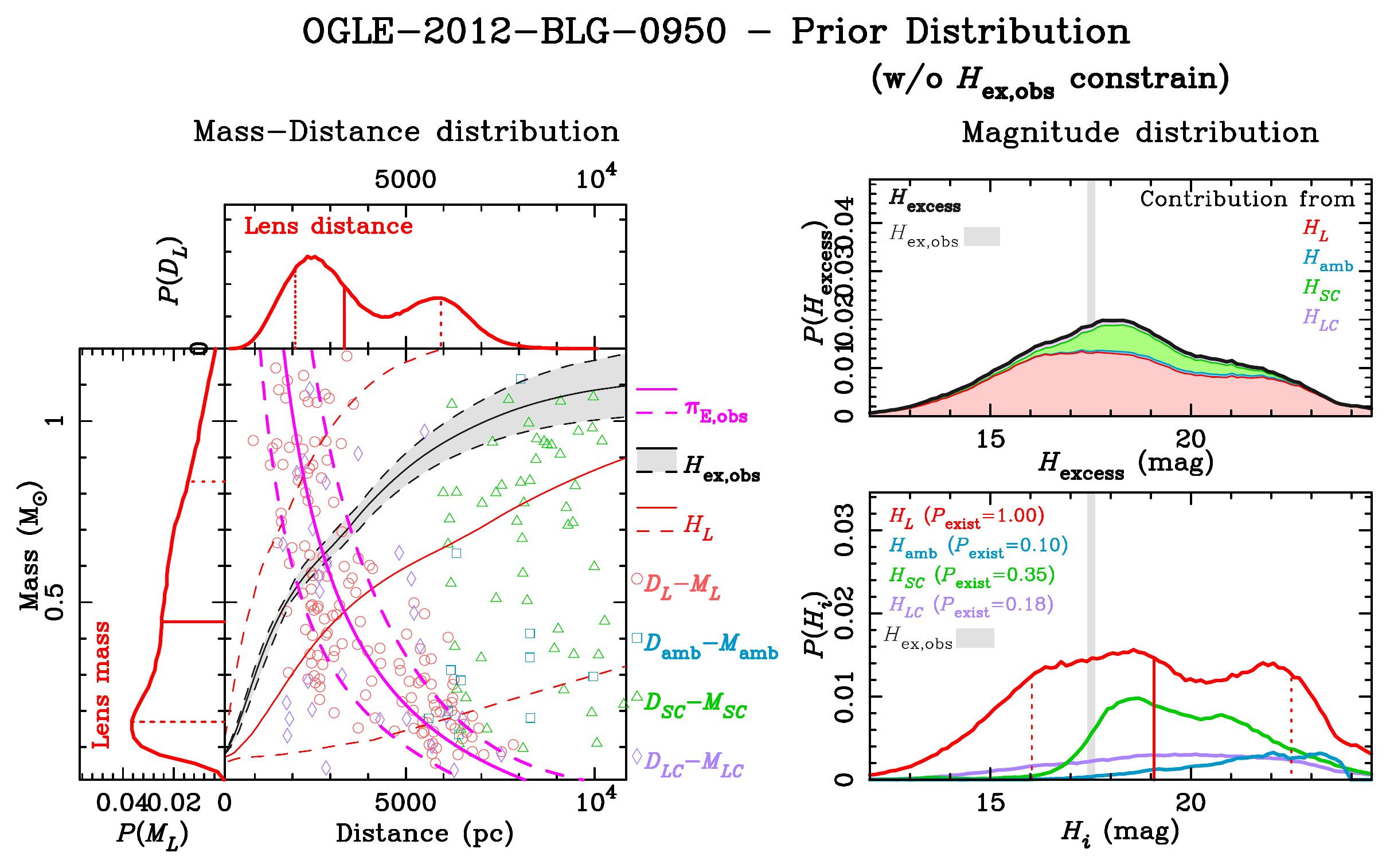}
 \label{fig-0950-pri}  
  }
  \hfill
  \subfigure[]{
   \includegraphics[width=140mm]{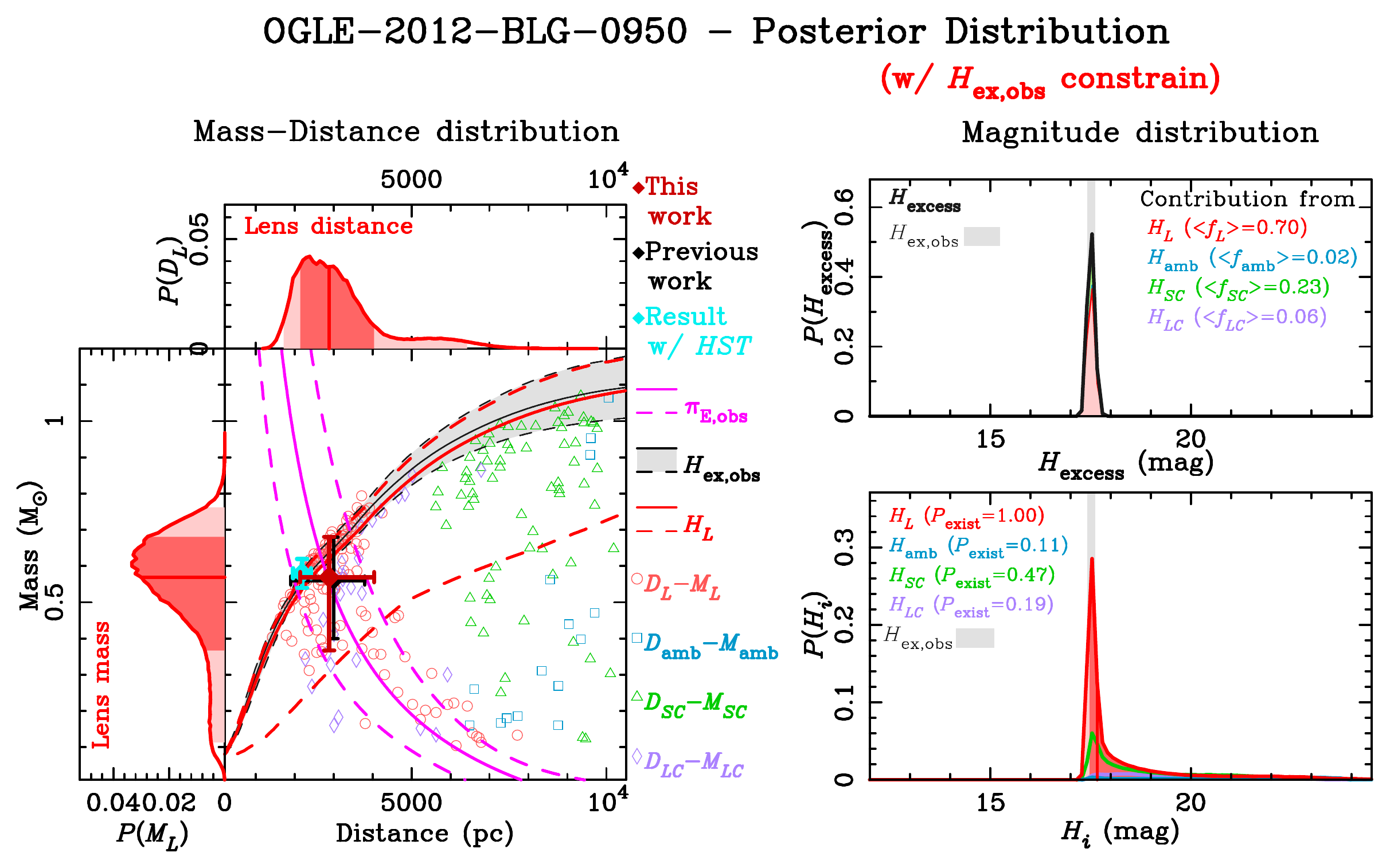}
    \label{fig-0950-post}
  }
  \end{center}
  \vspace{-0.6cm}
  \caption{Same as Fig. \ref{fig-227}, but for O120950. (a) Prior and (b) posterior probability distributions. 
  The lens mass and distance estimated by \citet{kos17b} and \citet{bha18} are indicated as ``Previous work'' and ``Result w/ {\it HST}'', respectively, 
  on the mass-distance plane in the posterior distribution.}
  \label{fig-0950}
\end{figure}

Possible origins of the observed excess flux can be found in all three panels in Figs. \ref{fig-227}--\ref{fig-0950} as described below.
In each of the left panels in Figs. \ref{fig-227}--\ref{fig-0950}, we plot part of the accepted combinations of mass and distance of the four contributors on the mass-distance plane, 
where the number of dots in each color is proportional to the $P_{\rm exist}$ values of each contributor.
We plot the distance to the ambient star, $D_{\rm amb}$, and its mass $M_{\rm amb}$ by assuming that they are located at the source distance in each step of the Monte Carlo simulation.
This is a crude assumption just for this plot, and has no effect on our result.
A part where a specific color is densely plotted indicates a high probability that the corresponding object has the mass and distance of the part.
In the same plane, we also show the mass-distance relations from $\theta_{E, \rm obs}$ or $\pi_{E, \rm obs}$, $H_{\rm ex, obs}$, and $H_L$.
Note that the value of $H_L$ is not the observed quantity; it is from the probability distribution obtained by our calculation.
The mass-distance relation from $H_{\rm ex, obs}$ is plotted by assuming that $H_{\rm ex, obs}$ comes from only one star.
This indicates that a contributor that has many dots on the $H_{\rm ex, obs}$ mass-distance relation curve is likely to be an origin of the observed excess flux.
This consideration is basically applicable to both (a) and (b) panels.
For example, we can see many green dots and fewer red dots on the $H_{\rm ex, obs}$ curve in the left panels of both Fig. \ref{fig-310} (a) and (b), which indicates that 
the most likely source of the excess flux is the source companion rather than the lens.

The origin of the excess flux can be discussed similarly with each of the magnitude distributions in the top right panels and 
bottom right panels in Figs. \ref{fig-227}--\ref{fig-0950}.
The bottom right panel of each figure shows the prior or posterior probability distributions of $H_i$ ($i = L, {\rm amb}, SC, LC$), where 
we can find which contributor is likely to be the main origin of the observed excess flux by comparing the $P (H_i)$ values in the gray shaded region of $H_{\rm ex,obs}$.
We note that the ratio among probabilities that a contributor has a brightness of $H_{\rm ex, obs}$, 
$P (H_i = H_{\rm ex, obs})/P (H_j = H_{\rm ex, obs})~ (i \ne j)$, in the prior PDF does not equal the ratio in its posterior PDF.
This is because the correlations among the parameters become stronger in the posterior PDF owing to the request of $F_{\rm excess} (\equiv F_L + F_{\rm amb} + F_{SC} + F_{LC}) = F_{\rm ex, obs}$
compared to the prior PDF where the parameters are nearly independent of each other except for the combination of $H_L$ and $H_{LC}$.

The black thick line in the top right panel in each of Figs. \ref{fig-227}--\ref{fig-0950} represents the $H_{\rm excess}$ probability distribution.
In the same panel, we divide the $H_{\rm excess}$ probability distribution into four color areas to visually clarify the average contribution from 
each contributor to the excess flux.
Let the fraction of each contributor's flux to the excess flux be $f_i \equiv F_i / F_{\rm excess}$ ($i = L, {\rm amb}, SC, LC$).
Then, the fraction of the vertical width of each color region to the height of $P (H_{\rm excess})$ at a given $H_{\rm excess}$ value equals the mean of $f_i$, where 
the mean is taken in scenarios in the Monte Carlo simulation whose excess flux corresponds to the magnitude bin of the given $H_{\rm excess}$.
In the top right panel of the (b) components, we show another mean value of $f_i$, which is taken in all the accepted scenarios in the posterior calculation, as $\VEV{f_i}$.
This value corresponds to the area of each color in the same panel. 
Each $\VEV{f_i}$ value indicates the average contribution of each contributor to the observed excess flux in various scenarios that 
are consistent with the observation.
Note that these are just average contributions; hence, it does not mean that a scenario with the brightness corresponding to 
these fractions (i.e., a scenario with $F_i = \VEV{f_i} F_{\rm excess}$) is highly likely.
We can determine which object is likely to contribute significantly to the excess flux from this panel, i.e., a contributor whose color occupies a large area 
in the gray shaded region of $H_{\rm ex,obs}$ in the prior distribution, or a contributor whose $\VEV{f_i}$ value is high in the posterior distribution.

\subsection{Lens properties constrained by excess fluxes}
Table \ref{tab-result} summarizes the lens mass $M_L$ and the distance to the lens, $D_L$, obtained from the posterior PDF, as well as the planet mass $M_{\rm p} = q\, M_L/(1+q)$ and
the projected separation between the host star and the planet, $a_{\perp} = s\, D_L \theta_{\rm E}$, for each event, where the host-planet mass ratio $q$ and 
the separation $s$ are given by the discovery paper of each event.
Because the events in consideration are all planetary events, we note that all the lens parameters listed in Table \ref{tab-result} depend on
the unknown prior of the planet hosting probability, which is assumed to be the same here regardless of the host star's property, but is, in fact,  probably different 
depending on the property.
We further discuss this point in Section \ref{sec-phost} and show that the change in the median value of $M_L$ is within the 1-$\sigma$ uncertainty shown in Table \ref{tab-result} for 
all the events if we consider a different assumption on it.

\subsubsection{Treatment for degenerate solutions}
All the events to which we applied our method, except for M16227, have two solutions of the close model and the wide models.

We combined the probability distributions of the projected separation $a_{\perp}$ calculated from each solution for events where 
both close and wide models have similar $s$ values, i.e., for M08310, which has $s = 1.085$ (wide model) and 
$s = 0.927$ (close model), and O120950, which has $s = 1.004$ (wide model) and $s = 0.895$ (close model).
We combine the two probability distributions with no weight, although there are $\chi^2$ differences between the close and wide models, 
i.e., $\Delta \chi^2 = 2.06$ for M08310 and $\Delta \chi^2 = 1.5$ for O120950. 
This is because the photometry data for the densest field such as the galactic bulge generally suffer from systematic errors; 
thus, we conservatively treated the two solutions equally with $\chi^2$ differences less than $\sim 2$.
Note that the relative difference of $s$ between the two solutions are 15\% for M08310 and 11\% for O120950, 
and these are less than the width of the 1-$\sigma$ confidence interval of the Einstein radius $R_{\rm E} = D_L  \theta_{\rm E}$ for the two events, i.e., 17\% and 21\%, respectively.
Therefore, any treatment of the weight between the two solutions negligibly affects the probability distribution of $a_{\perp} = s R_{\rm E}$ because the uncertainty of $R_{\rm E}$ is dominant.

Meanwhile, we show the two values separately as $a_{\perp, \rm close}$ and $a_{\perp, \rm wide}$ for events where the two $s$ values are 
largely separated, i.e., for M11293 and O120563.

\subsubsection{Comparison with previous studies}
For comparison, we show the values of the lens mass and the distance to the lens presented in the original paper for each event in the same table.
We also plot them on the mass-distance plane of each posterior distribution in Figs. \ref{fig-227} (b)--\ref{fig-0950} (b) using black dots with error bars\footnote{ 
In Fig. \ref{fig-293} (b), the black dot is slightly above even the mass-distance relation of $H_{\rm ex, obs}$ because of the difference in the $A_{H, \rm rc}$ value used. 
Whereas we use $A_{H, \rm rc} = 0.47 \pm 0.10$ from the extinction law of \citet{nis09}, 
they used a combination of $A_{H, \rm rc} = 0.65 \pm 0.12$ from the extinction law of \citet{car89} and the value from \citet{nis09}.
We do not combine them because \citet{nat16} reported that the extinction law toward the galactic bulge is clearly different from the law of \citet{car89}.},
and we plot our results using red dots with error bars.
Note that lens properties calculated with the assumption of $H_L = H_{\rm ex, obs}$, which is a common assumption in most of the previous studies,
are shown for M16227 instead of the values in the original paper because we presented the results of this method for the event previously in \citet{kos17}.

In Figs. \ref{fig-310} (b) and \ref{fig-0950} (b), we additionally plot recent results using {\it HST} and Keck follow-up observations
by \citet{bha17, bha18} on M08310 and O120950 in light blue dots, respectively.
The two observations were conducted after the excess measurements used in this analysis, when the lens stars were sufficiently separated from 
the source stars so that the lens stars were resolved.
Notably, our results of $M_L = 0.15^{+0.29}_{-0.08}\, M_{\odot}$ and $D_L = 7.2  \pm 1.1$ kpc for M08310 
and $M_L = 0.57^{+0.11}_{-0.20}\, M_{\odot}$ and $D_L = 2.62^{+0.53}_{-0.56}$ kpc for O120950 are both consistent with 
$M_L = 0.21^{+0.21}_{-0.09}\, M_{\odot}$ and $D_L = 7.7  \pm 1.1$ kpc obtained by \citet{bha17} and $M_L = 0.58 \pm 0.04\, M_{\odot}$ and $D_L = 2.19 \pm 0.23$ kpc obtained by \citet{bha18}, 
respectively, without their {\it HST} or Keck data.

Our lens mass estimates for M16227 ($M_L = 0.28^{+0.24}_{-0.15}\, M_{\odot}$), 
M08310 ($M_L = 0.14^{+0.27}_{-0.07}\, M_{\odot}$), and M11293 ($M_L = 0.41^{+0.35}_{-0.23}\, M_{\odot}$) are less massive and have 
larger uncertainty than the results reported in previous studies or the results with the assumption of $H_L = H_{\rm ex, obs}$, i.e., $M_L = 0.63 \pm 0.08 M_{\odot}$, 
$M_L = 0.67 \pm 0.14 M_{\odot}$, and $M_L = 0.86 \pm 0.06 M_{\odot}$, respectively.
Meanwhile, our results for O120563 ($M_L = 0.37 \pm 0.12\, M_{\odot}$) and O120950 ($M_L = 0.57^{+0.11}_{-0.20}\, M_{\odot}$) 
are similar to the previous results of the discovery papers, i.e., $M_L = 0.34^{+0.12}_{-0.20} M_{\odot}$ and $M_L = 0.56^{+0.12}_{-0.16} M_{\odot}$, respectively.

Because most of the previous studies derived the lens mass with the assumption of $H_L = H_{\rm ex, obs}$, whether our value is similar to theirs 
depends on the probability of the lens being the main origin of the excess flux.
If this probability is low, the probability of the lens flux being fainter than the excess flux increases, which makes the lens mass estimate less massive.
In such a case, there is no inconsistency regardless of how much the lens is fainter than the excess flux. 
Therefore, the posterior distribution of $H_L$ takes the shape of 
the prior distribution for the faint region, which results in a large uncertainty of the lens mass estimate.

\subsubsection{Probability that lens flux accounts for a certain fraction of excess flux}
For M16227, M08310 and M11293, the probabilities of the lens being the main origin of the excess flux are smaller than or comparable to the probabilities of other contaminants, as seen in the right bottom panels in Figs. \ref{fig-227}--\ref{fig-293}.
This is also known from the values $P(f_L > 0.1)$, $P(f_L > 0.5)$, and $P(f_L > 0.9)$ shown in Table \ref{tab-result}, which are 
the probabilities of the fraction of the lens flux to the excess flux, $f_L = F_L/F_{\rm excess}$, being larger than 0.1, 0.5, and 0.9, respectively.
The fraction $f_L$ is related to the difference of the magnitudes between the lens and the excess by $H_L - H_{\rm excess} = -2.5 \log f_L$, and 
these three probabilities are equivalent to the probabilities of $H_L - H_{\rm excess}$ being smaller than 2.5 mag, 0.75 mag, and 0.11 mag, respectively.

For example, the probabilities $P(f_L > 0.5)$ of M16227, M08310, and M11293 are 37.8\%, 25.2\%, and 31.3\%, respectively.
This indicates that the probabilities of $F_L \leq 0.5 \, F_{\rm excess}$, or equivalently, the probabilities of $H_L - H_{\rm excess} \geq 0.75~{\rm mag}$, are
higher than 60\% for these three events.
Meanwhile, the probabilities of $P(f_L > 0.5)$ are 99.94\% for O120563 and 70.5\% for O120950, which indicates that 
large parts of the observed excess flux for these events are likely to come from the lens stars.

The median and 1-$\sigma$ confidence interval values of the difference of magnitudes $H_L - H_{\rm excess}$ are also shown in the same table.
Because $H_L - H_{\rm excess}$ and $f_L$ are in one-to-one correspondence, a median value of $H_L - H_{\rm excess}$ close to 0 mag together with its small 1-$\sigma$
range indicates a high probability that $f_L$ is close to 1, i.e., a high probability of the lens being the origin of the excess flux, which results in a strong constraint on the 
lens properties.
For  M16227, M08310, and M11293, the 1-$\sigma$ ranges of $H_L - H_{\rm excess}$ are 3.02 mag, 6.09 mag, and 3.57 mag,
respectively. By contrast, they are 0.07 mag and 2.12 mag for O120563 and O120950, respectively.
The large 1-$\sigma$ ranges of $H_L - H_{\rm excess}$ for the former three events indicate much weaker constraints on their lens mass estimates compared to
the estimates where $f_L = 1$ is assumed.

\section{Interpretation of Results}\label{sec-interp}
We found that the probability of the lens being the main origin of the excess flux is not significant for M16227, M08310, and M11293, while 
it is significant for O120563 and O120950.
In this section, we investigate the causes for the different probabilities among these five events.
We find that when $\theta_{\rm E}$ is large or likely to be large, the lens is likely to be the main origin of the excess flux.
Otherwise, the probability of a large contribution from other contaminants, especially from the source companion, cannot be ruled out.

\subsection{Event with a small angular Einstein radius $\theta_{\rm E}$} \label{sec-sthetaE}
We find that the key parameter that determines whether we can impose a tight constraint on $H_L - H_{\rm excess}$ or on the lens properties 
is the angular Einstein radius $\theta_{\rm E}$ rather than the radius of the unresolvable circle, $\phi_{\rm wide}$, 
within which we have been considering all possible contamination scenarios.
The bottom right panels in Figs. \ref{fig-227} (b)--\ref{fig-293} (b) show 
high probabilities of ambient stars and the source companion being a possible origin of the observed excess for M16227 and M08310, 
and of the source companion being a possible origin of the observed excess for M11293, although they also show comparable probabilities for the lens star for all these events.
Thus, the source companion is the main contaminant in all the events where we cannot impose a strong constraint on the lens property.

\subsubsection{High angular resolution negligibly reduces probability of source companion}
To determine  how these contaminants work, we compare the results of M16227 with $\phi_{\rm wide} = 148$ mas and the results of M11293 with 
$\phi_{\rm wide} = 60$ mas, which is $\sim 2.5$ times smaller than the former value.
A small $\phi_{\rm wide}$ value effectively reduces the probability of existence of ambient stars because the average number of ambient stars within 
a circle of radius $\phi_{\rm wide}$, $\lambda_{\rm amb}$, given by Eq. (\ref{eq-lamamb}), is proportional to $\phi_{\rm wide}^2$.
The $P_{\rm exist}$ value for ambient stars in the prior distribution is 0.08 for M11293, while it is 0.39 for M16227, as shown in 
the bottom right panels in Figs. \ref{fig-293} (a) and \ref{fig-227} (a).

\begin{figure}
\centering
\epsscale{0.55}
\plotone{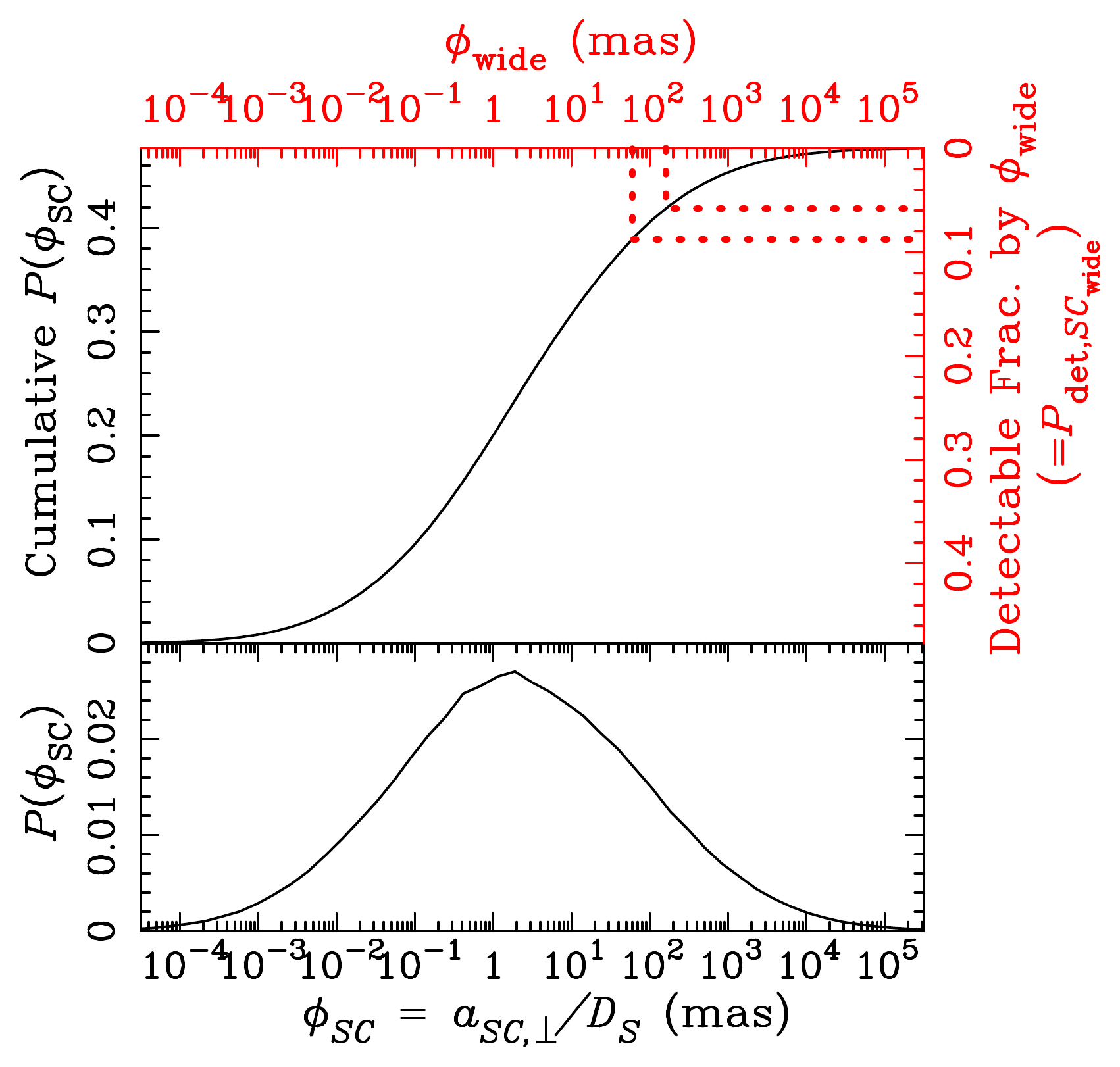}
\caption{Probability distribution of the angular separation of a source companion, $\phi_{SC} \equiv a_{\rm SC, \perp}/D_S$, 
in the full binary distribution of M16227 (bottom) and its cumulative distribution (top).
The right axis of the top panel corresponds to the detectable fraction of a source companion located 
outside of the undetectable circle with the radius $\phi_{\rm wide}$.
The two red dotted lines indicate the fraction of a source companion 
 $\phi_{\rm wide} = 60$ mas and $\phi_{\rm wide} = 160$ mas and their corresponding detectable fraction.
}
\label{fig-phiSC}
\end{figure}

Meanwhile, the log-normal distribution is used as the semi-major axis distribution of the binary system as described in Section \ref{sec-binpri} and 
summarized in Table \ref{tab-inputs}.
Therefore, a $\phi_{\rm wide}$ value that is even $\sim 2.5$ times smaller reduces the considered region of the semi-major axis by only $\sim 0.4$ dex on the log scale and negligibly reduces the probability of existence of the source companion. 
This is seen in a comparison of the far right value of the solid curves in the panel labeled $\log a_{SC}$ in Fig. \ref{fig-bindis} (a) with that in Fig. \ref{fig-bindis} (b).
Fig. \ref{fig-phiSC} shows this clearer by plotting the distribution of $\phi_{SC} \equiv a_{SC, \perp}/D_S$, the angular separation of companions around the source star of M16227.
The top panel shows the cumulative probability distribution of $\phi_{SC}$ and simultaneously the detectable fraction of source companions by $\phi_{\rm wide}$ by the red axes.
The two red dotted lines in the top panel indicate the detectable fractions by the maximum and minimum $\phi_{\rm wide}$ values in the five events, i.e., 160 mas and 60 mas.
The difference of the two fractions of $\sim 3$\% is small relative to the remained fraction.
Fig. \ref{fig-phiSC} indicates that it is difficult to drastically reduce the probability of existence of the source companion with any realistic $\phi_{\rm wide}$ value achieved by the current high-angular resolution imaging.

\subsubsection{Source companion flux prior probability is maximum at source flux} \label{sec-scpeak}
If the angular Einstein radius $\theta_{\rm E}$ is relatively small, as with M16227 ($\theta_{\rm E, obs} = 0.23 \pm 0.01$ mas), 
M08310 ($\theta_{\rm E, obs} = 0.16 \pm 0.01$ mas), and M11293 ($\theta_{\rm E, obs} = 0.26 \pm 0.02$ mas),
the prior probability of $H_L$ in a region brighter than $\sim 20$ mag, where the observed excess flux is distributed, is smaller than or comparable to 
the prior probability of $H_{SC}$, as shown in the bottom right panels in Figs. \ref{fig-227} (a)--\ref{fig-293} (a).

The same is also determined from the distributions of mass and distance in the left panels in the same figures.
As seen from the red dots of the $D_L$--$M_L$ distributions in the left panels in Figs.  \ref{fig-227} (a)--\ref{fig-293} (a), 
for events with small $\theta_{\rm E, obs}$,
the peak of the joint prior probability distribution of $D_L$ and $M_L$ is at a late- or mid-M dwarf close to the source star along the shape of
the $\theta_{\rm E}$ mass-distance relation.
This is because in the Galactic model, there are many more stars in the bulge than in the disk and there are many more low-mass stars 
than high-mass stars.
Moreover, there is another factor, namely the microlensing event rate ($\propto D_L^2\, \theta_{\rm E}\, \mu_{\rm rel}$), which increases with $D_L$ for given $\theta_{\rm E}$ and $\mu_{\rm rel} = \theta_{\rm E}/t_{\rm E}$.

Meanwhile, the distribution of the source companion is different.
The green dots in the same panels are distributed in a broad mass range at the source distance and 
they outnumber the red dots of the lens in $M \simgt 0.6\, M_{\odot}$, where the intersection of the two mass-distance relations of 
$\theta_{\rm E}$ and $H_{\rm ex, obs}$ is located for these three events.
The source companion mass distribution has a high probability at $M \simgt 0.6\, M_{\odot}$ because the peak of the $M_{SC}$ distribution
is at the source mass of each event, which is around $0.8-1.1\, M_{\odot}$ depending on the event, and the probability remains high at 
somewhat lower masses.
This distribution of the mass $M_{SC}$ is a reflection of the distribution of the mass ratio $q_{SC}$ in Fig. \ref{fig-bindis}, which has 
a peak at $q_{SC} = 1$ and a gentle slope toward $q_{SC} < 1$.
This property of the $q_{SC}$ distribution also indicates that the peak of the probability distribution of 
the source companion magnitude $H_{SC}$ is at the source magnitude $H_{S, {\rm obs}}$, and the probability remains high for somewhat fainter magnitudes.

Therefore, although the lens has an advantage of $P_{\rm exist} = 1$, in a case where an observed excess has the brightness of the source star or is slightly fainter, 
the probability of a source companion being the origin of the excess is comparable to or larger than the probability of the lens being the origin when $\theta_{\rm E}$ is small.

\subsection{Event with a large angular Einstein radius $\theta_{\rm E}$}
The result for O120563, which has a very large angular Einstein radius $\theta_{\rm E, obs} = 1.4 \pm 0.1$ mas, 
shows a very small uncertainty of  $H_L - H_{\rm excess} = 0.005^{+0.068}_{-0.004}$, which 
indicates that nearly the entire fraction of the excess flux is highly likely to originate from the lens flux.
This is because the angular Einstein radius $\theta_{\rm E}$ is proportional to $\sqrt{M_L \pi_{\rm rel}}$ and a large angular Einstein radius indicates 
a massive and/or close lens, i.e., a bright lens star.
In other words, a value of $\theta_{\rm E}$ can effectively impose a lower limit on the lens flux or an upper limit on the lens magnitude with the assumption that the lens is not a remnant\footnote{A mathematical 
lower limit of the lens flux for a given $\theta_{\rm E}$ is 0 when $M_L \rightarrow 0$ and $(D_L/D_S) \rightarrow 0$; however, the lensing probability for such an 
extremely close object is also 0.
Hence, we can effectively impose a lower limit on the lens flux within a likely range of $(D_L/D_S)$. We also note that this is true only when the lens is a single star. A remnant or binary lens could be fainter than the lower limit.}.
Thus, there is a critical value of $\theta_{\rm E}$ above which the corresponding lower limit of the lens flux is brighter than most of the source companions that are the main contaminants.

As discussed in Section \ref{sec-scpeak}, the peak of the probability of the source companion magnitude is at the source magnitude. 
Furthermore, considering that the mass-ratio distribution in Fig. \ref{fig-bindis} indicates a small probability of $q_{SC} > 1$, which is also the same for other events, 
the critical $\theta_{\rm E}$ value is roughly a value that gives a lower limit of the lens flux that equals the source flux.
We can constrain the lens mass by AO observations of the event with such a promising $\theta_{\rm E}$ value before the lens star is sufficiently separated from the source star.
For O120563 with $\theta_{\rm E, obs} = 1.4 \pm 0.1$ mas, the source magnitude of $H_{S, {\rm obs}} = 18.57 \pm 0.03$ is 
fainter than even the faint end of the prior probability distribution of the lens magnitude, as shown in the bottom right panel in Fig. \ref{fig-0563} (a).
This indicates that $\theta_{\rm E, obs}$ is higher than the critical value; thus, we can see that the probability of contamination 
from the source companion is very small without the calculation of the contamination probability.

Moreover, in this specific case, there is another reason why we can nearly completely rule out contributions from contaminants.
Although the left panel or the bottom right panel in the prior distribution of Fig. \ref{fig-0563} (a) shows a small probability of the source companion or ambient stars having the
brightness of the observed excess flux,  the posterior distribution of Fig. \ref{fig-0563} (b) does not show any possibility of such a case where a source companion or ambient stars are 
the origin of the excess flux.
This is because the observed excess flux is nearly the minimum limit of the lens flux expected from $\theta_{\rm E, obs}$, as seen in the bottom right panel in Fig. \ref{fig-0563} (a).
In this case, after subtraction of the lens flux from the observed excess flux, the remainder always becomes much fainter than the excess flux; 
thus, no contaminant can have a brightness of $H_{\rm ex, obs}$.

\subsection{Event with a long Einstein radius crossing time $t_{\rm E}$}
For O120950, where a microlens parallax of $\pi_{\rm E, obs} = 0.26 \pm 0.06$ was detected, the discussion is different.
Because the microlens parallax $\pi_{\rm E}$ is proportional to $\sqrt{\pi_{\rm rel}/M_L}$, the condition that $\pi_{\rm E}$ is constant requires smaller $\pi_{\rm rel}$ for 
smaller $M_L$, which makes the lens fainter, while it requires larger $\pi_{\rm rel}$ for larger $M_L$, which makes the lens brighter.
As a result, both the bright lens and the faint lens are approved and we cannot impose a lower limit on the lens flux from an observed $\pi_{\rm E}$ value 
in contrast to the constraint from $\theta_{\rm E}$.
In this case, one might expect that the red dots in the left panel in Fig. \ref{fig-0950} (a) should be distributed intensively around the bottom right part (i.e., low mass and distant lens) because such stars are the most common ones in our galaxy.
However, the figure shows that they are actually distributed more broadly in the plane along the $\pi_{\rm E}$ mass-distance relation compared to 
the prior distribution of the three events with small $\theta_{\rm E}$ in the left panels in Figs. \ref{fig-227} (a)--\ref{fig-293} (a).

To resolve the origin of this broad distribution, we need to recall another observed quantity, $t_{\rm E, obs}$, which has the information of $M_L$ but is not drawn explicitly on this plane.
Because $t_{\rm E} \propto \theta_{\rm E} \propto \sqrt{M_L}$, the lens of O120950, where a long event timescale of $t_{\rm E, obs} = 68$ days was observed,
is likely to have a relatively large mass.
Owing to the long $t_{\rm E, obs}$, the prior probability that the lens is sufficiently bright to explain the observed relatively bright excess flux increases; consequently, 
the estimated lens mass is similar to the value reported by the discovery paper of \citet{kos17b}, who derived the value by taking the mean of the solutions with $f_L = 1$ and $f_L = 0.5$.
Note that \citet{bha18} confirmed that $f_L$ is nearly 1 for this event by analyzing Keck and {\it HST} images taken in 2018, which indicates
an elongation caused by slightly separated source and lens. Their tightly constrained mass and distance are also plotted in light-blue in Fig. \ref{fig-0950} (b).

Because a long event timescale of $t_{\rm E} \simgt 50$ days is typically required to detect the annual parallax effect, 
we note that it is not by chance that we have a long $t_{\rm E}$ value for this event where $\pi_{\rm E}$ is detected.
It indicates that the probability of a lens being an origin of the excess flux is generally high when a bright excess flux is observed at the location of the 
event where the annual parallax effect is detected.

\subsection{Dependence on the angular Einstein radius and the angular resolution}\label{sec-HiHex}
To investigate how the constraint on the lens properties varies depending on the angular Einstein radius, we conduct the Bayesian analysis on 
hypothetical events with 7 different angular Einstein radius values, $\theta_{\rm E, obs} = (0.2, 0.4, 0.6, 0.8, 1.0, 1.2, 1.4)$ mas, with 10\% uncertainties for each.
We consider the case where the excess brightness of $H_{\rm ex, obs} = 18.0 \pm 0.2$ is measured on each event position.
For the radius of the unresolvable circle, we consider two values, $\phi_{\rm wide} = 60$ mas and $\phi_{\rm wide} = 240$ mas, the minimum value and a value even 1.5 times larger than the maximum value 
in the five events listed in Table \ref{tab-info}, respectively.
We use the values for M16227 for other input parameters, which means we use $n_{\rm amb} = 7.3 \, {\rm as}^{-2}$, the densest number density in the five events' fields, for these calculations.

\begin{figure}
\begin{minipage}{0.50\hsize}
\begin{center}
\includegraphics[width=80mm]{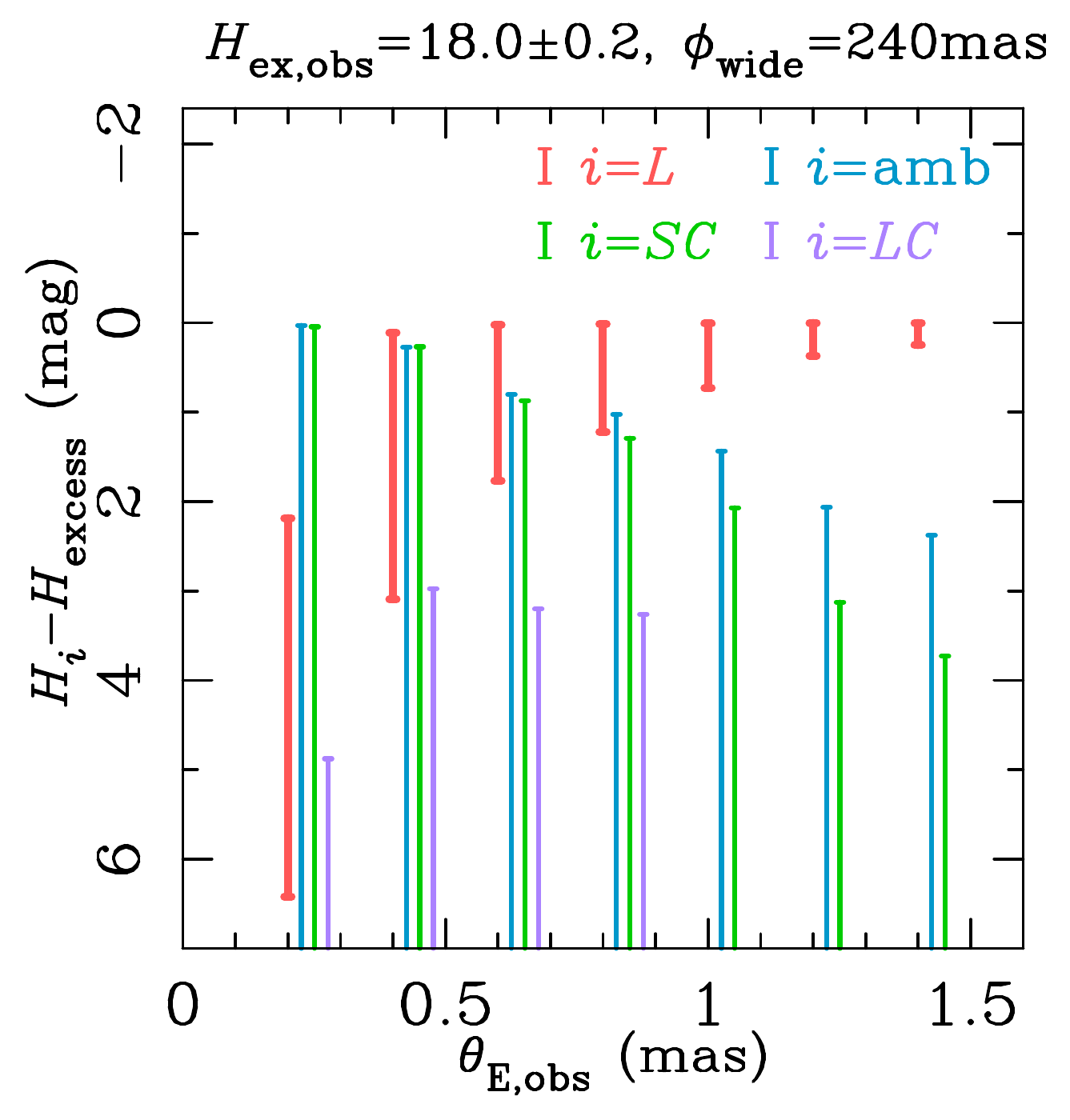}
\end{center}
\end{minipage}
\begin{minipage}{0.50\hsize}
\begin{center}
\includegraphics[width=80mm]{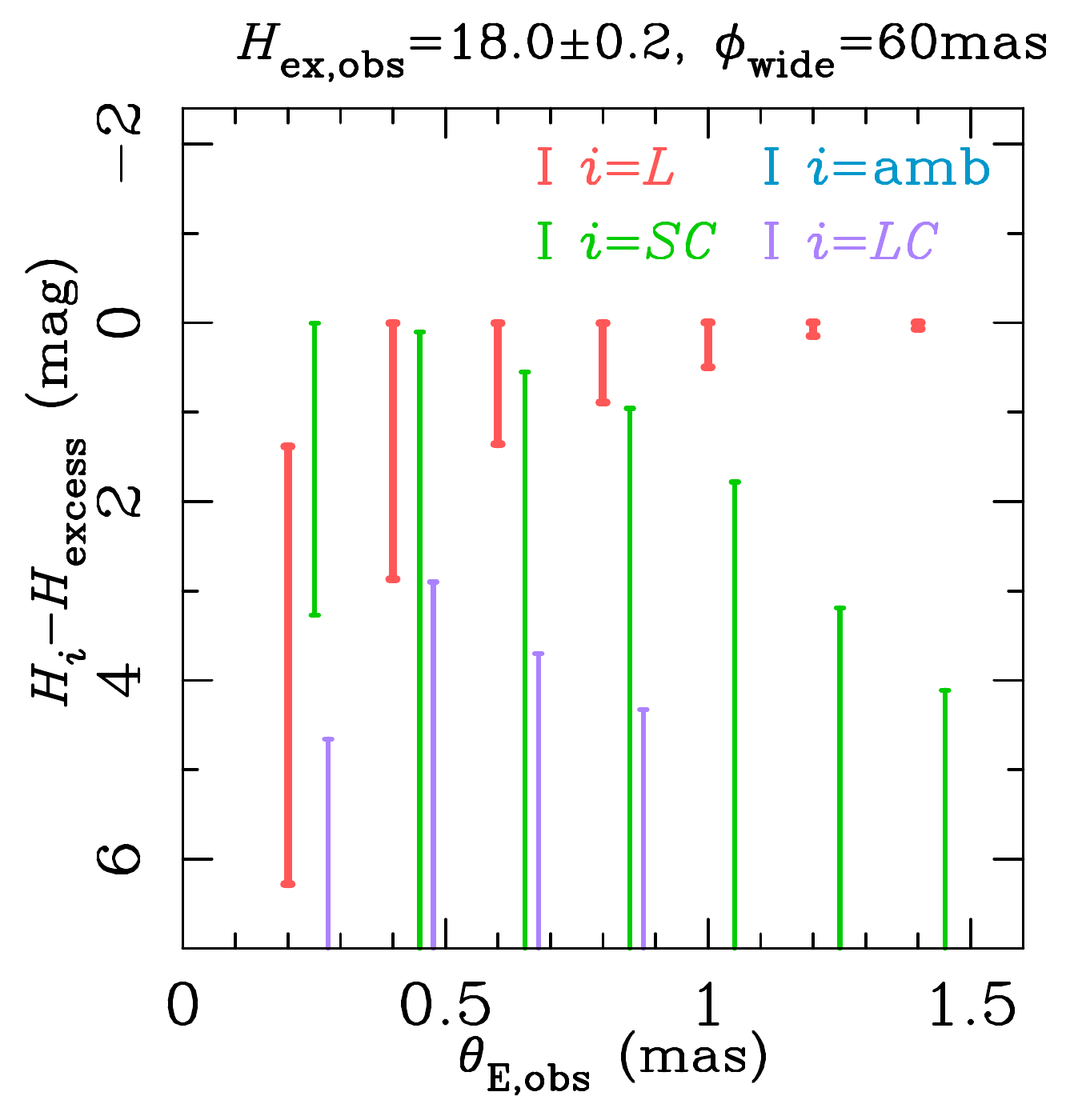}
\end{center}
\end{minipage}
\caption{1-$\sigma$ ranges of $H_i - H_{\rm excess} = -2.5 \log {f_i}$ ($i = L, {\rm amb}, SC, LC$) in the joint posterior PDF 
$f_{post} (F_L, F_{\rm amb}, F_{SC}, F_{LC} | F_{\rm excess} = F_{\rm ex, obs})$ for hypothetical events of 7 different $\theta_{\rm E, obs}$ values, on which
the excess brightness of $H_{\rm ex,obs} = 18.0 \pm 0.2$ is measured with two different $\phi_{\rm wide}$ values.
$H_i - H_{\rm excess} = 0$ is equivalent to $F_i = F_{\rm excess}$.
Plots for $i = {\rm amb}, SC, LC$ are horizontally shifted so that each bar is recognizable.
Note that most of the 1-$\sigma$ lower limits (i.e., 16th percentiles from the faintest) for $i = {\rm amb}, SC, LC$ or
some of the 1-$\sigma$ upper limits (i.e., 84th percentiles from the faintest) for $i = {\rm amb}, LC$ are not plotted in this magnitude 
range because they are too faint or correspond to non-existing cases.
}
\label{fig-HiHex_thE}
\end{figure}

Fig. \ref{fig-HiHex_thE} shows the resulting 1-$\sigma$ ranges (i.e., 16th to 84th percentiles) of $H_i - H_{\rm excess}$ $(i = L, {\rm amb}, SC, LC)$ in the posterior distributions from those analysis.
The left panel shows those with $\phi_{\rm wide} = 240$ mas while the right panel shows those with $\phi_{\rm wide} = 60$ mas.
These results confirm that the uncertainty of $H_L - H_{\rm excess}$ has a strong correlation with the size of $\theta_{\rm E, obs}$ and 
also that a source companion is the main source of contaminants regardless of the $\phi_{\rm wide}$ value.

Moreover, it is notable that the 1-$\sigma$ ranges of $H_L - H_{\rm excess}$ in the left panel are similar to those in the right panel even with the 4 times difference of $\phi_{\rm wide}$ values, or
equivalently 16 times difference of the mean number of ambient stars between the two panels.
In the case with $\phi_{\rm wide} = 240$ mas, ambient stars do contribute as a contaminant, but it does not lead to very different $H_L - H_{\rm excess}$ estimates.
This indicates that the ambient stars flux basically does not affect our conclusion whether the lens is likely to be the origin of the excess or not, as long as $\phi_{\rm wide}$ is within 
a typical range ($\simlt 200$ mas) for the AO observations.

\subsection{Lens companion as a contaminant}\label{sec-LC}
Finally, we discuss how a lens companion works as a contaminant.
Our results show that the probabilities of a lens companion being the origin of the excess flux are small for all the events we analyzed.
For example, the averaged contributions from the lens companions to the excess fluxes are $\VEV{f_{LC}} \leq 0.07$ for all the five events.
This is also seen in Fig. \ref{fig-HiHex_thE}, where the upper limits of $H_{LC} - H_{\rm excess}$ never exceed 2.5 mag in any case of hypothetical events.

We find that there are two main reasons for this small contribution from the lens companion.
First, the request for its location to be undetectable from the light curves, namely the request of $a_{LC,\perp}/D_L > \phi_{{\rm close}, LC}$,
effectively reduces the possibility of the lens companion, especially for high-magnification events with $u_{\rm 0, obs} \simlt 0.01$, as seen in the comparison of the full and the undetected 
binary distributions in the $q_{LC}$ and $\log a_{LC}$ panels in Fig. \ref{fig-bindis}.
Second, to be the main origin of the excess flux instead of the lens star, a binary system that consists of a lens companion with the brightness 
of an observed excess flux and a fainter lens that negligibly contributes to the excess is required.
This is equivalent to the requirement of $q_{LC}$ being moderately larger than 1.
However, the probability of the case where the lens is a secondary star (i.e., $q_{LC} > 1$) is low, as seen from the undetected binary distribution (solid curves) in 
the panels labeled $q_{LC}$ in Fig. \ref{fig-bindis}.
In summary, the probability of the lens companion being the main contaminant is effectively reduced by both the requirement on its semi-major axis (the first reason above) 
and the requirement on its mass ratio (the second reason above). Consequently, a confined parameter space remains.

This consideration of the possibility of a lens companion provides an important insight even when a lens candidate is detected with
separation from the source \citep{bat15,ben15,ben19,bha18,van19}.
Because such a candidate can be either the lens or a companion to the lens, one needs to distinguish these cases.
When the event has a 1D microlens parallax measurement from the light curve, it is distinguishable because both $\pi_{\rm E}$ and $\theta_{\rm E}$ can be determined 
independent of the measured brightness of the candidate \citep{bha18,ben19}.
Meanwhile, when the event does not have any microlens parallax measurement, the possibility of the lens companion cannot be ruled out in the same way \citep{van19}.
However, our calculation indicates that it is not highly likely that an unresolvable companion that is significantly brighter than the lens exists.
This is especially the case for a high-magnification event such as MOA-2013-BLG-220 \citep{van19} or MOA-2007-BLG-400 (Bhattacharya et al. 2019, in prep.) because of 
its high sensitivity to a companion, although the quantitative probability should be calculated in each case to show how unlikely it is.

\section{Prediction of Lens Detectability} \label{sec-pre}
It is difficult to impose a strong constraint on the lens flux or the lens mass if the prior probability of the lens flux having the brightness of the observed excess flux is 
smaller than or similar to that of other possible contributors.
Does this imply that we could have imposed a tight constraint on the lens mass if we had observed $H_{\rm ex,obs} \sim 22$ mag for M11293?
This is true if we had observed an excess flux of $\sim 22$ mag; however, it is impossible to detect such faint excess flux at the position of the much brighter source star.
Because the source flux $F_{S, {\rm obs}}$ has an uncertainty, we can claim that we detect the excess flux at the position of the source star only when the target flux 
(i.e., the source flux plus others' flux) is statistically significantly brighter than the source flux.
With the 3-$\sigma$ confidence limit, a detectable excess flux should have a greater brightness than 
$F_{\rm ex, 3 \sigma} = 3\, \sigma_{F_S}$, where $\sigma_{F_S}$ is the 1-$\sigma$ uncertainty of the source flux.

\begin{figure}
\begin{minipage}{0.32\hsize}
\begin{center}
\includegraphics[width=57mm]{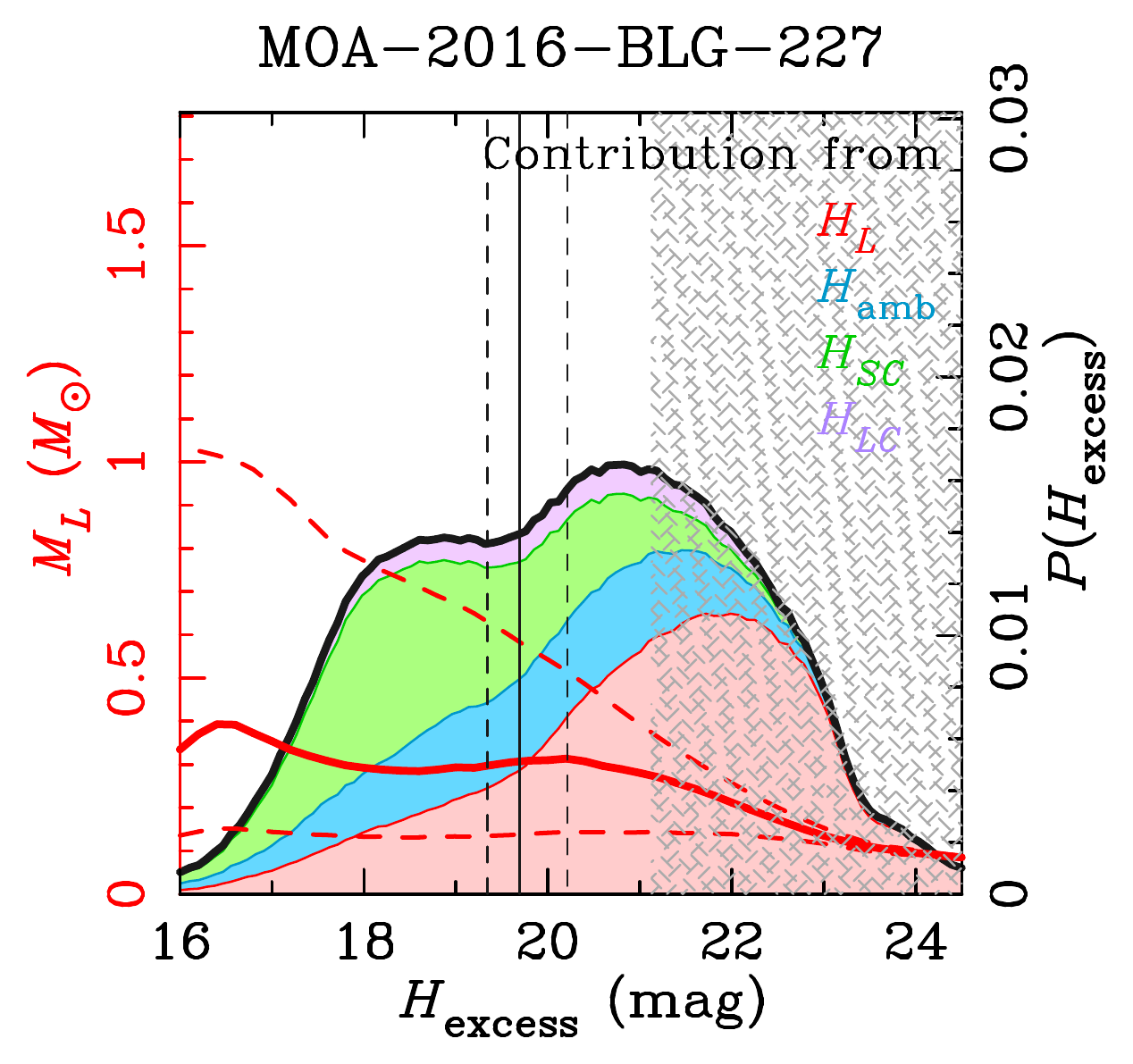}
\end{center}
\end{minipage}
\begin{minipage}{0.32\hsize}
\begin{center}
\includegraphics[width=57mm]{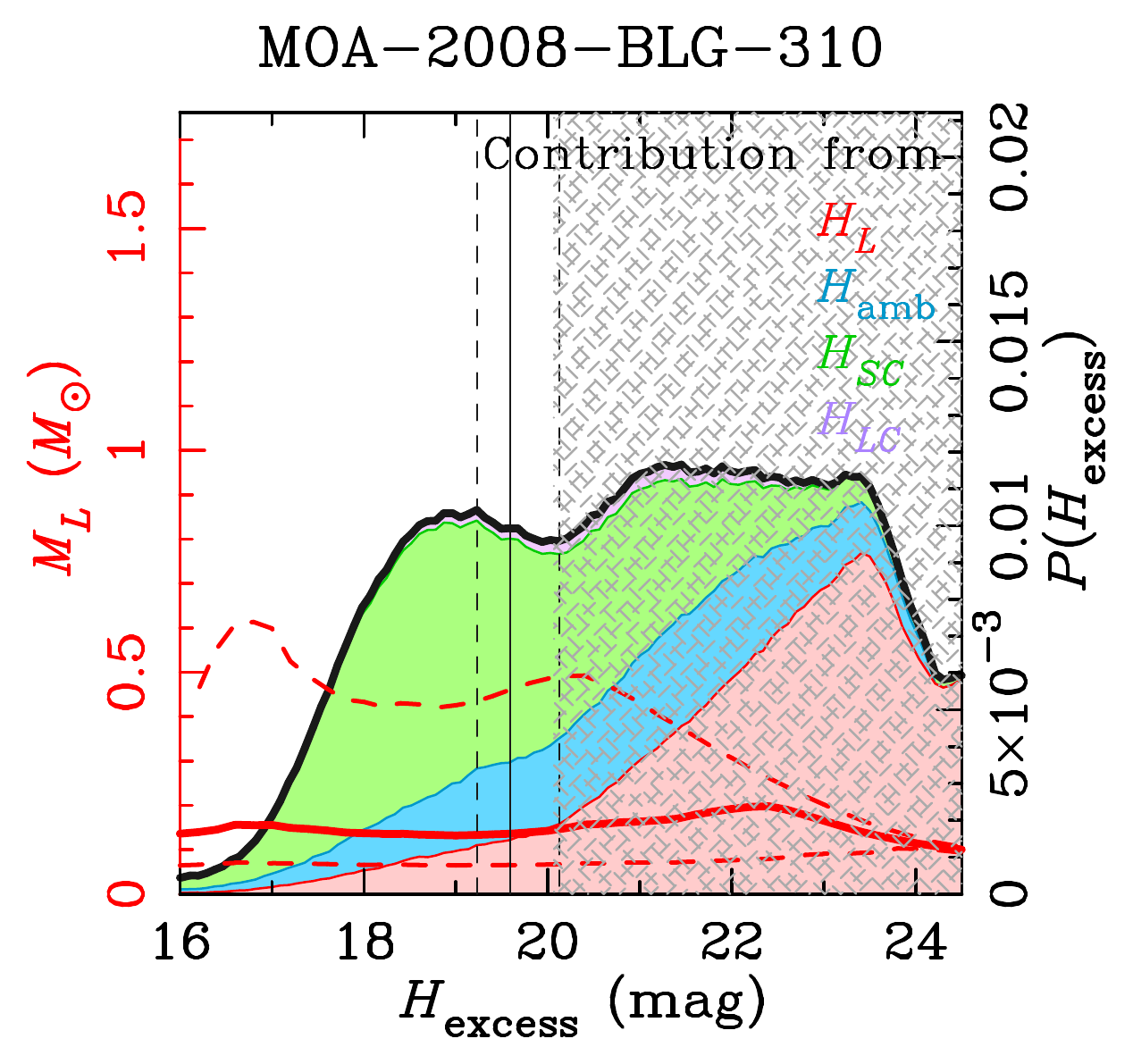}
\end{center}
\end{minipage}
\begin{minipage}{0.32\hsize}
\begin{center}
\includegraphics[width=57mm]{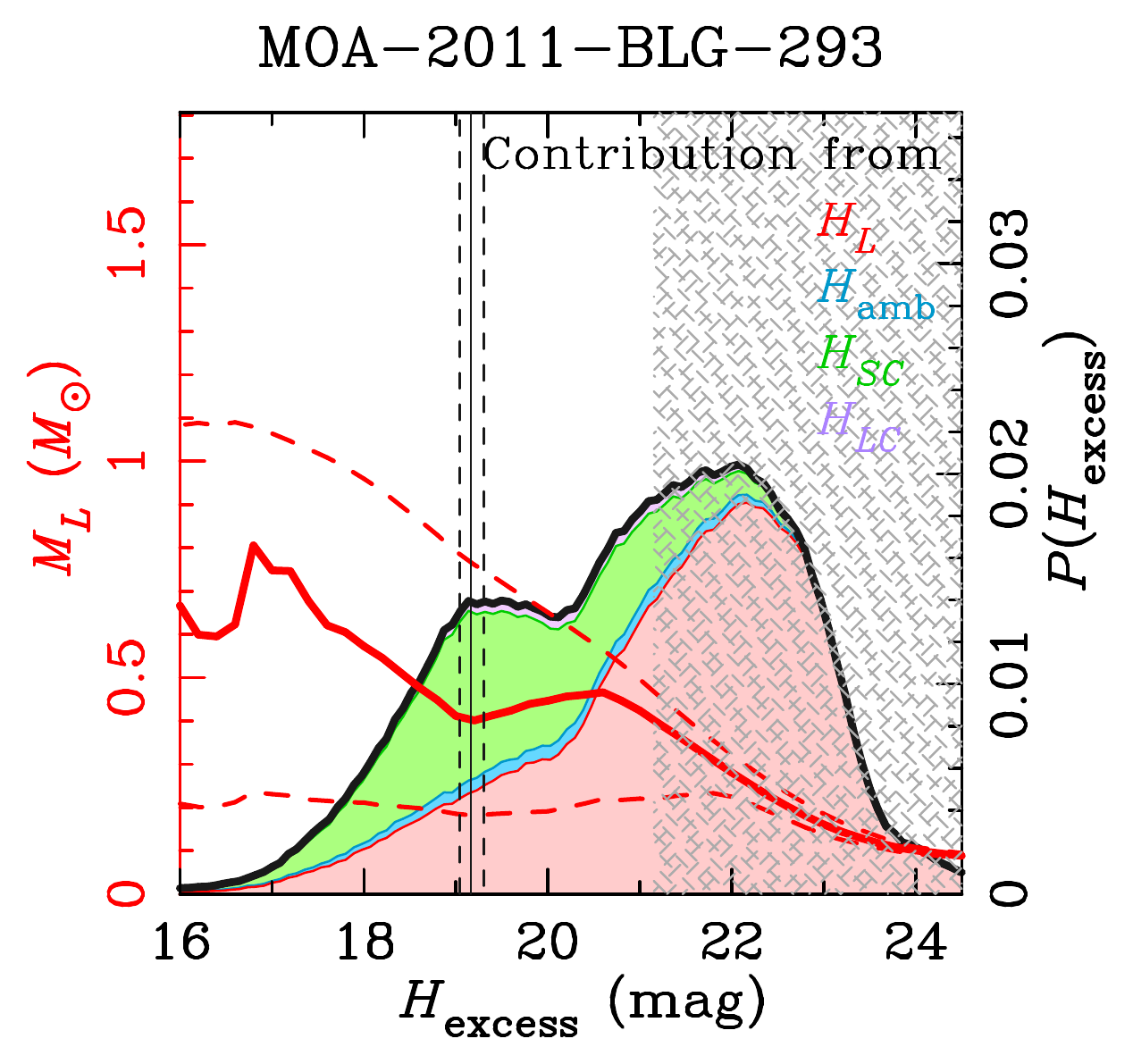}
\end{center}
\end{minipage}
\begin{minipage}{0.33\hsize}
\begin{center}
\includegraphics[width=57mm]{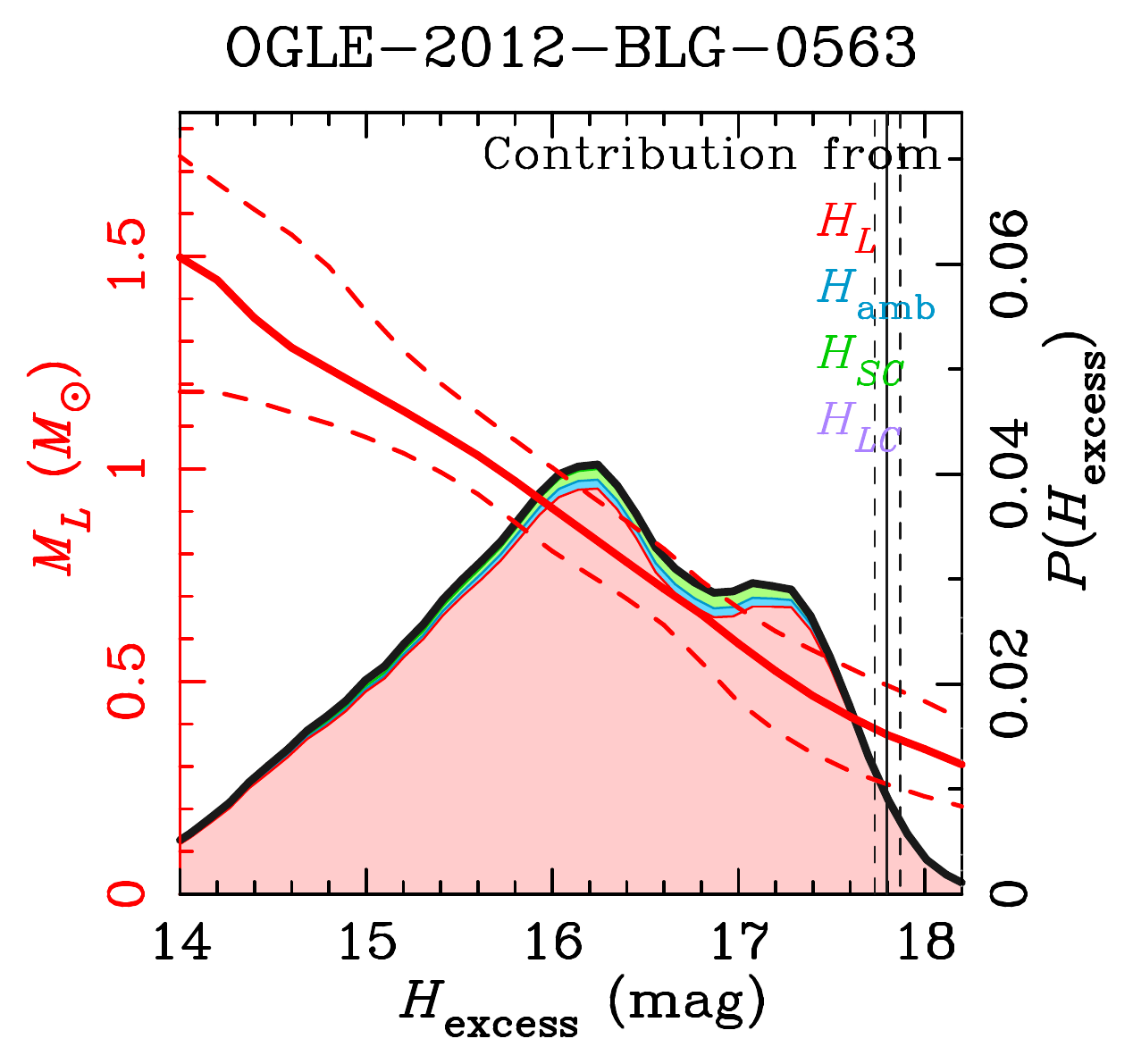}
\end{center}
\end{minipage}
\begin{minipage}{0.33\hsize}
\begin{center}
\includegraphics[width=57mm]{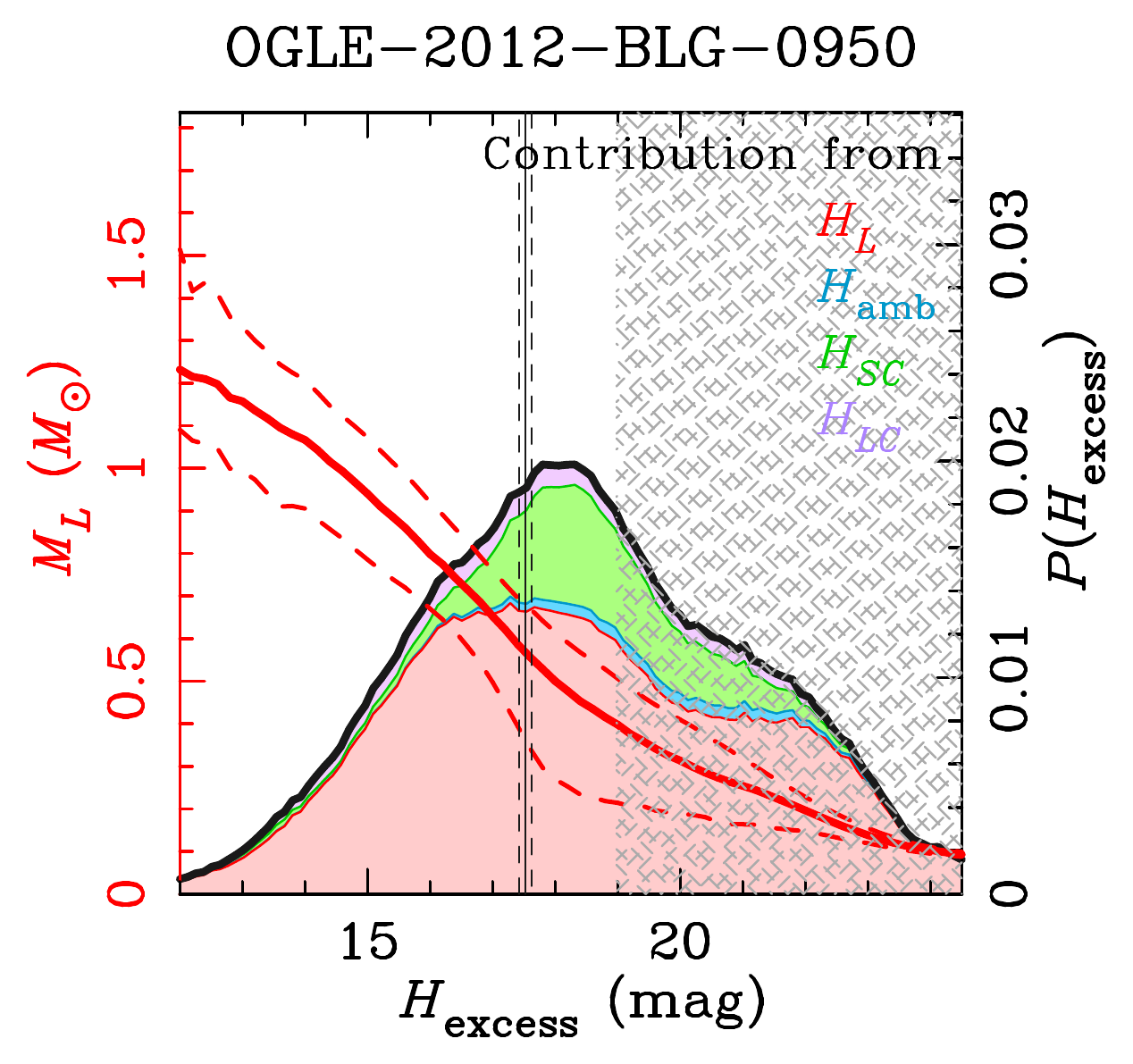}
\end{center}
\end{minipage}
\caption{Predictions for lens detectability. Each panel shows the prior probability distribution of the excess flux for each event.
They are repeated from the top right panels in the (a) components of Fig. \ref{fig-227}-\ref{fig-0950}, but with additional information.
The red solid and dashed lines indicate the median value and 1-$\sigma$ confidence limits of the posterior probability distribution of 
the lens mass $M_L$, which would be obtained if we were to observe $H_{\rm ex, obs} = H_{\rm excess} \pm 0.1~$mag, 
as a function of the magnitude of the excess flux $H_{\rm excess}$.
The gray hatched regions indicate 3-$\sigma$ undetectable regions of the $H_{\rm excess}$ value owing to the noise of the source flux, 
where we assumed unmagnified sources. The vertical black solid and dashed lines indicate the representative value and its 1-$\sigma$ uncertainty
of the actual observed excess flux $H_{\rm ex, obs}$, repectively.}
\label{fig-predict}
\end{figure}

Fig. \ref{fig-predict} shows the prior probability distribution of the excess flux for each event.
These distributions are repeated from the top right panels in the (a) components of Figs. \ref{fig-227}--\ref{fig-0950}, whereas additional information is plotted on the probability distributions.
The gray hatched areas in Fig. \ref{fig-predict} are the undetectable region $H_{\rm excess} > H_{\rm ex, 3 \sigma}$, where $H_{\rm ex, 3 \sigma}$ is 
the magnitude corresponding to $F_{\rm ex, 3 \sigma}$. We consider unmagnified source stars to derive $F_{\rm ex, 3 \sigma}$.
Because the region around $H_{\rm excess} \sim 22$ in the plot for M11293 is hatched, we cannot 
detect the excess flux value around $\sim 22$ mag where the prior probability of $H_L$ is the highest in our calculation.
The red solid and dashed curves represent the median and 1-$\sigma$ values of the posterior distribution of the lens mass obtained when 
the excess flux is measured as $H_{\rm ex, obs} = H_{\rm excess} \pm 0.1$, where $H_{\rm excess}$ is the value of the horizontal axis.
The black solid and dashed vertical lines represent the actually observed $H_{\rm ex, obs}$ values and the 1-$\sigma$ uncertainties.
Thus, the mass value at the intersection of the red curves and the black lines corresponds to the estimated mass value in Table \ref{tab-result}.

We emphasize that all the components of Fig. \ref{fig-predict} except for the black vertical lines can be obtained once the parameters derived 
from the microlens light curve are determined with an assumed value of $\phi_{\rm wide}$.
When one is planning high-angular-resolution follow-up observations, such a figure indicates the extent to which
one can expect to constrain the lens properties.
For example, it seems difficult to constrain the lens mass tightly for M16227 and M08310 because the width of the 68\% confidence interval of the
red curves is large in any detectable excess flux region in the unhatched area.
The situation is better for M11293, where a relatively faint source ($H_{S, {\rm obs}} = 19.20$ mag) and a small $\phi_{\rm wide}$ value (60 mas) were observed;
however, it still seems difficult to constrain the lens mass tightly unless one observes an excess flux with $\sim 21$ mag.
Meanwhile, for O120563 and O120950, it seems like we can expect to constrain the lens mass tightly once an excess flux is observed,
because the red shaded areas are large in their detectable $H_{\rm excess}$ region and the width of the 68\% confidence interval of the red dashed curves is small.

If one's goal is to measure the lens mass, such a prediction is informative for the selection of promising candidates for the planned follow-up observations.
When an event seems unlikely to be given a good mass measurement by the excess flux observation, then one should wait to observe it until the 
lens is sufficiently separated from the source star.
At this time, we no longer have the detection limit of $F_{\rm ex, 3 \sigma}$ due to the source star nor the high contamination probability from the source companion. 
There is still a possibility of contamination from the lens companion at this time; however, it is not highly likely as discussed in Section \ref{sec-LC}.

We recall that these predictions depend on the prior distributions that we used.
Conversely, one can test the prior distribution by comparing the observed excess flux with the detection probability of the 
excess flux expected from the prediction.
Hence, if our goal is to test the prior, we should not select targets for follow-up observations on the basis of the detectability of the lens flux because it causes a bias in the observed sample.
In particular, a prior that is currently unknown but is of interest is the planet hosting probability, 
which is assumed to be the same for all the stars in the above-mentioned calculations.
As described in Section \ref{sec-phost}, a different assumption on this probability changes the posterior distributions moderately, which indicates that we can statistically study the
planet hosting probability using excess flux measurements for a large number of planetary events.

\section{Dependence on Planet Hosting Probability} \label{sec-phost}
We have calculated the lens flux probability distributions under the assumption that all stars are equally likely to host a planet.
However, several studies have shown the dependence of the planet occurrence on the host mass, metallicity, and so on \citep{joh10,mon14, mul15}.
The dependence of the planet hosting probability on the host mass is often assumed to follow a power law $P_{\rm host} \propto M_h^{\alpha}$, 
where $M_h$ is the host mass.
\citet{joh10} used samples of the radial velocity (RV) method and found a linear relationship (i.e., $\alpha = 1$) between the host mass and the occurrence of giant planets 
within $\sim 2$AU around host stars whose masses range from 0.5 $M_{\odot}$ to 2.0 $M_{\odot}$.
This relationship is confirmed by a more recent study of \citet{ghe18}.
Meanwhile, the results of the transit method have shown the occurrence of more planets around the latter type of stars for low-mass planets close to their host, 
which indicates that $\alpha < 0$.
For example, \citet{mul15} found a higher occurrence rate of Earth- to Neptune-sized planets (1-4 $R_{\oplus}$) around
later type of stars in all orbital periods probed by $Kepler$. 
They reported that the planets around M-dwarfs occur twice as frequently as those around G-dwarfs and thrice as frequently as those around F-dwarfs.

As for the sensitivity region for the microlensing method, no study has found a statistically significant dependence of the planet occurrence on the host mass.
\citet{van19} and Bhattacharya et al. (2019, in prep.) measured the separation between the lens and the source stars of MOA-2013-BLG-220 and MOA-2007-BLG-400, respectively, using 
the Keck telescope. They showed that the host masses are both located at the $>90$ percentiles of the probability distribution calculated by Bayesian analysis with 
the assumption that all stars are equally likely to host a planet.
Because both planets are gas giants, this might indicate that such giant planets are more likely to be hosted by a more massive star, which is consistent with the RV results mentioned above.
Furthermore, microlensing has greater sensitivity (sub-kpc to $\sim$10 kpc) to the distance of planetary systems from the Sun compared to other methods ($<$ kpc).
It is also possible that the planet occurrence changes as a function of the distance of the microlensing planets.
In fact, \citet{pen16} suggested a possibility that planets in the bulge are less common than planets in the disk, which was not conclusive.

Because the planet hosting probability for a microlensing planet is still unknown, here, we simply show the results with $\alpha = 1$ and $\alpha = -1$ in Table \ref{tab-result2}, where 
the planet hosting probability $P_{\rm host} \propto M_h^{\alpha}$, as examples of variations of our results with different assumptions.
Note that we use $\alpha = 0$ for $M_L < 0.1 M_{\odot}$ in the model with $\alpha = -1$.
The estimated lens mass becomes more massive with $\alpha = 1$ than with $\alpha = 0$ because the probability of a large contribution of the lens flux to the excess flux 
increases while the mass estimate with $\alpha = -1$ becomes less massive because of the lower probability of the large contribution from the lens.

\section{Dependence on Other Priors} \label{sec-vali}
The results of this study depend on the choice of the prior distributions. Our fiducial choices are listed in Tables \ref{tab-inputs} and \ref{tab-info}.
The most influential uncertain prior is probably the planet hosting probability discussed in Section \ref{sec-phost}.
In this section, we review our choice of other priors and discuss how much their variation changes the estimates of the lens properties.
Table \ref{tab-Mdepend} summarizes all the mass estimates with the different choices of priors applied in this section, 
where all results are consistent with our fiducial values within $1\sigma$.

\subsection{Dependence on ambient star flux prior} \label{sec-valiamb} 
To calculate the $H_{\rm amb}$ prior distribution, we have used the number density of the target field $n_{\rm amb}$, 
the separation angle that is the boundary between resolved and unresolved stars, $\phi_{\rm wide}$, and 
the flux distribution of a single ambient star, $L_1 (F)$, as described in Section \ref{sec-priamb}.
Although there were some assumptions we made regarding $n_{\rm amb}$ and $L_1 (F)$, $\phi_{\rm wide}$ is the most influential parameter 
for ambient stars flux because the mean number of ambient stars is $\lambda_{\rm amb} \propto \phi_{\rm wide}^2$.

We have assumed $\phi_{\rm wide}$ as a constant regardless of brightness of a star and it is possible for a star fainter 
than the excess to be missed even outside of the circle of $\phi_{\rm wide}$, as mentioned in Section \ref{sec-ambdis}.
However, we showed that the ambient stars flux does not affect the $H_L - H_{\rm excess}$ estimates that much 
at least with $\phi_{\rm wide} < 240$ mas and $H_{\rm ex, obs} = 18.0 \pm 0.2$ mag in Section \ref{sec-HiHex} or Fig. \ref{fig-HiHex_thE}; hence, it is not likely the assumption on 
$\phi_{\rm wide}$ changes our results significantly.
Nevertheless, to confirm it for each event's parameters, we double each $\phi_{\rm wide}$ value listed in Table \ref{tab-info}, 
and calculate the posterior PDF for each event with the doubled $\phi_{\rm wide}$, or quadrupled mean number of ambient stars $\lambda_{\rm amb}$ under the assumption of $P_{\rm host} = $ const. (i.e., $\alpha = 0$) for all stars.
The line denoted ``$\phi_{\rm wide}$ doubled" in Table \ref{tab-Mdepend} shows the obtained mass estimates for the five events.
All of these are very similar to the fiducial values in Table \ref{tab-result}.

\subsection{Dependence on mass-luminosity relation} \label{sec-MLtest} 
We have used an empirical mass-luminosity relation for the wide mass range of $M = 0.1$ - $0.8 \, M_{\odot}$. 
To evaluate possible uncertainty due to this assumption, we apply the PARSEC isochrones model also for this mass range, 
and calculate the lens mass with it.
In the calculation, we also allow the age and metallicity to be wider than we assumed in Section \ref{sec-gal}.
We take the distributions from \citet{ben18a} who used ages of 1 Gyr $< T <$10 Gyr and 2 Gyr $< T < $ 12.6 Gyr for disk and bulge stars, respectively.
Metallicities between $-2.8 < \log Z < -1.3$ are used for both disk and bulge stars, but with different weight depending on the age for each component.
See \citet{ben18a} for the weights and more details.
The line denoted ``Isochrones for all mass range" in Table \ref{tab-Mdepend} shows the obtained mass estimates for the five events, which shows similar values to the fiducial values.

\subsection{Dependence on extinction distribution} \label{sec-AHtest} 
For extinction distribution toward a line of sight, we have used a simple dust disk model with the scale height of 100 pc, $A_H \propto 1- \exp\left[{-\frac{D}{0.1 ~ {\rm kpc}/\sin{|b|}}}\right]$, 
as described in Section \ref{sec-pricomp}.
To evaluate the effect of this assumption, we conduct our analysis with two extreme cases for the 3D extinction distribution, i.e., constant extinction and zero extinction.
In the calculation with constant extinction, we assume $A_{H, i} = A_{H, {\rm rc}}$ ($i = S, L$) while we assume $A_{H, i} = 0$ for the zero extinction, regardless of the distances.

The two lines denoted ``Constant extinction" and ``Zero extinction" in Table \ref{tab-Mdepend} show the resulting mass estimates for these two cases.
Although both cases are consistent with the fiducial values within 1-$\sigma$, the median mass for O120563 and O120950 are 
somewhat lighter than the fiducial values.
This is because as shown in Table \ref{tab-result}, the median of $H_L - H_{\rm excess}$ values for these two events are 0.005 mag and 0.08 mag, which are 
smaller than the $A_{H, {\rm rc}}$ values for these events; hence the $H_L - H_{\rm excess}$ estimates are sensitive to the variation of $A_{H, {\rm rc}}$ for these two events.
Meanwhile, the median of $H_L - H_{\rm excess}$ values for the other three events are much larger than the $A_{H, {\rm rc}}$ values for them; hence it is not sensitive to the variation of $A_{H, {\rm rc}}$.

\subsection{Dependence on Galactic model} \label{sec-galtest}
To calculate the joint prior PDF of $M_L$, $D_L$, $D_S$ and $v_t$, $f_{pri}' (M_L, D_L, D_S, v_{\rm t})$, we have 
used the S11 model in \citet{kos19}, as described in Section \ref{sec-gal}.
As seen in Fig. \ref{fig-compLF}, the LF calculated with the S11 model shows a good agreement with the observed LF by \citet{zoc03}.
Although it does not indicate a good agreement with the velocity distribution that is not related to brightness,
\citet{sum11} showed consistency between the Galactic model and the observation using the $t_{\rm E}$ distribution, which does depends on velocity distribution.
They showed that a simulated $t_{\rm E}$ distribution using nearly the same Galactic model as the S11 model is in good agreement with the observed $t_{\rm E}$ distribution from their two-year survey 
in the range of $t_{\rm E} \simgt 2$ days that covers all events' $t_{\rm E}$ values in this paper.
This is why we chose the model as our fiducial model.

Nevertheless, we calculate the lens properties with the other two models used in \citet{kos19}, i.e., the B14 model and the Z17 model, which are 
slightly modified versions of the \citet{ben14} and \citet{zhu17} models, respectively. 
The B14 model uses similar distributions to the S11 model for the mass function and bulge density.
They use a disk density model with a hole of the scale length 1320 pc in the galactic center that is used by \citet{rob03}. They include a solid bar rotation of 50/km/s/kpc in their velocity 
distribution for the bulge stars, rather than streaming motion used by the S11 model.
The Z17 model uses the \citet{kro01} MF with $1.3\, M_{\odot}$ cutoff. 
They adopt the same bulge density profile of Eq. (\ref{eq-rhob}), but with a 1.46 times smaller $y_0$ value of 424 pc and
1.82 times larger $\rho_{0, B}$ value of $3.76~M_{\odot}~$pc$^{-3}$, i.e., more centralized and totally 1.27 times more massive bulge model.
For bulge kinematics, they assume the mean velocity of 0 and faster velocity dispersion along all axes, 120 km/s, than the other two models.
These are main differences of these two models from the S11 model. More details of these models are seen in \citet{kos19} or each original paper.

The two lines denoted ``B14 Galactic model" and ``Z17 Galactic model" in Table \ref{tab-Mdepend} show the lens mass estimates with the two models.
The result with the B14 model is almost same as the fiducial one while the result with the Z17 model shows larger median values 
for M16227, M08310, and M11293.
This is because the bulge the Z17 model has 1.27 times more massive and centralized mass distribution than the other two models, which makes
the lens distance probability distribution shift toward the galactic center for these three events.
Because of the upward trend of the mass-distance relation with a given $\theta_{\rm E}$ on the mass-distance planes in Figs. \ref{fig-227} - \ref{fig-293}, 
the shift of the distance corresponds to the shift of the lens mass toward more massive side.
However, the shift level with the Z17 model is still small compared to their $1~\sigma$ range, and totally consistent with our fiducial values with the S11 model.
We note that \citet{kos19} showed the Z17 model's bulge mass must be multiplied by $0.75 \pm 0.05$ times to be consistent with a dynamic model of the bulge by \citet{por17}.

\subsection{Dependence on binary distribution}\label{sec-depbin}
We have applied the undetected binary distribution constructed in Section \ref{sec-bindis}, which combines the detection efficiencies and 
full binary distribution $f_{\rm arb}(q,a \,|\, M)$ of Eq. (\ref{eq-fqac}) based on studies of nearby stellar binary systems \citep{duc13} 
and the stellar present-day mass function, to the source system and lens system, which are likely to be in the bulge or disk.
Our choice of using the nearby binary distribution is because no study thus far 
had explored the binary distribution for each spectral-type star with wide coverage of semi-major axes in the galactic bulge.
With these assumptions, a source companion is the main origin of contamination possibility as described in Sections \ref{sec-sthetaE} and \ref{sec-HiHex}, 
in contrast a lens companion does not affect our lens mass estimates as described in Section \ref{sec-LC}.
We discuss variations of our results due to possible uncertainties in the source companion prior below.
Note that the small contributions from lens companions are attributed to high detection efficiency $\epsilon_{LC}$ for wide range of companions and negative slopes of the IMF.
We show that our assumption on $\phi_{{\rm close}, LC}$, which dominantly determines $\epsilon_{LC}$, is reasonable in Section \ref{sec-pdetlow} 
while the negative slopes of the IMF have been confirmed in the bulge field \citep{zoc00, cal15}.

Given the discussions in Sections \ref{sec-sthetaE} and \ref{sec-HiHex}, the large contributions of source companions as 
contaminants originate from the following features in the undetected binary distribution:
(i) a positive slope $\gamma > 0$ in $f_{\rm prim}(q,a\,|\,M) \propto q^\gamma$ that makes probability of a bright companion high, 
(ii) the detection efficiency $\epsilon_{SC}$ that is insensitive to a few orders of magnitude in the semi-major axis distribution,
and (iii) a moderate binary fraction over the insensitive region, typically $0 \simlt \log [a/{\rm AU}] \simlt 3$.

\subsubsection{Slope of mass-ratio function $\gamma$}
Evidence of $\gamma > 0$ for the bulge stars was discovered by \citet{shv16} who studied a sample of 224 microlensing events including $\sim 20$ binary events to derive $\gamma = 0.32 \pm 0.38$.\footnote{
Although their binary sample was contaminated by disk lenses, it is likely to be dominated by bulge lenses as ordinary microlensing events because 
they found a consistent frequency with the study of nearby stars under the assumption that the binary frequency is identical in the disk and bulge.
If the binary events were dominated by disk lenses, the consistency would indicate several times larger binary fraction for disk stars than nearby stars' one, which is not very likely 
because the nearby stars also belong to the disk.}
However, the uncertainty is still large and $\gamma < 0$ is also possible within 1$\sigma$. 
Table \ref{tab-Mdepend} shows the lens mass estimates with $\gamma = -0.1, -0.5,$ and -0.9, which are approximately corresponding to 
the 1$\sigma$, 2$\sigma$, and 3$\sigma$ lower limits on $\gamma$ by \citet{shv16}, respectively.
In the calculations, we use  $\gamma$ values for binary distribution of a non-secondary star $f_{\rm prim}(q,a\,|\,M) \propto q^\gamma$ regardless of the mass $M$.
The median values are all similar to the fiducial values except for M11293. 
This is because the result of M11293 is sensitive to the source companion prior because ambient stars as the excess origin is almost completely ruled out.
Nevertheless, their $1\sigma$ ranges are almost same as the fiducial one.

\subsubsection{Detection efficiency $\epsilon_{SC}$}
The detection efficiency is given by $\epsilon_{SC} =  \Theta[(\phi-\phi_{\rm wide})(\phi-\phi_{{\rm close},SC})]$.
In Sections \ref{sec-sthetaE} and \ref{sec-HiHex}, we discussed the difficulty of reducing the source companion probability
by any reasonable $\phi_{\rm wide}$ values, but not for $\phi_{{\rm close},SC}$.
Our assumption of $\phi_{{\rm close},SC} = \theta_{\rm E}/4$ was originally set by \citet{bat14}.

\citet{sko09} found projected separations of 2 to 7 times of $\theta_{\rm E}$ in the binary source fittings on their 19 candidates of repeating events, but 
the average detection efficiency is 0.0105, which is very small.
We can also detect a source companion through the xallarap effect, but the sensitivity is up to $\sim 1000$ days period, i.e., $\simlt \theta_{\rm E}$ \citep{poi05}, and 
the detection efficiency for them is close to 0 unless the event timescale is comparable to the period.
Thus, the assumption of $\epsilon_{SC} = 1$ for companions closer than $\theta_{\rm E}/4$ is not likely to overestimate the source companion probability significantly.
To show the robustness of our results on this assumption, we conduct our analysis with $\phi_{{\rm close},SC} = 5 \, \theta_{\rm E}$, i.e., 20 times larger $\phi_{{\rm close},SC}$ than the fiducial one, 
and show the consistent result with the fiducial one in Table \ref{tab-Mdepend}.

\subsubsection{Binary fraction in the insensitive region $0 \simlt \log [a/{\rm AU}] \simlt 3$}
A moderate binary fraction in $0 \simlt \log [a/{\rm AU}] \simlt 3$ is probably the case also in the bulge because star forming processes in the bulge and disk are not likely to be different from each other significantly, 
given their similar IMFs.
\citet{kro95} proposed a star formation process where all stars are born as a binary member and then some of them become single stars due to dynamic interactions within a star cluster.
\citet{mar11} applied the model to star clusters under various environments including different types of galaxies
and show that moderate fraction of companions always exists in $0 \simlt \log [a/{\rm AU}] \simlt 3$ after the dynamic evolution, which 
indicates a high dynamic stability.
Their high stability is also supported by the fact that nearby stars in all mass ranges take the highest binary frequencies in $0 \simlt \log [a/{\rm AU}] \simlt 3$ (see Fig. \ref{fig-nonsec}).

From the observation side, \citet{sko09} analyzed 19 candidates of repeating microlensing events. 
From 12 promising candidates, they found 28 \% binary frequency in $1 < s < 36$, roughly corresponding to $0 \simlt \log [a/{\rm AU}] \simlt 2$, 
assuming a uniform distribution in $\log q$ and $\log a$. 
Although their assumption of the distribution is different from ours, 
their result indicates even higher frequency than ours in the region because the corresponding frequency is 18\% in Fig. \ref{fig-phiSC}.
Furthermore, our result with $\phi_{{\rm close},SC} = 5 \, \theta_{\rm E}$ in Table \ref{tab-Mdepend} shows 
the robustness of our result on a smaller binary fraction because 20 times larger $\phi_{{\rm close},SC}$ corresponds to 1.3 orders of magnitude narrower range
of the insensitive region ($1.3 \simlt \log [a/{\rm AU}] \simlt 3$) in $a_{SC}$  compared to the fiducial model.

\subsection{Remnant or close-binary lens} \label{sec-rem}
We have not considered the cases of a remnant lens in our calculations.
This corresponds to an assumption of $P_{\rm host} = 0$ for such remnants. 
The assumption is probably true for neutron stars given the null result of the 11-year study on 45 pulsars by \citet{beh19}.
If this rareness of planets around neutron stars is attributed to their past explosion (i.e., supernova) and/or their very massive 
initial mass, this could be also true for blackholes.
Although no planet detection has been reported around white dwarfs so far, evidence of disintegrated rocky objects has been 
reported around or on the surface of them \citep{zuc03, koe14}.

If their planet hosting probability is comparable to that of late-type stars, 
including remnants possibility affects our lens mass estimates little for M16227, M08310 and M11293, i.e., the three with small $\theta_{\rm E}$, 
but more largely for O120563 and O120950.
It is beyond the scope of this paper because this work is focusing on how to deduce the lens flux posterior PDF from 
excess flux measurements considering the contamination probabilities, whereas the PDF for a remnant lens is independent from such measurements.
We note that it is not straightforward to include the remnant possibility in the PDF for planetary events because 
the planet hosting probability for them is even more uncertain than that for late-type stars.
Thus, one should consider the possibility only when they have strong evidence for the lens to be a remnant, e.g., too faint 
lens flux or excess flux compared to expectation from its microlensing parameters under the assumption of a stellar lens.

We have also assumed $P_{\rm host} = 0$ for tightly close binary systems which have the gravitational lensing effect that closely resembles 
that of a single star (the (ii) assumption in the top of Section \ref{sec-bindis}).
This is not true because such a circumbinary planet has been detected by microlensing \citep{ben16}.
The corresponding region is approximately given by Eq. (\ref{eq-solcau1}), which is $\simlt 0.1$ mas in angular separation for a relatively 
low magnification event M16227 and less for high-magnification events.
The top panel of Fig. \ref{fig-phiSC} shows that $\simlt$ 10\% of the source star has a companion $\simlt 0.1$ mas.
Because we are considering the lens system, which is closer and usually less massive than the source star, this percentage gets even smaller because 
a less massive star is less likely to host a secondary star (see $P_{\rm prim} (M)$ in Fig. \ref{fig-pmult}) and also the physical distance 
corresponding to $\simlt 0.1$ mas gets shorter with the closer distance.
Therefore, considering the close-binary systems only adds up to several percent population of them to posterior PDF, which negligibly 
changes our lens mass estimates.

\section{Comparison of Predicted and Detected Fractions of Binary Companions}\label{sec-valiSLC}
In Section \ref{sec-depbin}, we showed robustness of our lens mass estimates on the various different priors for the undetected binary distribution.
In this section, we test our fiducial distribution by comparing the number of detectable companions predicted by the distribution with actuall detected number of companions.

The detectable fraction of companions in our model is given by solving Eq. (\ref{eq-exist}) in Section \ref{sec-undetbin} for  $P_{{\rm det}, i}$ $(i = SC, LC)$,
\begin{align}
P_{{\rm det}, i} = \frac{P_{{\rm exist, full}, i} - P_{{\rm exist, undet}, i}}{1 - P_{{\rm exist, undet}, i}}, \label{eq-det}
\end{align}
where $P_{{\rm exist, undet}, i}$ and $P_{{\rm exist, full},i}$ are the $P_{\rm exist}$ values for object $i$ in the undetected and full binary distributions (Fig. \ref{fig-bindis}), respectively.
The flowchart in Fig. \ref{fig-flow} shows the practical way to calculate $P_{{\rm det}, i}$ in our Monte Carlo simulation.
$P_{{\rm det}, i}$ includes both fractions of the companions located at $\phi_{i}  < \phi_{{\rm close}, i}$ and $\phi_{i} > \phi_{\rm wide}$ 
($\phi_{SC} =  a_{SC, \perp}/D_S$, $\phi_{LC} =  a_{LC, \perp}/D_L$), where the former region is assumed to be detectable through the light curve while the latter region is assumed to be detectable through AO imaging.
To distinguish them, we divide $P_{{\rm det}, i}$ into $P_{{\rm det}, i_{\rm close}}$ and $P_{{\rm det}, i_{\rm wide}}$ such that the two fractions include 
companions only in the former region and only in the latter region, respectively. 
These fractions are shown in Table \ref{tab-fracrej} for each event, in addition to the parameters related to the detection efficiency $\epsilon_i$.

Below we compare the $P_{{\rm det}, i_{\rm close}}$ with the actually detected fraction of binary companions through light curves, and find that
there is a possible discrepancy between them for lens companions, which might indicate
lower stellar companion frequency in the planetary systems in the microlensing field than that for nearby random stars and/or the existence of several missed binary and
planet events thus far.

\subsection{Sample for comparison} \label{sec-sample}
Because this comparison does not aim to have a statistically strong claim, we simply use a sample of published events, 
which allows the possibility of incompleteness of the sample in our test.
However, we try to avoid the effect of publication bias at a certain level
by the following requests.
First, we focus on planetary events because planetary events are always attempted to be published regardless of the events' characteristics, such as a binary source 
or stellar binary lens feature in the light curve.
Nevertheless, there is a publication delay due to difficulty in modeling, impact of the discovery, etc.
To avoid the effect of such delays as much as possible, we use a sample consisting of planetary events that were identified by 2014, where we assume 
that most recognizable planetary events have been published thus far.

These two requests yield a sample of 49 planetary events ($q < 0.03$) that were identified by 2014 and published as of April 2019.
We note that recognition as a planetary event might be affected by whether the event light curve has a binary source or binary lens feature.
This indicates that the effect of publication bias is inevitable even if our requests work very well as intended. Therefore,
we cannot exclude the possibility of missed binary and planet events from the interpretations of the possible discrepancy found in Section \ref{sec-valiLC} below.
Nevertheless, we believe that it is worthwhile to conduct this test to determine whether there is clear evidence of discrepancy between the model expectation and the number of published events.

\subsection{Comparison for source companion}\label{sec-valiSC}
Table \ref{tab-fracrej} shows that the fraction of source companions detectable through
 the light curve is $P_{{\rm det}, SC_{\rm close}} = $ 6\%--14\%.
We compare this fraction to the fraction of actually detected binary source events.
In our sample of 49 planetary events, there are two events \citep{sum10, ben18} where the source star is in a binary system and 
one possible event \citep{fur13} where the source star might be in a binary system.
As a result, the fraction of actually detected binary source events in planetary events is 4\% to 6\%, which is slightly less than but consistent with the fraction of $P_{{\rm det}, SC_{\rm close}}$ = 6\%--14\% in Table \ref{tab-fracrej}.
Thus, no clear evidence of contradiction between our model and observations is found thus far.

\subsection{Comparison for lens companion} \label{sec-valiLC}
We find that the fraction of lens companions detectable through the light curve is $P_{{\rm det}, LC_{\rm close}} = $ 25\%--36\%.
In this case, where only planetary events are in the sample, an average fraction of $P_{{\rm det}, LC_{\rm close}}$ should correspond to 
the fraction of binary and planet (i.e., a planet in a binary system) events with respect to all planetary events detected via microlensing thus far.
However, there are only three planetary systems where a stellar companion is detected through microlens light curves \citep{gou14, pol14, ben16} in 
our sample of the 49 published planetary events identified by 2014.
Out of the three events, the stellar companion in OGLE-2008-BLG-092 \citep{pol14} was discovered through a very separated caustic, which is different from the central caustic, 
owing to the fortuitous geometry in the sky.
Therefore, only two binary and planet events have been discovered in the situation we have considered in this work, where the companion is detected through its central caustic.
This indicates that their observational fraction is only 2/49 $\sim$ 4\% and is much smaller than $P_{{\rm det}, LC_{\rm close}}$ = 25\%--36\% shown in Table \ref{tab-fracrej}.

In fact, we have used some crude assumptions for simplicity in our model, which could lead to overestimation of the fraction $P_{{\rm det}, LC_{\rm close}}$.
Below we discuss these assumptions and reassess the $P_{{\rm det}, LC_{\rm close}}$ estimate.

\subsubsection{Reassessment of lens companions fraction detectable via central caustic} \label{sec-pdetlow}
\begin{figure}
\centering
\epsscale{1.0}
\plotone{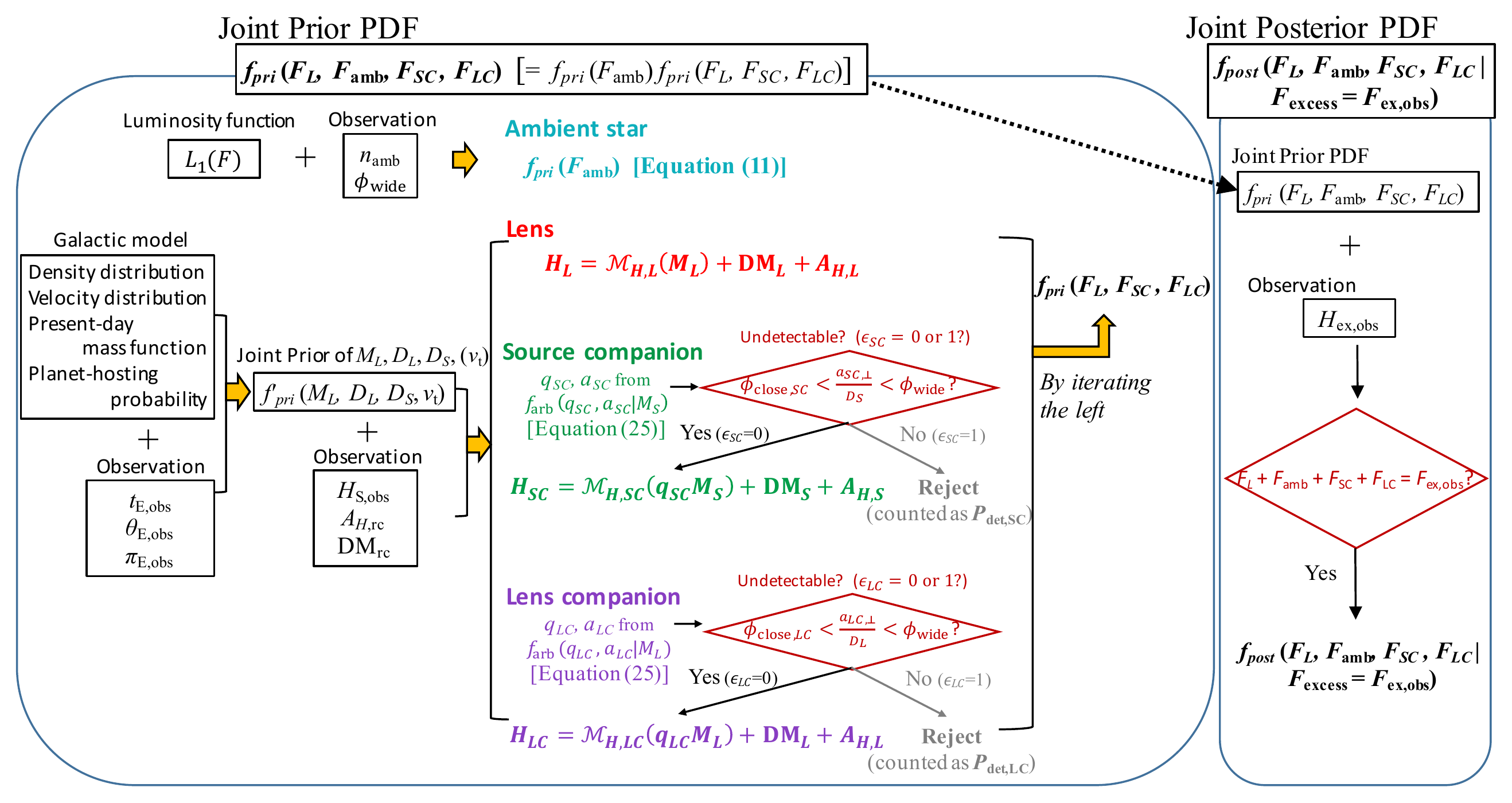}
\caption{Flowchart of our calculation method. We determine the  joint prior PDF through a Monte Carlo method
using the calculations outlined in the left panel, which determine $f_{pri} (F_L, F_{\rm amb}, F_{SC}, F_{LC})$.
Source and lens companions that are not compatible with the prior separation constraints are rejected and
excluded from consideration. Then, the calculation of the joint posterior PDF, 
$f_{post} (F_L, F_{\rm amb}, F_{SC}, F_{LC} | F_{\rm excess} = F_{\rm ex, obs})$, is preformed as shown
in the right panel, rejecting all combinations that do not match the measured brightness.}
\label{fig-flow}
\end{figure}

There are some factors in our calculation shown in Fig. \ref{fig-flow} that clearly or potentially cause overestimation of $P_{{\rm det}, LC_{\rm close}}$ for planetary events:
 (i) inner undetectable region regarded as detectable, 
 (ii) companions always generated from $f_{\rm arb} (q_{LC}, a_{LC} | M_L)$, 
 (iii) the approximated formula for the central caustic size that is used to determine the close limit of the undetectable region $\phi_{{\rm close},LC}$,
(iv) existence of the detected planet.
The factors (i), (iii) and (iv) might cause overestimation of the ``detectable" region while 
the factor (ii) might cause overestimation of binary fraction in the lens systems.
Here, we review these factors and reassess $P_{{\rm det}, LC_{\rm close}}$ by dealing with those of possible overestimations.

\begin{description}
\item  (i) {\it Inner undetectable region regarded as detectable}

We have regarded a very close stellar companion in the inner undetectable region approximately given by Eq. (\ref{eq-solcau1}) in Section \ref{sec-undetbin} as detectable (i.e., counted in $P_{{\rm det}, LC_{\rm close}}$) although 
they are in fact too close to be detected via light curve modeling.
This is because a correct treatment for this region, where $M_L$ refers to the total mass of a binary system rather than an arbitrary star mass, requires
a significant modification to our algorithm shown in Fig. \ref{fig-flow}, whereas the effect of this region on the lens mass estimates is little as discussed in Section \ref{sec-rem}.
Now that we are focusing on $P_{{\rm det}, LC_{\rm close}}$ located beyond the arrow of ``No ($\epsilon_{LC} = 1$)" in Fig. \ref{fig-flow}, 
the overestimation due to this region can be simply corrected by regarding them as undetectable, or putting them beyond the arrow of ``Yes ($\epsilon_{LC} = 0$)". 

\item  (ii) {\it Companions always generated from $f_{\rm arb} (q_{LC}, a_{LC} | M_L)$}

Because the lens companions have been always generated from $f_{\rm arb} (q_{LC}, a_{LC} | M_L)$, 
$P_{{\rm det}, LC_{\rm close}}$ refers to the integral of $f_{\rm arb} (q_{LC}, a_{LC} | M_L)$ with $0 < q_{LC} < \infty$ over the detectable parameter spaces.
However, it should be generated from $f_{\rm arb} (q_{LC}, a_{LC} | \frac{M_L}{1+q_{LC}})$
at least in a case where a counted companion is located close to the inner undetectable region but is still marginally detectable.
This causes an overestimation of $P_{{\rm det}, LC_{\rm close}}$ first because the integral of $f_{\rm arb} (q_{LC}, a_{LC} | \frac{M_L}{1+q_{LC}})$ should be conducted with $0 < q_{LC} < 1$ to
avoid the double counting, and second because $M_L/(1+q_{LC})$ is always smaller than $M_L$. 
Comparing $P_{\rm prim} (M/2)$ and $P_{\rm prim} (M) + P_{\rm second} (M)$ in Fig. \ref{fig-pmult} might help to understand this.
Whether we should use $M = M_L$ or $M = M_L/(1+q_{LC})$ probably depends on the $s_{LC}$ value and the source trajectory; however, 
it is beyond the scope of this study to examine an accurate way to deal with it.
Here, we decide to always use $f_{\rm arb} (q_{LC}, a_{LC} | M_L/2)$ to generate lens companions, and limit $q_{LC} < 1$ when $s_{LC} < 1$ 
but keep $q_{LC} < \infty$ when $s_{LC} > 1$ to conservatively estimate $P_{{\rm det}, LC_{\rm close}}$.

\item (iii) {\it Use of approximated formula for the central caustic size}

We have used an approximated formula of the central caustic size, $w_{LC} =  4q_{LC}/(s_{LC} - s_{LC}^{-1})^2$, which is true only when $q_{LC} \ll 1$, 
to calculate the inner and outer undetectable regions given by Eqs. (\ref{eq-solcau1}) and (\ref{eq-solcau2}), respectively.
In fact, we find that this formula overestimates the caustic size in units of the angular Einstein radius by a maximum factor of $\sim$3 when $q_{LC} \sim 1$.
This seems to lead to overestimation of $\phi_{{\rm close}, LC}$ of Eq. (\ref{eq-phicloselc}) by a factor of $\sim \sqrt{3}$. 
However, we also find that this factor is roughly canceled by the factor $\sim (1 + q_{LC})^{3/4}$ ($\sim 1.68$ when $q_{LC} = 1$) for a wide ($s_{LC} > 1$) stellar companion that 
originates in an unconsidered expansion of the angular Einstein radius by $\sqrt{1 + q_{LC}}$ owing to the additional wide stellar companion.

We confirm this by calculating the detection efficiencies for M16227 and O120950 as shown in Fig. \ref{fig-DEs}.
To calculate the detection efficiency, we follow \citet{suz16} but apply some modifications to deal with the expansion of the angular Einstein radius.
Appendix \ref{sec-DE} describes the details of the modifications.
The right region of the black dashed line in the figure corresponds to the analytical outer undetectable region given by Eq. (\ref{eq-solcau2}) while the left region of gray dashed line 
corresponds to the analytical inner undetectable region given by Eq. (\ref{eq-solcau1}).
Both panels show that the black line lies nearly along the right edge of the red region of the detection efficiency $\epsilon_{LC}= 1$ while the gray line is 
located slightly to the left of the left edge of $\epsilon_{LC}= 1$ when $q_{LC} \sim 1$.
This indicates that using Eq. (\ref{eq-solcau2}) as the outer undetectable region slightly underestimates the detectability of a stellar companion because both panels show moderate 
detection efficiencies even in the right region of the black dashed line.
Meanwhile, using Eq. (\ref{eq-solcau1}) as the inner undetectable region slightly overestimates the detectability of a stellar companion with $q_{LC} \sim 1$.
This is because the expansion of the angular Einstein radius does not work at $s_{LC} \ll 1$; thus, the factor of the overestimated caustic size is not canceled.
Because Eqs. (\ref{eq-solcau2}) and (\ref{eq-solcau1}) misestimate the detectability in opposite directions, the misestimation is mitigated by using both of them.
Nevertheless, using the simulated detection efficiency is clearly better than just using Eqs. (\ref{eq-solcau1})--(\ref{eq-solcau2}) as the undetectable regions.
Thus, we use the detection efficiencies shown in Fig. \ref{fig-DEs} to calculate the $P_{{\rm det}, LC_{\rm close}}$ values for M16227 and O120950.

\item (iv) {\it Existence of the detected planet}

We have assumed that the existence of the detected planet makes no difference in the detection efficiency of a stellar companion $\epsilon_{LC}$, and also that
the detection of the planet is robust independent from the existence of a hypothetical stellar companion.
This is probably true if the binary signal is small when $s_{LC} \gg 1$ or $s_{LC} \ll 1$ 
because it is known that the triple-lens signal is well-assumed by the superposition of 
two binary-lens signals when each signal is a small perturbation like the one due to a planet \citep{han05}.
However, when $s_{LC}$ is close to 1, the very large binary signal might make the planetary signal undetectable, which is inconsistent with the fact.

We decided to go with the following two options for this.
One is to remove all stellar companions in $\theta_{\rm E}/3  < \phi_{LC} < 3 \, \theta_{\rm E} \, \sqrt{1+q_{LC}}$ from the full binary distribution 
assuming these companions make the planetary signal undetectable, where $\sqrt{1+q_{LC}}$ is to deal with the expansion of the angular Einstein radius.
The other is to keep the original full binary distribution assuming the detection of the planet is anyway robust.
Note that the two binary and planet events actually detected both have their stellar companions out of the removed range in the former option.

\end{description}

The recalculated $P_{{\rm det}, LC_{\rm close}}$ values are shown in Table \ref{tab-fracrej}, where 
``--w/ factors (i)-(iv)'' is results with the option in the factor (iv) that removes  $\theta_{\rm E}/3  < \phi_{LC} < 3 \, \theta_{\rm E} \, \sqrt{1+q_{LC}}$,
while ``--w/ factors (i)-(iii)'' is results with the other option that keeps the full binary distribution.
The $P_{{\rm det}, LC_{\rm close}}$ values of ``--w/ factors (i)-(iii)'' are very similar to the previous estimates, while those of ``--w/ factors (i)-(iv)'' are very different; hence, 
the choice in the factor (iv) dominantly determines the detectable fractions of lens companion.

In each option, the two relatively low-magnification events (M16227 and O120950) and 
the other three high-magnification events (M08310, M11293, and O120563)
show similar values among each of the groups, which indicates 
that $P_{{\rm det}, LC_{\rm close}}$ highly depends on $u_{\rm 0}$, but less on other event parameters, such as $\theta_{\rm E}$.
Thus, we apply these values to all the planetary events in our sample.
Because around 40\% of our sample of the 49 planetary events are high-magnification events with $u_0 < 0.01$, we 
roughly calculate a mean value of $P_{{\rm det}, LC_{\rm close}}$ in the sample as 
$(0.085 \times 0.6 + 0.16 \times 0.4) = 0.115$ for the case of (i)-(iv) and $(0.25 \times 0.6 + 0.30 \times 0.4) = 0.27$ for the case of (i)-(iii).
The binomial distribution with trial number 49 and success probability  
0.115 and 0.27 yields the probability of success (i.e., detection of binary and planet events) of less than 2 as
0.068 and $3.6 \times 10^{-5}$, respectively.

This might imply a discrepancy between the predicted and detected numbers of binary and planet events, 
but not conclusive because of the variation of the $p$-values depending on the assumption on the effect of a stellar companion on the planet detectability.
We discuss this results and possible implications in Section \ref{sec-lessLC}.

\section{Discussion} \label{sec-dis}
\subsection{Need for resolving lens star} 
We found that it is difficult to conclude whether the excess flux comes from the lens for events with small $\theta_{\rm E}$ ($\simlt 0.3$ mas).
This difficulty of inconclusive results for events with small $\theta_{\rm E}$ originates from the method itself, where excess flux is required to be detected at the source position that is unresolved.
To be detected at the position of a source star, the excess flux has to be significantly brighter than the uncertainty of the source flux, e.g., $F_{\rm ex, obs} > 3\, \sigma_{F_S}$ for 3-$\sigma$ detection, 
as discussed in Section \ref{sec-pre}.
This requirement imposes a lower limit on the brightness of the detectable excess flux regardless of the imaging quality, which depends only on the source flux error and is typically $\sim 20$ mag in 
$H$-band for the current ground-based optical survey.
For events with small $\theta_{\rm E} \simlt 0.3$ mas, the probability of the lens brightness being $H_L \sim 20$ or brighter is smaller than or comparable to the probability of 
a source companion being $H_{SC} \sim 20$ or brighter, as shown in the bottom right panels in the (a) components in Figs. \ref{fig-227}--\ref{fig-293}.
This always leads to inconclusive interpretations for the origin of excess fluxes detected at the position of events with small $\theta_{\rm E}$, even with a perfect AO correction.

One might think that color measurements of the excess flux would be useful; however, in fact, they are not useful in the case of small $\theta_{\rm E}$ because a lens with the brightness of such a 
detectable excess flux ($H_L \simlt 20$) corresponds to a star located very close to the source star.
This can be seen in the shape of the mass-distance relation of small $\theta_{\rm E}$, i.e., the blue lines in the left panels in Figs. \ref{fig-227}--\ref{fig-293}.
Thus, we cannot distinguish a case where the excess originates in the lens from a case where the excess originates in a source companion even with color measurements
because in both cases, the excess origins are located at similar distances and thus have similar colors.
Meanwhile, in the case of large $\theta_{\rm E}$, color measurements would help distinguish the two cases because the color of the excess should be different depending on which is true.
For example, if we had a color measurement of the excess for O120950, we might have completely excluded the low possibility of the source companion although 
it was already proved that the detected excess was certainly from the lens by \citet{bha18}.

To overcome the difficulty associated with events with small $\theta_{\rm E}$, we need to wait for several years after the event until the lens 
becomes resolvable from the source, where the detectable lens brightness is limited by the limiting magnitude of each imaging 
rather than the error of the source flux, and the probability of contamination from the source companion becomes much lower.
Even when a candidate is found at a position expected from the lens-source relative proper motion $\mu_{\rm rel}$ measured via light curve modeling, 
the resolved candidate can still be a lens companion.
However, such a probability is low, especially for a high-magnification event, as discussed in Section \ref{sec-LC}. Furthermore,
whether the candidate is the lens or a lens companion is distinguishable if we have a 1D or 2D microlens parallax measurement from the light curve \citep{ben19}.

Part of our results is already confirmed by such observations conducted several years after each event to resolve the lens star.
\citet{bha17} observed M08310 using the {\it HST} and found that the lens star is not the source of the excess detected by \citet{jan10}.
They found that the excess was likely due to a nearby unrelated star if the excess was solely provided by one star.
Our calculation also indicates moderate probability of ambient stars as the origin of the excess, as shown in Fig. \ref{fig-310-post}.
Furthermore, \citet{bha18} found that the origin of the excess flux at the location of O120950 was actually the lens itself.
Our calculation also indicates the highest probability of this scenario, as shown in Fig. \ref{fig-0950-post}.

\subsection{Study of planet hosting probability by excess flux measurements}
Although resolving the lens star from the source star is required for robust lens detection, especially for events with small $\theta_{\rm E}$, excess flux measurements can still be used
to statistically determine the dependence of the planet hosting probability on the stellar mass or location in our galaxy.
Because the prior distribution of the excess flux depends on the planet hosting probability, 
we can find which dependence gives the maximum likelihood for the measurements by excess flux observations for many planetary events.
Although the same study could be conducted using a sample of resolved lens stars, one obvious advantage of this method is that we do not need to wait for the lens to become
resolvable; thus, it is easier to increase the number of samples.

In particular, the discovery rate of planetary events has been increasing since the Korean Microlensing
Network \citep[KMTNet, ][]{kim16} started their survey using three 1.6-m telescopes in 2015.
However, it is not scientifically meaningful to wait for several years to resolve such lens stars in dozens of newly discovered planetary events because such a opportunity 
will come in the era of the {\it WFIRST} survey, where the discovery of $\sim 1400$ planetary events and mass measurements for most of them are expected \citep{pen19}.
We would rather propose that excess flux measurements be conducted for the newly discovered planetary events soon after their discovery so 
that a statistically study of these samples can reveal the planet hosting probability through our method.

In addition, the time needed to resolve the lens depends on the $\mu_{\rm rel}$ value and on the contrast between the source and lens stars flux.
Some events with very slow $\mu_{\rm rel}$ and/or much brighter source are not likely to be resolved even after several years.
One such example is MACHO-97-BLG-28, where the lens could not be resolved even by Keck AO imaging conducted 16 years after the event's peak \citep{bla19}.
A campaign of high-angular-resolution follow-up observations for the 30 planetary events in \citet{suz16}, the largest statistical sample of planetary 
events thus far, is ongoing and most of them have already been observed \citep{bat15,ben15,ben19,bha17,bha18,van19}.
Although some lens stars can be identified in the images, there are some lens stars that cannot 
be resolved, to which the method developed in this paper can be applied for analysis.
Because excluding such events without lens identifications causes a bias in the sample, they must also be included and correctly treated 
using our method in a statistical study with the results of the follow-up campaign.

\subsection{Less stellar companion in planetary system?} \label{sec-lessLC}
In Section \ref{sec-valiLC}, we found a possible discrepancy between the fraction of detectable lens companions expected from our model and 
the fraction of actually detected lens stellar companions in planetary events.
If this discrepancy is real, there are two possibilities to interpret this result.

The first possibility is that the frequency of a stellar companion in a planetary system located in the galactic disk or bulge is smaller than 
the binary frequency of nearby randomly selected stars.
Some studies have investigated the effect of the existence of stellar companions on planet frequency \citep{wan14, wan15, ngo17, zie18}; however, all of them are for 
planets close to their host star and not for planets beyond the snow line, such as microlensing planets.
In the close region that they explored, they found clues that planet formation is suppressed by the existence of stellar companions, which might support this possibility.

The second possibility is the existence of some detectable binary and planet events that have been misclassified.
In this case, there should be some detectable binary and planet events that are classified as just a binary or a planetary event owing to the omission of triple lens model fitting.
This idea is supported by \citet{gou14}, who found the preference of the best-fit binary and planet model to the best-fit binary model by $\Delta \chi^2 = 216$ just over 
the peak of OGLE-2013-BLG-0341, which appears as just a binary event at first glance.

However, the level of discrepancy depends on our assumption on the 
region where existence of a stellar companion makes the planet detected in each event undetectable.
That is, $p = 3.6 \times 10^{-5}$ if we assume there is no such region, and it gets a marginally acceptable $p$-value of $p = 0.068$ if we assume the region is 
$\theta_{\rm E}/3  < \phi_{LC} < 3 \, \theta_{\rm E} \, \sqrt{1+q_{LC}}$ and remove it from the full-binary distribution.
The $p$-value gets more larger if the range of the removal region is larger.
Therefore, more careful study of the detection efficiency for an additional stellar component in a binary-lens system is needed to conclude.

\subsection{Application to other studies} \label{sec-lessLC}
\subsubsection{Constraint on the lens properties}
The method developed in this study should be employed for all analyses that includes constraints from excess flux mesurements.
We have already applied our method to several studies of event analysis \citep[Poleski et al. 2020, submitted]{kos17, ben18, bea18, nag19, fuk19} to 
determine the lens properties or calculate the contamination probabilities in the excess flux before this paper is published.
Because the estimated lens properties by our method get quite different from previous calculations for events with small $\theta_{\rm E}$, as seen in Table \ref{tab-result} or Figs. \ref{fig-227}(b)-\ref{fig-293}(b),
future statistical studies that use lens properties of planetary events estimated by Bayesian analysis like the one by \citet{cas12} or \citet{pen16} should use 
the properties estimated by our method for events with excess flux measurement.

The largest unbiased statistical sample of events with excess flux measurements will be provided by {\it WFIRST}, and our method can be applicable to all of them including single lens events.
Although more than half of the lens stars can be resolved during the {\it WFIRST} survey, there are still moderate fraction of events with slow relative proper motion
where our method would give the most tightest constraint.

\subsubsection{Study of binary distribution undetectable by microlensing} \label{sec-binstudy}
The binary distribution in the galactic bulge is still unknown. 
Microlensing can reach companions located in $\log [a/{\rm AU}] < 2$ while high-angular resolution imaging  
can reach those in $\log [a/{\rm AU}] > 3$ as seen in the undetectable binary distributions in Fig. \ref{fig-bindis}; hence, 
companions in $2 < \log [a/{\rm AU}] < 3$ cannot be accessed by either of them.
An interesting feature of the excess flux is that those undetectable binary companions contributes to it.
Thus, we can study binary fraction in the inaccessible region in the galactic bulge by comparing our prediction of the excess flux distribution 
with the detected excess flux distribution in the {\it WFIRST} sample.
 
Moreover, the binary distributions optimized for the bulge field can be applicable for a binary correction in the study of the stellar IMF.
There are many studies that investigated the IMF in the galactic bulge, such as by the observed luminosity function \citep[e.g.,][]{zoc00,cal15} or by the 
observed $t_{\rm E}$ distribution in microlensing survey \citep{sum11,weg17,mro17}.
An important procedure to convert the observed luminosity or $t_{\rm E}$ distribution into the mass function is the binary correction.
Most of the studies just applied a binary distribution for solar-type stars \citep{duq91, rag10}.
We have developed the binary distribution for an arbitrary star as a function of its mass, $f_{\rm arb} (q, a | M)$, and also that for a non-secondary star in Section \ref{sec-bindis}.
These two distributions optimized for the bulge field can be applied to the binary correction for the $t_{\rm E}$ distribution and the luminosity function.

\section{Summary} \label{sec-summ}
We developed a Bayesian approach for calculating the prior and posterior probability distributions of the flux for four possible origins of the excess flux.
The four possible origins are the lens, unrelated ambient stars, and companions to the source and lens.
Although this probability has been considered in some previous studies, they have not always treated the prior and posterior constraints consistently, which has always led to a claim that the lens star is the likely origin of the excess flux, regardless of the extent to which the lens flux is likely to be faint a priori.
Such flawed treatment has been performed to avoid the calculation of the prior probability distribution of the lens flux, which requires us to assume the unknown planet hosting probability
in the case of planetary microlensing events.
However, calculating the prior probability for the lens flux is inevitable to correctly calculate the posterior probability of each possible origin.

Then, we assumed that the planet hosting probability was the same for all stars in our galaxy, and we applied our method to five planetary events where the excess flux 
had been detected by previous studies.
We found that the probability of the lens being the main origin of the observed excess is smaller than or comparable to the total probability of 
the other contaminants being the main origin for the three events with small $\theta_{\rm E}$ ($\simlt 0.3$ mas), namely M16227, M08310, and M11293.
Consequently, our lens mass estimates for M08310 and M11293 were more uncertain than the estimates of previous studies that
assumed the excess flux to be the lens flux \citep{jan10,bat14}.
Meanwhile, for O120563 with a large angular Einstein radius of $\theta_{\rm E} = 1.4$ mas, a large part of the excess flux is highly likely to be from the lens itself.
This is also the case for O120950 with a long Einstein radius crossing time of $t_{\rm E} = 68$ days, which indicates that $\theta_{\rm E}$ is likely to be large.
Thus, our lens mass estimates for these two events are consistent with previous studies that treated a large part of the excess flux as the lens flux \citep{fuk15,kos17b}.
This qualitative interpretation for the origin of each excess flux does not change even if we apply a different prior for the dependence of the planet hosting probability on the host mass.
We recommend using our estimates of the lens properties for M11293 and O120563, whereas 
one should use the properties for M08310 and O120950 estimated by \citet{bha17} and \citet{bha18}, respectively.

To robustly detect the lens star, especially for events with small $\theta_{\rm E}$, we need to resolve the lens star from the source star by 
additional high-angular-resolution imaging conducted when the two systems become resolvable.
Our interpretations of the excess fluxes for M08310 and O120950 were already confirmed by such observations \citep{bha17,bha18}.
However, such observations typically require a waiting time of several years after the event discovery.
Although the KMTNet survey \citep{kim16} has significantly increased the discovery rate of planetary events, 
this requirement makes it unlikely to resolve the lens stars in dozens of newly discovered events by the time of the {\it WFIRST} survey.
The method developed in this study can be used to statistically study the dependence of the planet hosting probability on the host's property 
by only excess flux measurements for planetary events.
Because there is no requirement with regard to the observation time for excess flux measurements, all the newly discovered events can be added to the sample by conducting  
high-angular-resolution follow-up imaging.
This would provide us with the largest sample of planetary events to study the dependence of the planet hosting probability on the host's properties before the {\it WFIRST} era.

Our method can be applied to estimate the lens property of all the {\it WFIRST} events although more than half of them should be updated after the lens star is resolved.
Also we can study the binary fraction in the galactic bulge located in $2 < \log [a/{\rm AU}] < 3$, an inaccessible region by other methods, by comparing 
the predicted excess flux distributions to the observed one in the {\it WFIRST} sample.

\acknowledgments
We are grateful to J.~P.~Beaulieu and Y. Shvartzvald, who provided us with the data of Keck images taken in their observations.
We are also grateful to H.~Shibai, T.~Sumi, and C.~Ranc for fruitful discussions.
We would like to thank Editage (www.editage.com) for English language editing.
The work of N.K. is supported by JSPS KAKENHI Grant Number JP18J00897.
D.P.B. was supported by NASA through grant NASA-80NSSC18K0274.

\appendix

\section{How to Calculate Detection Efficiency for A Stellar Companion}\label{sec-DE}
\begin{figure}
\centering
\epsscale{0.4}
\plotone{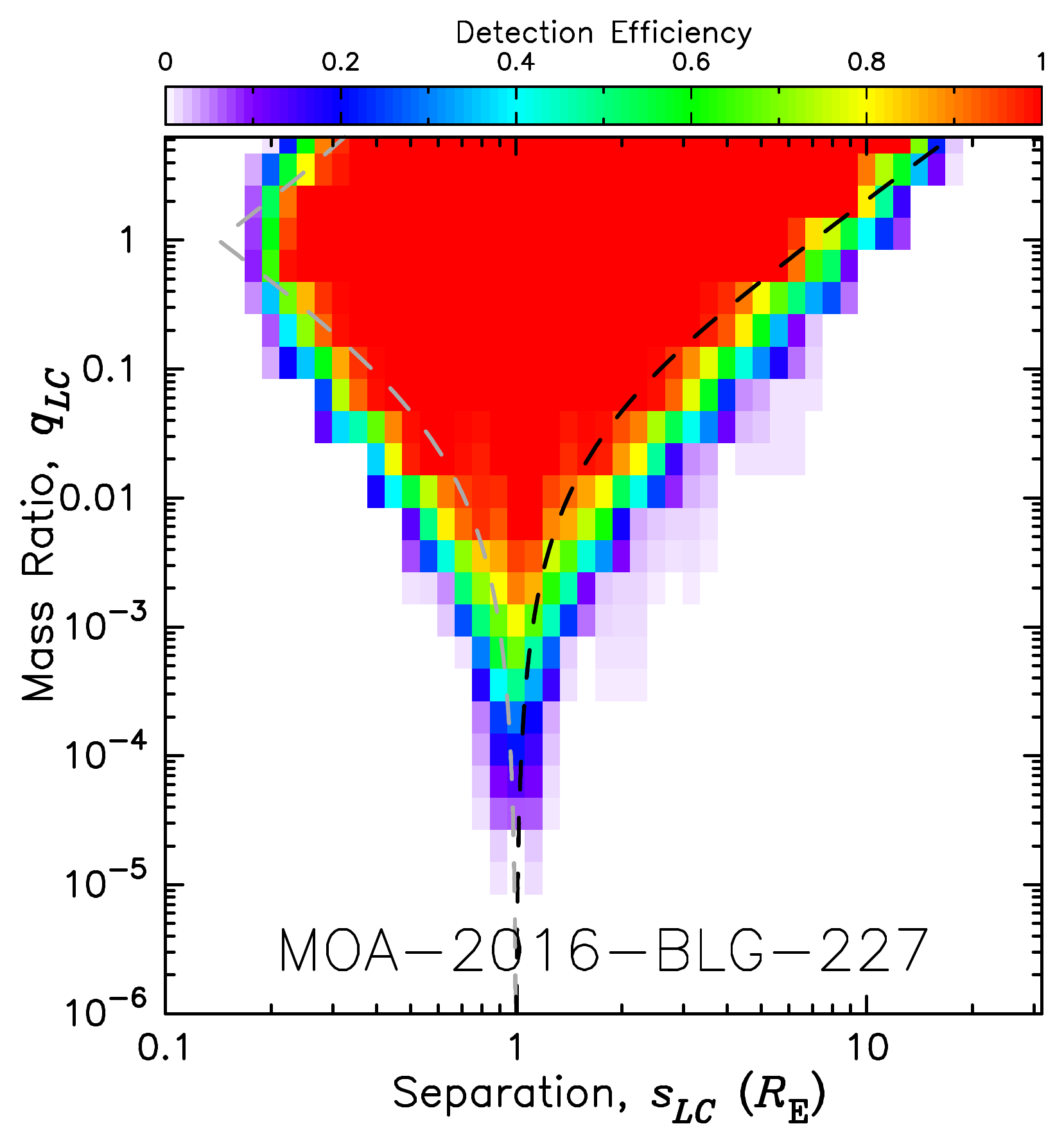}
\plotone{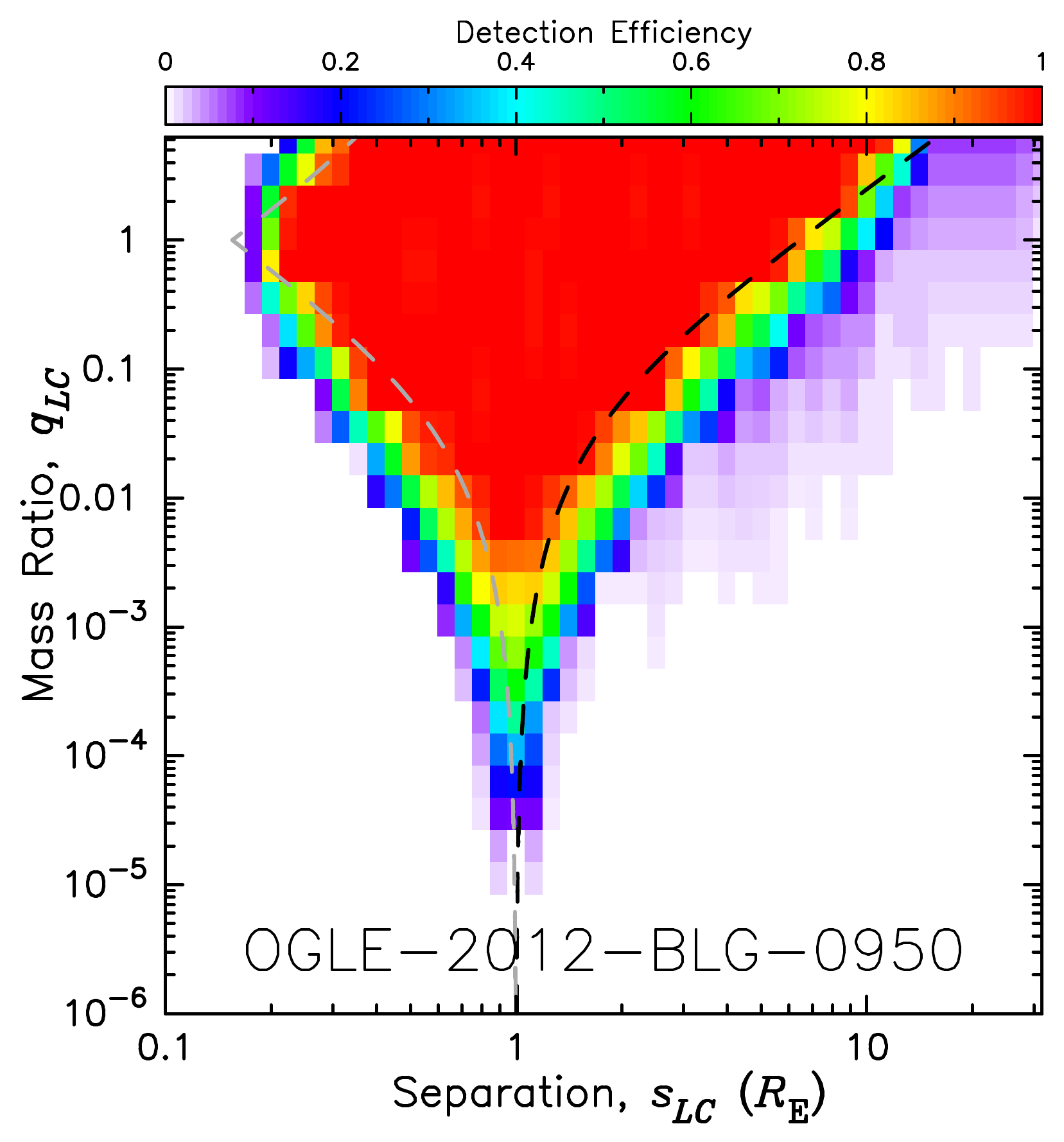}
\caption{Detection efficiencies for additional lens companions assuming binary lens model for M16227 (left) and O120950 (right).
See Appendix \ref{sec-DE} for how they are calculated. 
The black and gray dashed lines represent two solutions of $w_{LC} = u_{\rm 0, obs}$, where an approximated formula of the central caustic size for $q_{LC} \ll 1$, $w_{LC} = 4q_{LC}/(s_{LC} - s_{LC}^{-1})^2$, is used. 
The black line has been used as the border between the detection efficiency $\epsilon_{LC}= 1$ and $\epsilon_{LC}= 0$ in our calculation for the prior and posterior PDFs.}
\label{fig-DEs}
\end{figure}

In Section \ref{sec-pdetlow}, we calculated the detection efficiency $\epsilon_{LC}$ for a planetary companion to a stellar companion around the lens in
M16227 and O120950, as shown in Fig. \ref{fig-DEs} (we used only the result for $q_{LC} > 0.1$ in this study).
In this section, we focus on the difference between the ways to calculate the detection efficiency for a planetary companion and a stellar companion.
This difference mainly arises from the expansion of the angular Einstein radius by considering a stellar companion, which does not matter for a planetary companion.
Readers may refer to the work of \citet{suz16} and the references therein for details on the way to calculate the detection efficiency for a planetary companion.

For simplicity, we ignore the existence of a detected planet and consider a situation where we observed a single lens event 
with $t_{\rm E, 1L}$, $u_{\rm 0, 1L}$, $\rho_{\rm 1L}$, and $\theta_{\rm E, 1L}$, where $\rho$ is the angular source star radius in
units of the angular Einstein radius, $\rho \equiv \theta_*/\theta_{\rm E}$.
In this situation, we consider the detection efficiency for a lens companion with a mass ratio $q_{LC}$ located at an angular separation of 
$s_{LC} \, \theta_{\rm E, 1L}$ in the lens system. Note that $s_{LC}$ is the separation defined in units of $\theta_{\rm E, 1L}$.

When considering the detection efficiency for a stellar companion with mass ratio $q_{LC}$ that is not negligible compared to 1, 
an expansion of the angular Einstein radius by a factor of $\sqrt{q_{LC} + 1}$ has to be considered.
This is because most quantities observable by microlensing are defined in the units of the angular Einstein radius.
In this situation, we consider the parameters that we should use for artificial light curves to simulate hypothetical observations.
Although it seems to be complicated to consider this effect when the separation $s_{LC} \sim 1$, 
easy approximations are applicable for $s_{LC} \gg 1$ and $s_{LC} \ll 1$.

\begin{figure}
\centering
\epsscale{0.45}
\plotone{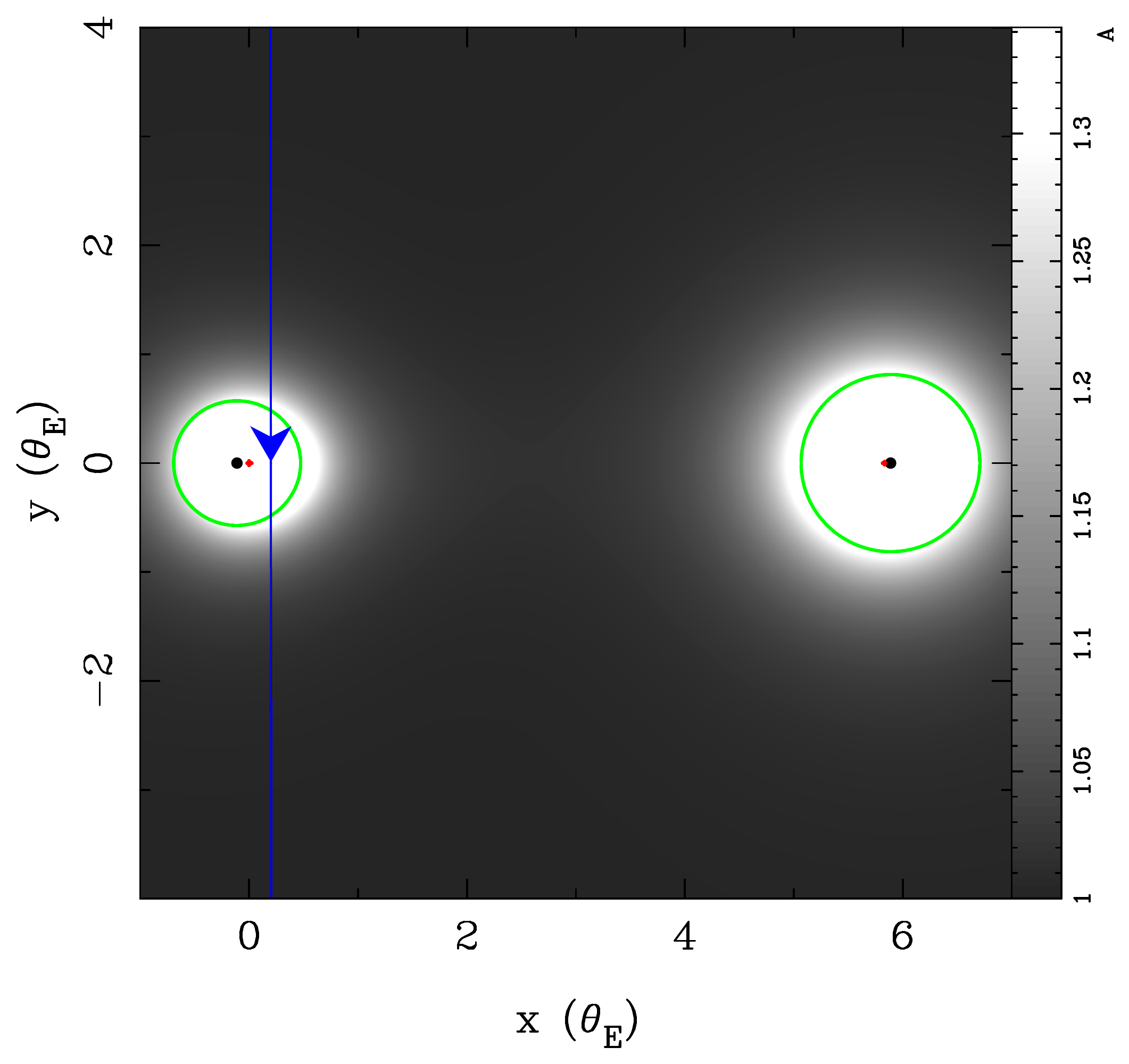}
\plotone{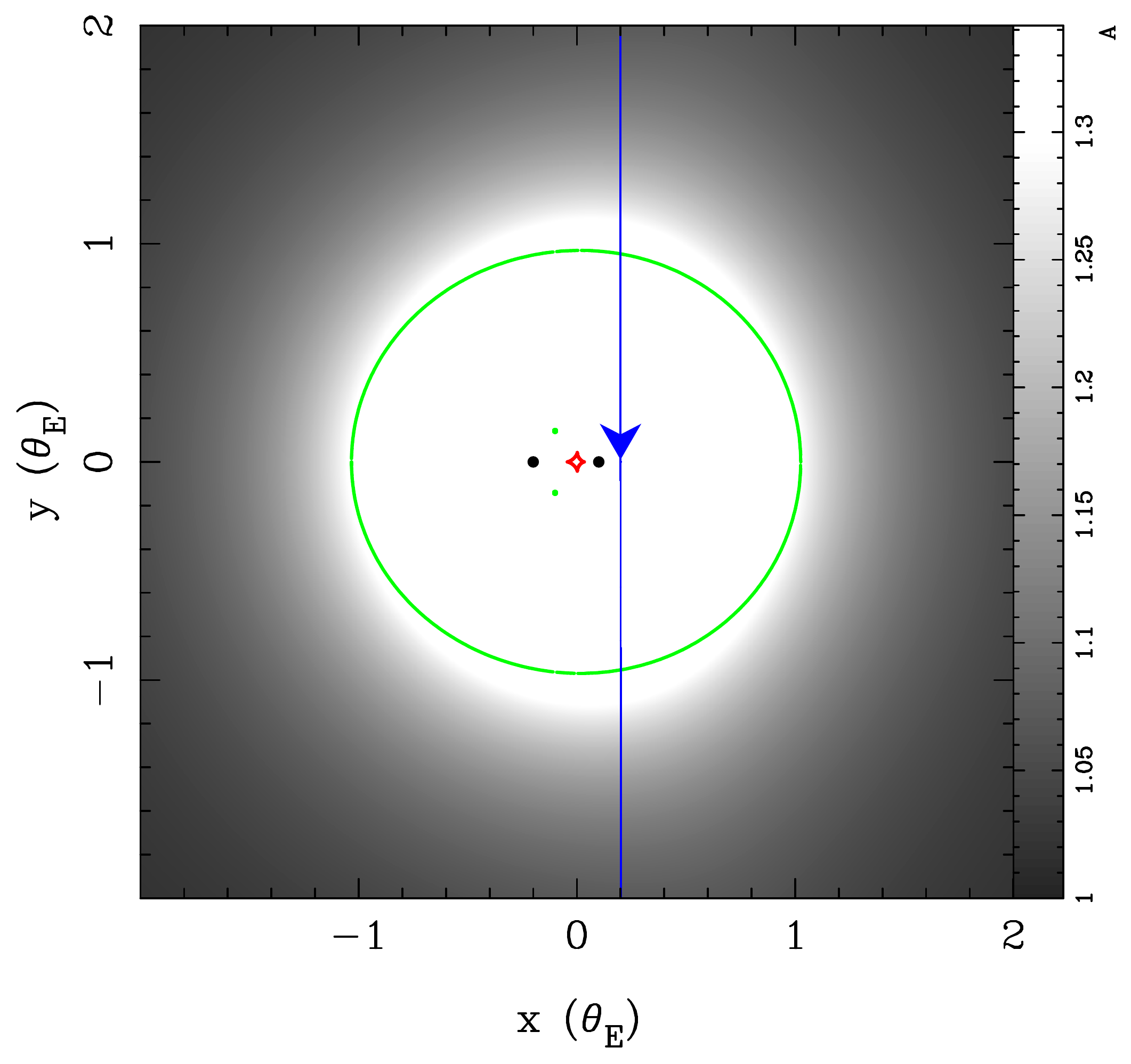}
\caption{Two examples of geometry with an undetected companion.
Left: Magnification map for a binary lens with mass ratio $q = 2$ and separation $s = 6$.
The two black dots indicate the lens objects, the small red structures indicate caustics, and the green closed curves indicate critical curves.
The regions in pure white indicate the area with magnification $A > 1.34$, which corresponds to the region of $u < 1$ in a single lens event.
An observer who observes a magnification curve of the source star passing the blue arrow
misestimates the microlens parameters as $\theta_{\rm E, 1L} = \theta_{\rm E}/\sqrt{1 + q} = \theta_{\rm E}/\sqrt{3}$, $t_{\rm E, 1L} =  t_{\rm E}/\sqrt{3}$, $u_{0,0} = \sqrt{3}\, u_0$, and $\rho_0 = \sqrt{3} \rho$.
Furthermore, in a lens plane with the units of the observed $\theta_{\rm E, 1L}$, the separation of this undetected companion becomes $s_{LC} = \sqrt{3}\, s = 6 \sqrt{3}$.
Right: Magnification map for a binary lens with mass ratio $q = 0.5$ and separation $s = 0.3$. 
An observer who observes a magnification curve of the source star passing the blue arrow 
correctly estimates $\theta_{\rm E, 1L} = \theta_{\rm E}$, $t_{\rm E, 1L} =  t_{\rm E}$, $u_{0,0} = u_0$, and $\rho_0 = \rho$,
but does not notice that the lens actually consists of two stars.
In this case, the separation in the units of $\theta_{\rm E, 1L}$ is $s_{LC} = s$.}
\label{fig-caus}
\end{figure}

As shown in the left panel of Fig. \ref{fig-caus}, two separated central caustics are created by a wide binary system with the separation 
$s_{LC} \gg 1$ and a stellar mass ratio $q_{LC} \simgt 1$, and the magnification map is separated into two nearly distinct areas.
Although the angular Einstein radius for the entire lens system is $\theta_{\rm E}$, 
each of the separated areas behaves as if it is created by a single lens with the angular Einstein radius $\theta_{\rm E}/\sqrt{q_{LC} + 1}$ (the left area) or 
$\theta_{\rm E}\sqrt{q_{LC}/(q_{LC} + 1)}$ (the right area) centered at each of the caustics.
In the considered situation where we observed only a magnification caused by the left area in the left panel and identified it as a single lens with $\theta_{\rm E, 1L}$ (but actually,  
$\theta_{\rm E, 1L} = \theta_{\rm E}/\sqrt{q_{LC} + 1}$),
we have to calculate the detection efficiency of a stellar companion by making artificial light curves with $t_{\rm E} = \sqrt{q_{LC} + 1} \, t_{\rm E, 1L}$, 
$u_0 = u_{\rm 0, 1L}/\sqrt{q_{LC} + 1}$,  $\rho = \rho_{\rm 1L}/\sqrt{q_{LC} + 1}$, $q = q_{LC}$, and $s = s_{LC}/\sqrt{q_{LC} + 1}$ to reproduce 
the observed apparent single lens event.
Thus, the detection efficiency $\epsilon_{LC}$ at a grid ($q_{LC}$, $s_{LC}$) with $s_{LC} \gg 1$ in Fig. \ref{fig-DEs} is actually calculated
by putting a stellar companion with $q = q_{LC}$ and $s = s_{LC}/\sqrt{q_{LC} + 1}$ to deal with the expansion of the angular Einstein radius.

Meanwhile, as shown in the right panel of Fig. \ref{fig-caus}, the magnification map around the central caustic that is created by 
a close binary system with the separation $s_{LC} \ll 1$ and mass ratio $q_{LC} \simgt 1$ behaves like a single lens event centered at the central caustic except for 
the region close to the caustic.
In this case, the components of the binary system contributed to the observed magnification that appears as a single lens light curve with $\theta_{\rm E, 1L}$.
Therefore, in contrast to the case for $s_{LC} \gg 1$, the angular Einstein radius is not expanded compared to the observed size, i.e., $\theta_{\rm E} = \theta_{\rm E, 1L}$.
In other words, we should make the artificial light curves with $t_{\rm E} = t_{\rm E, 1L}$, $u_0 = u_{\rm 0, 1L}$,  $\rho = \rho_{\rm 1L}$, and $s = s_{LC}$ to calculate detection efficiencies.
For the mass ratio $q$, we should use $q = q_{LC}$ for $q_{LC} \leq 1$ and $q = 1/q_{LC}$ for $q_{LC} > 1$. 
This is because we cannot distinguish the light curve with $q_{LC}$ from that with $1/q_{LC}$ in contrast to the case of $s_{LC} \gg 1$.
Therefore, the gray line in Fig. \ref{fig-DEs} has a turn-off point while the black line does not.

The events analyzed in this study have high sensitivity to a stellar companion because all of them have moderate coverage on their light curves with $u_0 \simlt 0.1$.
Therefore, separation values that correspond to the borders between $\epsilon_{LC} = 1$ and $\epsilon_{LC} < 1$ with a given $q_{LC} (> 0.1)$ should be located in 
a region satisfying the condition of $s_{LC} \gg 1$ or $s_{LC} \ll 1$, where the approximations mentioned above can be applied.
If this is true, the detection efficiency is expected to be $\epsilon_{LC} = 1$ in the medium region of $s_{LC} \sim 1$, regardless of how 
the detection efficiency for $s_{LC} \sim 1$ and $q_{LC} \simgt 1$ is calculated.
We use the parameters of $s_{LC} \gg 1$ for all companions with $s_{LC} > 1$ while we use the parameters of $s_{LC} \ll 1$ for all companions 
with $s_{LC} \leq 1$ to create artificial light curves in the detection efficiency calculation.
Fig. \ref{fig-DEs} shows the calculation results. 
In both events, we find that the separation values that correspond to borders between $\epsilon_{LC} = 1$ and $\epsilon_{LC} < 1$ with $q_{LC} > 0.1$ are 
sufficiently large or small to be applied to the approximation of $s_{LC} \gg 1$ or $s_{LC} \ll 1$, respectively.
Thus, the color maps shown in Fig. \ref{fig-DEs} are correct if the detection efficiency $\epsilon_{LC}$ always increases when $s$ approaches 1.

\clearpage

\clearpage

\clearpage

\clearpage

\clearpage

\clearpage

\clearpage

\clearpage

\clearpage

\clearpage

\clearpage

\clearpage

\clearpage

\clearpage

\begin{landscape}
{\tabcolsep = 1.2mm
\footnotesize
\catcode`?=\active \def?{\phantom{0}}
 \tblcaption{Probability density functions (PDFs) and parameters that are modeled in our calculation.}
\vspace{-0.4cm}
\begin{center}
\begin{threeparttable}
   \begin{tabular}{llllrlllllccc}\hline\hline
                                                              & Notation                                                                         & Note  \\\hline
{\bf Modeled PDFs}                                            &                                                                                  &       \\     
??Joint posterior PDF of $F_L, F_{\rm amb}, F_{SC}, F_{LC}$   & $f_{post} (F_L, F_{\rm amb}, F_{SC}, F_{LC} | F_{\rm excess} = F_{\rm ex, obs})$ & Eq. (\ref{eq-postPDF}), modeled by iteration of the right box in Fig. \ref{fig-flow}     \\
??Joint prior     PDF of $F_L, F_{\rm amb}, F_{SC}, F_{LC}$   & $f_{pri} (F_L, F_{\rm amb}, F_{SC}, F_{LC})$                                     & $= f_{pri} (F_L, F_{SC}, F_{LC}) f_{pri} (F_{\rm amb})$ \\
????-- Prior PDF of $F_{\rm amb}$                             & $f_{pri} (F_{\rm amb})$                                                          & Eq. (\ref{eq-priFamb}) \\
????-- Joint prior PDF of $F_L, F_{SC}, F_{LC}$               & $f_{pri} (F_L, F_{SC}, F_{LC})$                                                  & Modeled by iteration of the left box in Fig. \ref{fig-flow} \\\hline
{\bf Modeled parameters}                                      &                                                                                  &       \\  
??\bf{Brightness}                                             &                                                                                  &       \\
????Flux (mag) of the lens in $H$-band                        & $F_L$, $H_L$                                                                     & $H_L = {\cal M}_{H,L} (M_L) + 5 \log\frac{D_L}{\rm 10 pc} + A_{H,L}$\\
????Flux (mag) of ambient stars in $H$-band                   & $F_{\rm amb}$, $H_{\rm amb}$                                                     & Follows Eq. (\ref{eq-priFamb})\\
????Flux (mag) of a source companion in $H$-band              & $F_{SC}$, $H_{SC}$                                                               & $H_{SC} = {\cal M}_{H,SC} (q_{SC}M_{S}) + 5 \log\frac{D_S}{\rm 10 pc} + A_{H,S}$\\
????Flux (mag) of a lens companion in $H$-band                & $F_{LC}$, $H_{LC}$                                                               & $H_{LC} = {\cal M}_{H,LC} (q_{LC}M_{L}) + 5 \log\frac{D_L}{\rm 10 pc} + A_{H,L}$\\
????Excess flux (mag) in $H$-band                             & $F_{\rm excess}$, $H_{\rm excess}$                                               & $F_{\rm excess} \equiv F_L + F_{\rm amb} + F_{\rm SC} + F_{\rm LC}$\\
????Fraction of each flux to the excess                       & $f_L, f_{\rm amb}, f_{SC}, f_{LC}$                                               & $f_i \equiv F_i/F_{\rm excess}$ ($i = L, {\rm amb}, SC, LC$)\\
??\bf{For $\boldsymbol{F_L}$, $\boldsymbol{F_{LC}}$}          &                                                                                  &       \\
????Lens mass                                                 & $M_L$                                                                            & Follows $f_{pri}' (M_L, D_L, D_S, v_{\rm t})$\tnote{a} \, in the prior\\
????Lens distance                                             & $D_L$                                                                            & Follows $f_{pri}' (M_L, D_L, D_S, v_{\rm t})$\tnote{a} \, in the prior\\
????Mass ratio of a lens companion                            & $q_{LC}$                                                                         & Follows $f_{\rm arb}(q_{LC},a_{LC} \,|\, M_L)$ combined with $\epsilon_{LC}$ \tnote{b} \, in the prior\\
????Semi-major axis of a lens companion                       & $a_{LC}$                                                                         & Follows $f_{\rm arb}(q_{LC},a_{LC} \,|\, M_L)$ combined with $\epsilon_{LC}$ \tnote{b} \, in the prior\\
????Extinction for the lens system                            & $A_{H, L}$                                                                       & $A_{H, L} = A_{H, {\rm rc}} \frac{1- \exp\left[{-D_L/(0.1 ~ {\rm kpc}/\sin{|b|})}\right]}{1- \exp\left[{-D_{\rm rc}/(0.1 ~ {\rm kpc}/\sin{|b|})}\right]}$\\
??\bf{For $\boldsymbol{F_{SC}}$}                                 &                                                                                  &       \\
????Source mass                                               & $M_S$                                                                            & $M_S = M_S({\cal M}_{H,S})$, where ${\cal M}_{H,S} = H_{S, {\rm obs}} - A_{H,S} -  5 \log\frac{D_S}{\rm 10 pc}$\\
????Source distance                                           & $D_S$                                                                            & Follows $f_{pri}' (M_L, D_L, D_S, v_{\rm t})$\tnote{a} \, in the prior\\
????Mass ratio of a source companion                          & $q_{SC}$                                                                         & Follows $f_{\rm arb}(q_{SC},a_{SC} \,|\, M_S)$ combined with $\epsilon_{SC}$ \tnote{b} \, in the prior\\
????Semi-major axis of a source companion                     & $a_{SC}$                                                                         & Follows $f_{\rm arb}(q_{SC},a_{SC} \,|\, M_S)$ combined with $\epsilon_{SC}$ \tnote{b} \, in the prior\\
????Extinction for the source system                          & $A_{H, S}$                                                                       & $A_{H, S} = A_{H, {\rm rc}} \frac{1- \exp\left[{-D_S/(0.1 ~ {\rm kpc}/\sin{|b|})}\right]}{1- \exp\left[{-D_{\rm rc}/(0.1 ~ {\rm kpc}/\sin{|b|})}\right]}$\\
??\bf{To be compared in \S \ref{sec-valiSLC}}                     &                                                                                  &       \\
????Fraction of detectable source companions                   & $P_{{\rm det},SC}$                                                               & Fraction of SC that gives $\epsilon_{SC} = 1$ (Fig. \ref{fig-flow})\\
??????-- Fraction of detectable close source companions        & $P_{{\rm det},SC_{\rm close}}$                                                   & Fraction of SC that gives $\epsilon_{SC} = 1$ because $\frac{a_{SC,\perp}}{D_S} < \phi_{{\rm close}, SC}$ \\
??????-- Fraction of detectable wide  source companions        & $P_{{\rm det},SC_{\rm wide}}$                                                    & Fraction of SC that gives $\epsilon_{SC} = 1$ because $\frac{a_{SC,\perp}}{D_S} > \phi_{\rm wide}$ \\
????Fraction of detectable lens companions                     & $P_{{\rm det},LC}$                                                               & Fraction of LC that gives $\epsilon_{LC} = 1$ (Fig. \ref{fig-flow})\\
??????-- Fraction of detectable close lens companions          & $P_{{\rm det},LC_{\rm close}}$                                                   & Fraction of LC that gives $\epsilon_{LC} = 1$ because $\frac{a_{LC,\perp}}{D_L} < \phi_{{\rm close}, LC}$ \tnote{c} \\
??????-- Fraction of detectable wide  lens companions          & $P_{{\rm det},LC_{\rm wide}}$                                                    & Fraction of LC that gives $\epsilon_{LC} = 1$ because $\frac{a_{LC,\perp}}{D_L} > \phi_{\rm wide}$ \\\hline
      \end{tabular}
     \begin{tablenotes}
     \small
      \item[a]  Abbreviation of $f_{pri} (M_L, D_L, D_S, v_t | t_{\rm E} = t_{\rm E, obs}, \theta_{\rm E} = \theta_{\rm E, obs}, \pi_{\rm E} = \pi_{\rm E, obs})$.
      \item[b]  Binary distribution combined with the detection efficiency is referred to as the undetected binary distribution.
      \item[c]  This is reassessed in Section \ref{sec-pdetlow}.
     \end{tablenotes}
      \end{threeparttable}
    \end{center}
    \label{tab-outputs}
  \hfill
}
\end{landscape}

\begin{landscape}
{\tabcolsep = 1.2mm
\footnotesize
\catcode`?=\active \def?{\phantom{0}}
 \tblcaption{The models need to calculate the prior PDF and our choices for them.}
\vspace{-0.4cm}
\begin{center}
\begin{threeparttable}
   \begin{tabular}{lllllrrrrrrcccccclllrlllllccc}\hline\hline      
  Model                                        & Notation                                         &  To model                                                &  Our choice                                                                      & Section            \\\hline
\bf{For $\boldsymbol{F_{\rm amb}}$}               &                                                  &                                                          &                                                                                  &       \\ 
??Luminosity function (LF)                            & $L_1 (F)$                                        &  $f_{pri}  (F_{\rm amb})$                                &  LF by \citet{zoc03}                                                             & \S \ref{sec-ambdis}   \\ 
??Detection efficiency for ambient stars            & $\epsilon_{\rm amb}$                             &  $f_{pri}  (F_{\rm amb})$                                &  $\Theta(\phi - \phi_{\rm wide})$ \tnote{a}                                      & \S \ref{sec-ambdis}   \\
\bf{For $\boldsymbol{F_L, F_{SC}, F_{LC}}$}         &                                                  &                                                          &                                                                                  &       \\ 
??Prior PDF of $M_L$, $D_L$, $D_S$ and $v_{\rm t}$  & $f_{pri}' (M_L, D_L, D_S, v_{\rm t})$\tnote{b}   &  $M_L$, $D_{L}$, $D_{S}$                                 & Galactic model\tnote{c} $\times$ ${\cal L}$($t_{\rm E, obs}$, $\theta_{\rm E, obs}$, $\pi_{\rm E, obs}$)\tnote{d} & \S \ref{sec-priML}, \S \ref{sec-gal}   \\ 
????-- Density distribution                         & $\rho_{B}$, $\rho_{D}$                           &  $f_{pri}' (M_L, D_L, D_S, v_{\rm t})$\tnote{b}          & Eqs. (\ref{eq-rhob})-(\ref{eq-rhod})                                             & \S \ref{sec-gal}   \\
????-- Velocity distribution                        & --                                               &  $f_{pri}' (M_L, D_L, D_S, v_{\rm t})$\tnote{b}          & S11 model in \citet{kos19}                                                       & \S \ref{sec-gal}   \\
????-- Present-day mass function                    & $\Phi_{\rm PD}(M)$                               &  $f_{pri}' (M_L, D_L, D_S, v_{\rm t})$\tnote{b}          & IMF + age and metallicity distributions                                          & \S \ref{sec-gal}   \\
??????-- Initial mass function (IMF)                & $\Phi_{\rm IMF}(M)$                              &  $\Phi_{\rm PD}(M)$                                      & Eq. (\ref{eq-IMF})                                                               & \S \ref{sec-gal}   \\
??????-- Age distribution                           & --                                               &  $\Phi_{\rm PD}(M)$, ${\cal M}_{H} (M)$, $M ({\cal M}_{H})$ & Bulge: ${\cal N} (9 {\rm Gyr}, 1 {\rm Gyr})$\tnote{e}\,, ? disk: ${\cal N} (5 {\rm Gyr}, 2 {\rm Gyr})$\tnote{e} & \S \ref{sec-gal}   \\
??????-- Metallicity distribution                   & --                                               &  $\Phi_{\rm PD}(M)$, ${\cal M}_{H} (M)$, $M ({\cal M}_{H})$ & Solar metallicity                                                                & \S \ref{sec-gal}   \\
????-- Planet hosting probability                   & $P_{\rm host}$                                   &  $f_{pri}' (M_L, D_L, D_S, v_{\rm t})$\tnote{b}          & $P_{\rm host} \propto M_h^{\alpha}$ w/ $\alpha = 0, \pm 1$, $P_{\rm host}   = 0$ for remnant & \S \ref{sec-priML},  \S \ref{sec-phost} \\
??Mass--luminosity relation                         & ${\cal M}_{H} (M)$, $M ({\cal M}_{H})$           &  $H_L$, $H_{LC}$, $H_{SC}$, $M_{S}$                      & Empirical (0.1 - 0.8 $M_{\odot}$), isochrones (otherwise)                    & \S \ref{sec-masslumi}   \\  
??Extinction distribution along the line of sight   & --                                               &  $A_{H, S}$, $A_{H, L}$                                  & $A_H \propto 1- \exp\left[{-\frac{D}{0.1 ~ {\rm kpc}/\sin{|b|}}}\right]$         & \S \ref{sec-pricomp}   \\
\bf{For $\boldsymbol{F_{SC}, F_{LC}}$}               &                                                  &                                                          &                                                                                  &       \\
??Binary distribution for an arbitrary star         & $f_{\rm arb}(q,a \,|\, M)$                       &  $q_{SC}, a_{SC}, q_{LC}, a_{LC}$                        &  Eq. (\ref{eq-fqac})                                                             & \S \ref{sec-binarb}   \\
????-- Binary distribution for a non-secondary star  & ${\cal F}_{\rm mult} (M)$, $f_{\rm prim}(q,a\,|\,M)$ &  $f_{\rm arb}(q,a \,|\, M)$                          & $f_{\rm prim} (q, a \, | \, M) \propto q^{\gamma} \, \Lambda (a; \eta_{\log a}, \sigma_{\log a}^2)$\tnote{f}& \S \ref{sec-binpri}   \\
??????-- Multiplicity fraction                       & ${\cal F}_{\rm mult} (M)$                        &  $f_{\rm arb}(q,a \,|\, M)$                              &  ${\cal F}_{\rm mult} = 0.196 + 0.255 \, M$                                      & \S \ref{sec-binpri}   \\
??????-- Slope of mass-ratio function                & $\gamma$                                         &  $f_{\rm prim}(q,a \,|\, M)$                             &  $\gamma_c$ ($\log [a/{\rm AU}] < \eta_{\log a}$), $\gamma_w$ ($\log [a/{\rm AU}] > \eta_{\log a}$) & \S \ref{sec-binpri}   \\
????????-- $\gamma$ for close binary                 & $\gamma_c$                                       &  $f_{\rm prim}(q,a \,|\, M)$                             &  $\gamma_c = 1.2 - 2.8 \, \log M$                                                & \S \ref{sec-binpri}   \\
????????-- $\gamma$ for wide binary                  & $\gamma_w$                                       &  $f_{\rm prim}(q,a \,|\, M)$                             &  $\gamma_w = 0$ ($M \geq 0.34$), $\gamma_w = -3.1 -6.7 \, \log M$ ($M < 0.34$)   & \S \ref{sec-binpri}   \\
??????-- Mean of $\log [a/{\rm AU}]$ distribution               & $\eta_{\log a}$                                  &  $f_{\rm prim}(q,a \,|\, M)$                             &  $\eta_{\log a} = 0.57 + 1.02 \, M$                                              & \S \ref{sec-binpri}   \\
??????-- Standard deviation of $\log [a/{\rm AU}]$ distribution & $\sigma_{\log a}$                                &  $f_{\rm prim}(q,a \,|\, M)$                             &  $\sigma_{\log a} =  1.6 + 1.2  \, M$                                            & \S \ref{sec-binpri}   \\
??Detection efficiency for a source companion        & $\epsilon_{SC}$                                  &  $q_{SC}, a_{SC}$                                        &  $\Theta[(\phi-\phi_{\rm wide})(\phi-\phi_{{\rm close},SC})]$ \tnote{a}          & \S \ref{sec-undetbin}  \\ 
????-- Close limit of undectable source companion    & $\phi_{{\rm close},SC}$                          &  $\epsilon_{SC}$                                         &  $\theta_{\rm E}/4$                                                              & \S \ref{sec-undetbin}   \\ 
??Detection efficiency for a lens  companion         & $\epsilon_{LC}$                                  &  $q_{LC}, a_{LC}$                                        &  $\Theta[(\phi-\phi_{\rm wide})(\phi-\phi_{{\rm close},LC})]$ \tnote{a}          & \S \ref{sec-undetbin}   \\ 
????-- Close limit of undectable lens companion      & $\phi_{{\rm close},LC}$                          &  $\epsilon_{LC}$                                         &  $\theta_{\rm E} (\sqrt{q_{LC}/u_{\rm 0} + 1} + \sqrt{q_{LC}/u_{\rm 0}})$         & \S \ref{sec-undetbin}   \\\hline
      \end{tabular}
     \begin{tablenotes}
      \item[a]  $\Theta$ is the Heaviside step function.
      \item[b]  Abbreviation of $f_{pri} (M_L, D_L, D_S, v_t | t_{\rm E} = t_{\rm E, obs}, \theta_{\rm E} = \theta_{\rm E, obs}, \pi_{\rm E} = \pi_{\rm E, obs})$.
      \item[c]  Galactic model consists of the density distribution, the velocity distribution, the present-day mass function, and the planet hosting probability for stellar objects in our galaxy.
      \item[d]  Likelihood of ($t_{\rm E, obs}$, $\theta_{\rm E, obs}$, $\pi_{\rm E, obs}$), where 3D normal distribution with reported values and errors (listed in Table \ref{tab-info}) is applied. No correlation among the three is assumed.
      \item[e]  ${\cal N} (\overline{T}, \sigma_T)$ is the normal distribution with the mean $\overline{T}$ and the standard deviation $\sigma_T$.
      \item[f]  $\Lambda (a; \eta_{\log a}, \sigma_{\log a}^2)$ is the log-normal distribution.
     \end{tablenotes}
      \end{threeparttable}
    \end{center}
    \label{tab-inputs}
  \hfill
}
\end{landscape}

\begin{landscape}
{\tabcolsep = 1.2mm
\footnotesize
\catcode`?=\active \def?{\phantom{0}}
 \tblcaption{Input parameters needed to calculate the prior/posterior PDFs and our choices for them.}
\vspace{-0.4cm}
\begin{center}
\begin{threeparttable}
   \begin{tabular}{lllcccccccccc}\hline\hline
   Input parameter                                & Notation                    &  To model                                               &  M16227 \tnote{(1)}   & M08310 \tnote{(2)}  & M11293 \tnote{(3, 4)}     & O120563 \tnote{(5)} & O120950 \tnote{(6)}  \\\hline
   For prior PDF                                  &                             &                                                         &                       &                     &                            &                    &                   \\
   ??Einstein radius crossing time                & $t_{\rm E, obs}$ (days)     & $f_{pri}' (M_L, D_L, D_S, v_{\rm t})$\tnote{a}          &  $17.0 \pm 0.2$       &   $11.1 \pm 0.5$    & $21.7 \pm 0.1$             &  $78 \pm 2$        &  $68 \pm 2$        \\
   ??Angular Einstein radius                      & $\theta_{\rm E, obs}$ (mas) & $f_{pri}' (M_L, D_L, D_S, v_{\rm t})$\tnote{a}          &  $0.23 \pm 0.01$      &  $0.16 \pm 0.01$    &  $0.26 \pm 0.02$           &   $1.4 \pm 0.1$    &  --              \\
   ??Microlensing parallax                        & $\pi_{\rm E, obs}$          & $f_{pri}' (M_L, D_L, D_S, v_{\rm t})$\tnote{a}          &      --               &    --               &   --                       &  --                &  $0.26 \pm 0.06$ \\
   ??Impact parameter                             & $u_{\rm 0, obs}$            & $\epsilon_{LC}$                                         &  0.08                 &   0.003             &  0.0035                    &  0.001             &  0.10            \\
   ??Source flux                                  & $H_{S, {\rm obs}}$   (mag)  & $M_{S}$                                                 &  17.81 $\pm$ 0.02     &  17.73 $\pm$ 0.05   &  19.20 $\pm$ 0.06          & 18.57 $\pm$ 0.03   & 17.78 $\pm$ 0.12 \\
   ??Mean extinction for red clump in the field   & $A_{H, {\rm rc}}$    (mag)  & $A_{H, S}$, $A_{H, L}$                                  &  0.19 $\pm$ 0.02      & 0.33 $\pm$ 0.10     & 0.47 $\pm$ 0.10 \tnote{b}  & 0.26 $\pm$ 0.02    &  0.25 $\pm$ 0.03 \\
   ??Mean distance modulus to the red clump stars & ${\rm DM}_{\rm rc}$         & $A_{H, S}$, $A_{H, L}$                                  &  14.48 $\pm$ 0.24     &   14.70 $\pm$ 0.28  & 14.52 $\pm$ 0.19           &  14.48 $\pm$ 0.24  & 14.50 $\pm$ 0.26  \\
   ??Number density of ambient stars in the field & $n_{\rm amb}$ (as$^{-2}$)   & $f_{pri}  (F_{\rm amb})$                                &  7.3                  &     5.5             &    7.3                     &   7.3              &  4.0              \\
   ??Wide limit of undetectable stars in AO image & $\phi_{\rm wide}$ (mas)     & $\epsilon_{\rm amb}$, $\epsilon_{LC}$, $\epsilon_{SC}$  &  148                  &      132            &       60                   &   160              &     90             \\\hline
   For posterior PDF                              &                             &                                                         &                       &                     &                            &                    &                    \\
   ??Excess flux measured by AO imaging           & $H_{\rm ex, obs}$   (mag)   & $f'_{post} (F_L, F_{\rm amb}, F_{SC}, F_{LC})$\tnote{c} &  $19.7 \pm 0.4$       &  $19.6 \pm 0.4$     &  $19.16 \pm 0.13$          & $17.80 \pm 0.07$   & $17.52 \pm 0.10$  \\\hline
      \end{tabular}
     \begin{tablenotes}
     \small
      \item{\bf Notes.}
      All values of parameters other than ${\rm DM}_{\rm rc}$ and $n_{\rm amb}$ are from the indicated references. ${\rm DM}_{\rm rc}$ is from \citet{nat13} while we derived $n_{\rm amb}$ in \S \ref{sec-namb}. 
              The normal distribution is assumed for each parameter, but in flux unit for $H_{S, {\rm obs}}$ and $H_{\rm ex, obs}$.
     The blank (--) for $\theta_{\rm E, obs}$ or $\pi_{\rm E, obs}$ indicates no constraint on the parameter, i.e., an infinite uncertainty.
      \item{\bf References.} (1) \citet{kos17}; (2) \citet{jan10}; (3) \citet{yee12}; (4) \citet{bat14}; (5) \citet{fuk15}; (6) \citet{kos17b}.
      \item[a]  Abbreviation of $f_{pri} (M_L, D_L, D_S, v_t | t_{\rm E} = t_{\rm E, obs}, \theta_{\rm E} = \theta_{\rm E, obs}, \pi_{\rm E} = \pi_{\rm E, obs})$.
      \item[b]  A value given by \citet{bat14} using the extinction law of \citet{nis09}.
                They also used $A_H = 0.65 \pm 0.12$ estimated with the extinction law of \citet{car89}, which is not used in our analysis.
      \item[c]  Abbreviation of $f_{post} (F_L, F_{\rm amb}, F_{SC}, F_{LC} | F_{\rm excess} = F_{\rm ex, obs})$.
     \end{tablenotes}
      \end{threeparttable}
    \end{center}
    \label{tab-info}
  \hfill
}
 \end{landscape}

\clearpage

\footnotesize
\catcode`?=\active \def?{\phantom{0}}
 \tblcaption{Lens properties calculated from the posterior probability distributions assuming $\alpha = 0$.}
\vspace{-0.4cm}
\begin{center}
\begin{threeparttable}
\begin{tabular}{lllllllllllccccccccc}\hline\hline 
Event                                     & M16227                            &  M08310                      & M11293                            & O120563                              & O120950       \\
References                                &   1                               &  2, 7                        &   3, 4                            &       5                              &  6, 8  \\\hline
This work ($\alpha = 0$)                  &                                   &                              &                                   &                                      &   \\
??$M_L$         ($M_{\odot}$)             & $0.28^{+0.24}_{-0.15}$            & $0.14^{+0.27}_{-0.07}$       &  $0.41^{+0.35}_{-0.23}$           &  $0.37 \pm 0.12$                     &  $0.57^{+0.11}_{-0.20}$ \\
??$M_{\rm p}$                             & $2.7^{+2.4}_{-1.4} ~M_{\rm Jup}$  & $14^{+28}_{-7}  ~M_{\oplus}$ &  $2.3^{+1.9}_{-1.3} ~M_{\rm Jup}$ &  $0.43 \pm 0.13 ~M_{\rm Jup}$        &  $38^{+10}_{-14} ~M_{\oplus}$  \\
??$D_L$           (kpc)                   & $6.6^{+1.0}_{-0.9}$               & $7.3^{+1.2}_{-1.1}$          &  $7.3^{+0.9}_{-1.0}$              &  $1.6 \pm 0.5$                       &  $2.9^{+1.2}_{-0.7}$ \\
??$a_{\perp}$              (AU)                & $1.42 \pm 0.21$                   & $1.16^{+0.23}_{-0.20}$       &         --                        &              --                      &  $2.62^{+0.53}_{-0.56}$ \\
??$a_{\perp, \rm close}$   (AU) \tnote{a} &          --                       &          --                  &  $1.04 \pm 0.16$                  &  $0.82^{+0.23}_{-0.24}$              &            --          \\
??$a_{\perp, \rm wide}$    (AU) \tnote{a} &          --                       &          --                  &  $3.49^{+0.52}_{-0.54}$           &  $4.82^{+1.35}_{-1.39}$              &            --          \\
??$P(f_L > 0.1)$\tnote{b}                 &         74.9\%                   &            41.9\%           &           58.0\%                 &            99.99\%                 &    86.1\% \\
??$P(f_L > 0.5)$\tnote{b}                 &         37.8\%                   &            25.2\%           &           31.3\%                 &            99.94\%                  &    70.5\%  \\
??$P(f_L > 0.9)$\tnote{b}                 &         21.8\%                   &            19.3\%           &           21.9\%                 &            ?89.7\%                  &    52.3\% \\
??$H_L - H_{\rm excess}$ (mag) \tnote{c}  &     $1.30^{+1.72}_{-1.30}$        &   $3.12^{+2.97}_{-3.12}$     &   $2.12^{+1.46}_{-2.11}$          &    $0.005^{+0.068}_{-0.004}$         & $0.08^{+2.04}_{-0.08}$ \\\hline
Previous work                             &                                   &                              &                                   &                                      &   \\
??$M_L$         ($M_{\odot}$)             & $0.63 \pm 0.08$\tnote{d}          & $0.67 \pm 0.14$              &   $0.86 \pm 0.06$                 &  $0.34^{+0.12}_{-0.20}$              & $0.56^{+0.12}_{-0.16}$ \\
??$D_L$           (kpc)                   & $7.4 \pm 1.1$\tnote{d}            & $8.3^{+1.5}_{-1.2}$\tnote{e} &   $7.72 \pm 0.44$                 &  $1.3^{+0.6}_{-0.8}$                 & $3.0^{+0.8}_{-1.1}$ \\\hline
Result w/ {\it HST}                       &                                   &                              &                                   &                                      &   \\
??$M_L$         ($M_{\odot}$)             &          --                       & $0.21^{+0.21}_{-0.09}$       &         --                        &              --                      & $0.58 \pm 0.04$ \\
??$D_L$           (kpc)                   &          --                       & $7.7 \pm 1.1$                &         --                        &              --                      & $2.19 \pm 0.23$ \\\hline
    \end{tabular}
     \begin{tablenotes}
     \small
     \item{\bf Notes.}  Values given in the form of the median with 1-$\sigma$ uncertainty.
      \item{\bf References.} (1) \citet{kos17}; (2) \citet{jan10}; (3) \citet{yee12}; (4) \citet{bat14}; (5) \citet{fuk15}; (6) \citet{kos17b}; (7) \citet{bha17}; (8) \citet{bha18}.
     \footnotesize
    \item[a]  For events with degenerate models with largely different $s$ values, we give the projected separation value separately as $a_{\perp, \rm close}$ and $a_{\perp, \rm wide}$.
    \item[b]  Probabilities that the fraction of the lens flux to the excess flux,  $f_L \equiv F_L/F_{\rm excess}$, is larger than the indicated values. 
    \item[c]  Difference in magnitude, $H_L - H_{\rm excess} = -2.5~\log f_L$.
    \item[d]  Assumed $H_L = H_{\rm ex, obs}$. The values do not come from \citet{kos17}.
    \item[e]  Assumed $D_S = 8.6^{+1.5}_{-1.2}$ kpc, which is the same source distance that we used. \citet{jan10} just provided the $D_S -D_L$ value as  $\sim 0.3$ kpc.
    \end{tablenotes}
    \end{threeparttable}
    \end{center}
\label{tab-result}

\clearpage

\footnotesize
\catcode`?=\active \def?{\phantom{0}}
\tblcaption{Lens properties calculated from the posterior probability distributions assuming $\alpha = + 1 /- 1$.}
\vspace{-0.5cm}
\begin{center}
\begin{threeparttable}
    \begin{tabular}{lllllllllllccccccccc}\hline\hline 
Event                          & M16227                            &  M08310                       & M11293                           & O120563                             & O120950       \\
References                     &   1                               &  2                            &   3, 4                           &       5                             &  6  \\\hline
This work  ($\alpha = 1$)      &                                   &                               &                                  &                                     &   \\
??$M_L$         ($M_{\odot}$)  & $0.42^{+0.18}_{-0.21}$            & $0.34^{+0.30}_{-0.22}$        & $0.67^{+0.12}_{-0.37}$           & $0.41 \pm 0.11$                     & $0.60^{+0.09}_{-0.13}$        \\
??$M_{\rm p}$                  & $4.1^{+1.8}_{-2.0} ~M_{\rm Jup}$  & $36^{+33}_{-23} ~M_{\oplus}$  & $3.7^{+0.7}_{-2.0} ~M_{\rm Jup}$ & $0.47^{+0.12}_{-0.13} ~M_{\rm Jup}$ & $41^{+9}_{-10} ~M_{\oplus}$  \\
??$D_L$           (kpc)        & $6.9 \pm 0.9$                     & $7.8^{+1.2}_{-1.0}$           & $7.6 \pm 0.9$                    & $1.7 \pm 0.5$                       & $2.9^{+0.8}_{-0.7}$ \\
??$a_{\perp}$    (AU)          & $1.49^{+0.21}_{-0.19}$            & $1.25^{+0.23}_{-0.20}$        &            --                    &               --                    & $2.75^{+0.50}_{-0.48}$ \\
??$a_{\perp, \rm close}$ (AU)  &          --                       &          --                   &  $1.10^{+0.15}_{-0.14}$          &  $0.89^{+0.21}_{-0.22}$             &            --          \\
??$a_{\perp, \rm wide}$  (AU)  &          --                       &          --                   &  $3.66^{+0.49}_{-0.47}$          &  $5.25^{+1.22}_{-1.29}$             &            --          \\
??$P(f_L > 0.1)$               &         89.1\%                   &            69.1\%            &           80.2\%                &          100.00\%                  &    94.5\% \\
??$P(f_L > 0.5)$               &         55.0\%                   &            44.9\%            &           52.1\%                &          ?99.97\%                  &    80.9\%  \\
??$P(f_L > 0.9)$               &         31.8\%                   &            31.2\%            &           37.3\%                &           ?90.7\%                  &    61.5\% \\
??$H_L - H_{\rm excess}$ (mag) &   $0.59^{+1.55}_{-0.59}$          &      $1.09^{+2.69}_{-1.08}$   &    $0.61^{+2.12}_{-0.61}$        &   $0.003^{+0.064}_{-0.002}$         &  $0.02^{+0.94}_{-0.02}$ \\\hline
This work  ($\alpha = -1$)     &                                   &                               &                                  &                                     &   \\
??$M_L$         ($M_{\odot}$)  & $0.18^{+0.20}_{-0.08}$            & $0.096^{+0.096}_{-0.040}$     & $0.23^{+0.33}_{-0.11}$           & $0.33^{+0.13}_{-0.12}$              & $0.48^{+0.16}_{-0.33}$ \\
??$M_{\rm p}$                  & $1.7^{+1.9}_{-0.8} ~M_{\rm Jup}$  & $10^{+10}_{-4} ~M_{\oplus}$   & $1.3^{+1.8}_{-0.6} ~M_{\rm Jup}$ & $0.38^{+0.14}_{-0.13} ~M_{\rm Jup}$ & $32^{+13}_{-21} ~M_{\oplus}$  \\
??$D_L$           (kpc)        & $6.3 \pm 1.0$                     & $7.0^{+1.1}_{-1.0}$           & $6.8^{+1.0}_{-1.2}$              & $1.4 \pm 0.5$                       & $3.2^{+2.6}_{-1.0}$ \\
??$a_{\perp}$    (AU)          & $1.34 \pm 0.22$                   & $1.11^{+0.21}_{-0.19}$        &            --                    &               --                    & $2.36^{+0.62}_{-0.93}$ \\
??$a_{\perp, \rm close}$ (AU)  &          --                       &          --                   &  $0.96^{+0.17}_{-0.18}$          &  $0.74^{+0.25}_{-0.24}$             &            --          \\
??$a_{\perp, \rm wide}$  (AU)  &          --                       &          --                   &  $3.20^{+0.57}_{-0.59}$          &  $4.33^{+1.46}_{-1.40}$             &            --          \\
??$P(f_L > 0.1)$               &         56.7\%                   &            26.6\%            &           31.0\%                &           99.99\%                  &    66.6\% \\
??$P(f_L > 0.5)$               &         23.2\%                   &            16.7\%            &           12.7\%                &           99.88\%                  &    51.1\%  \\
??$P(f_L > 0.9)$               &         14.1\%                   &            14.6\%            &           ?8.7\%                &           ?88.5\%                  &    37.2\% \\
??$H_L - H_{\rm excess}$ (mag) & $2.21^{+1.63}_{-1.97}$            &  $4.27^{+2.26}_{-3.75}$       &    $3.20^{+1.07}_{-1.85}$        &   $0.005^{+0.075}_{-0.004}$         &  $0.68^{+4.42}_{-0.68}$ \\\hline 
     \end{tabular}
     \begin{tablenotes}
     \small
     \item{\bf Notes.}  Same as Table \ref{tab-result}, but for $\alpha = 1$ and $\alpha = -1$, where $\alpha$ is the slope of the planet hosting probability, $P_{\rm host} = M_h^{\alpha}$. See Table \ref{tab-result} for the references or the other notes.
    \end{tablenotes}
    \end{threeparttable}
    \end{center}
\label{tab-result2}

\normalsize
\catcode`?=\active \def?{\phantom{0}}
\tblcaption{Dependence of lens mass estimates on various prior choices when $\alpha = 0$.}
\vspace{-0.5cm}
\begin{center}
\begin{threeparttable}
    \begin{tabular}{lllllllllllccccccccc}\hline\hline
Difference from Tables \ref{tab-inputs}-\ref{tab-info}          & M16227                            &  M08310                       & M11293                           & O120563                             & O120950       \\\hline
?? None (fiducial) \tnote{a}                                    & $0.28^{+0.24}_{-0.15}$            & $0.14^{+0.27}_{-0.07}$        & $0.41^{+0.35}_{-0.23}$           & $0.37^{+0.12}_{-0.12}$              & $0.57^{+0.11}_{-0.20}$ \\
?? $\phi_{\rm wide}$ doubled                                    & $0.24^{+0.21}_{-0.11}$            & $0.13^{+0.21}_{-0.06}$        & $0.39^{+0.36}_{-0.21}$           & $0.37^{+0.12}_{-0.11}$              & $0.56^{+0.12}_{-0.23}$ \\
?? Isochrones for all mass range \tnote{b}                      & $0.28^{+0.25}_{-0.14}$            & $0.13^{+0.27}_{-0.07}$        & $0.43^{+0.32}_{-0.24}$           & $0.35^{+0.09}_{-0.10}$              & $0.56^{+0.11}_{-0.20}$ \\
?? Constant extinction ($A_{H,S} = A_{H,L} = A_{H, {\rm rc}}$)  & $0.28^{+0.24}_{-0.15}$            & $0.14^{+0.26}_{-0.07}$        & $0.41^{+0.35}_{-0.22}$           & $0.38^{+0.12}_{-0.12}$              & $0.57^{+0.11}_{-0.20}$ \\
?? Zero extinction ($A_{H,S} = A_{H,L} = 0$)                    & $0.27^{+0.23}_{-0.14}$            & $0.14^{+0.27}_{-0.08}$        & $0.43^{+0.25}_{-0.24}$           & $0.31^{+0.11}_{-0.10}$              & $0.52^{+0.12}_{-0.24}$ \\
?? B14 Galactic model                                           & $0.29^{+0.25}_{-0.15}$            & $0.15^{+0.28}_{-0.08}$        & $0.43^{+0.34}_{-0.23}$           & $0.38^{+0.12}_{-0.12}$              & $0.58^{+0.11}_{-0.19}$ \\
?? Z17 Galactic model                                           & $0.39^{+0.18}_{-0.17}$            & $0.23^{+0.29}_{-0.12}$        & $0.55^{+0.22}_{-0.27}$           & $0.36^{+0.12}_{-0.12}$              & $0.56^{+0.11}_{-0.15}$ \\
?? $\gamma = -0.1$ \tnote{c}                                    & $0.29^{+0.25}_{-0.15}$            & $0.14^{+0.28}_{-0.07}$        & $0.52^{+0.26}_{-0.32}$           & $0.37^{+0.12}_{-0.12}$              & $0.58^{+0.10}_{-0.16}$ \\
?? $\gamma = -0.5$ \tnote{c}                                    & $0.30^{+0.24}_{-0.16}$            & $0.15^{+0.30}_{-0.08}$        & $0.63^{+0.16}_{-0.41}$           & $0.38^{+0.12}_{-0.12}$              & $0.59^{+0.10}_{-0.14}$ \\
?? $\gamma = -0.9$ \tnote{c}                                    & $0.31^{+0.24}_{-0.17}$            & $0.15^{+0.32}_{-0.08}$        & $0.70^{+0.09}_{-0.46}$           & $0.38^{+0.12}_{-0.12}$              & $0.60^{+0.09}_{-0.13}$ \\
?? $\phi_{\rm close, SC} = 5 \, \theta_{\rm E}$                    & $0.30^{+0.24}_{-0.16}$            & $0.14^{+0.30}_{-0.08}$        & $0.56^{+0.23}_{-0.35}$           & $0.37^{+0.12}_{-0.12}$              & $0.59^{+0.10}_{-0.16}$ \\\hline
     \end{tabular}
     \begin{tablenotes}
     \item{\bf Notes.}  All values are given in $M_{\odot}$.
     \small
      \item[a]  Values from Table \ref{tab-result}.
      \item[b]  Age and metallicity distributions different from the fiducial ones are also applied. See Section \ref{sec-MLtest}.
      \item[c]  $\gamma =$ const. is applied to the binary distribution for a non-secondary star $f_{\rm prim}(q,a\,|\,M) \propto q^{\gamma}$. -0.1, -0.5, and -0.9 are approximately $1\sigma$, $2\sigma$, and $3\sigma$ lower limits on $\gamma$ by \citet{shv16}
     \end{tablenotes}
    \end{threeparttable}
    \end{center}
\label{tab-Mdepend}

\clearpage
\normalsize
\catcode`?=\active \def?{\phantom{0}}
 \tblcaption{Removed fractions of binary companions considered as detectable when considering the prior probability distribution.}
\vspace{-0.4cm}
\begin{center}
\begin{threeparttable}
    \begin{tabular}{lrrrrrrccccccccc}\hline\hline
Event                                      & M16227  &  M08310  &  M11293  & O120563  & O120950   \\\hline
$\phi_{\rm wide}$ (mas)                    &    148  &     132  &      60  &     160  &     90    \\
$u_0$                                      &   0.08  &   0.003  &  0.0035  &   0.001  &   0.10    \\
$\theta_{\rm E}$ (mas)                     &   0.23  &    0.16  &    0.26  &     1.4  &  $\sim 0.5$ \\\hline
$P_{{\rm det}, SC}$                        &  0.135  &   0.135  &   0.160  &   0.203  &  0.193    \\
??--$P_{{\rm det}, SC_{\rm close}}$        &  0.066  &   0.058  &   0.077  &   0.141  &  0.106    \\
??--$P_{{\rm det}, SC_{\rm wide}}$         &  0.069  &   0.077  &   0.083  &   0.062  &  0.087    \\\hline
$P_{{\rm det}, LC}$                        &  0.273  &   0.357  &   0.372  &   0.442  &  0.334    \\
??--$P_{{\rm det}, LC_{\rm close}}$        &  0.254  &   0.348  &   0.336  &   0.342  &  0.261    \\
???? --w/ factors (i)-(iv)\tnote{a}        &  0.085  &   0.169  &   0.163  &   0.199  &  0.085    \\
???? --w/ factors (i)-(iii)\tnote{a}       &  0.257\tnote{b}  &   0.336  &   0.330  &   0.309  &  0.269\tnote{b}    \\
??--$P_{{\rm det}, LC_{\rm wide}}$         &  0.019  &   0.009  &   0.036  &   0.100  &  0.073    \\\hline
    \end{tabular}
     \begin{tablenotes}
     \small
     \item{\bf Notes.}  Fractions are given with respect to all the scenarios including both rejected and accepted scenarios in the Monte Carlo simulation for each event.
     \footnotesize
     \item[a] The factors (i)-(iv) are explained in Section \ref{sec-pdetlow}.
     \item[b] These are larger than the original $P_{{\rm det}, LC_{\rm close}}$ values because of the simulated detection
              efficiencies in Fig. \ref{fig-DEs} that have wider sensitivity than the approximated formula.
    \end{tablenotes}
    \end{threeparttable}
    \end{center}
\label{tab-fracrej}

\end{document}